\documentclass[aps,prd,final,groupedaddress,showpacs,showkeys,floatfix,preprintnumbers]{revtex4}

\usepackage[latin1]{inputenc}                    % Codepage latin1
\usepackage{graphicx}                            % Graphics package
\usepackage{latexsym}                            % Load additional symbols
\usepackage{amsfonts}                            % Use AMS fonts,
\usepackage{amssymb}                             % symbols,
\usepackage{amsmath}                             % and definitions
\usepackage[mathscr]{eucal}                      % Euler font symbols
\usepackage{dcolumn}                             % Decimal-dot aligned
\usepackage{theorem}                            % Allow use of theorems
\usepackage{hyperref}                            % References as links

\graphicspath{{.}{Graphics/}}                    % Path for graphics
\newcommand{\imag}{i}                     %  imaginary unit
\newcommand{\msbar}{$\overline{\mbox{\rm MS}}$}  % MS-bar
\newcommand{\fslash}[1]{\slash\!\!\!{#1}}        % Feynman slash
\newcommand{\ts}{\tau}                           % Timeslice
\newcommand{\dlangle}{\langle\!\langle}          % Double langle
\newcommand{\drangle}{\rangle\!\rangle}          % Double rangle
\newcommand{\bea}{\begin{eqnarray}}
\newcommand{\eea}{\end{eqnarray}}
\newcommand{\Dlr}{\stackrel{\leftrightarrow}{D}}

\hypersetup{%
     debug=false,                                % Don't be verbose
     a4paper=true,                               % Use A4 paper
     pdfpagemode={UseOutline},                   % Open outline in
     pdftitle={Nucleon Generalized Parton Distributions from Full Lattice QCD},                  % Title of publication
     pdfauthor={LHPC collaboration},         % Group of authors
     pdfsubject={Phys Rev D publication},        % Type of document
     pdfkeywords={12.38.Gc,13.60.Fz,Generalized parton distribution,
                  lattice matrix elements}       % Keywords
     }

\theoremstyle{plain}
\theorembodyfont{\slshape}
\theoremheaderfont{\scshape}

\begin{document}

% Setup title page
\preprint{JLAB-THY-07-651, MIT-CTP-3841, TUM-T39-07-09} 
\title[GPDs]{Nucleon Generalized Parton Distributions from Full Lattice QCD}

  \author{Ph.~H{\"a}gler}
  \affiliation{Institut f\"ur Theoretische Physik T39, Physik-Department der TU M\"unchen, James-Franck-Stra\ss{}e,
D-85747 Garching, Germany}
\author{W.~Schroers\footnote{Present address: Department of Physics, Center for Theoretical Sciences,
 National Taiwan University, Taipei 10617, Taiwan}}
  \affiliation{John von Neumann-Institut f\"ur Computing NIC/DESY, D-15738 Zeuthen, Germany}
\author{ J.~Bratt} 
 \affiliation{Center for Theoretical Physics, Massachusetts Institute of Technology, Cambridge, MA 02139}
\author{ R.G.~Edwards} 
  \affiliation{Thomas Jefferson National Accelerator Facility, Newport News, VA 23606}
\author{M.~Engelhardt}
 \affiliation{Physics Department, New Mexico State University, Las Cruces, NM 88003-8001}
\author{G.T.~Fleming}
  \affiliation{Sloane Physics Laboratory, Yale University, New Haven, CT 06520}
\author{B.~Musch}
 \affiliation{Institut f\"ur Theoretische Physik T39, Physik-Department der TU M\"unchen, James-Franck-Stra\ss{}e,
D-85747 Garching, Germany }
\author{J.W.~Negele}
  \affiliation{Center for Theoretical Physics, Massachusetts Institute of Technology, Cambridge, MA 02139} 
\author{K.~Orginos}
  \affiliation{Department of Physics, College of William and Mary,
Williamsburg VA 23187-8795 \\
and Thomas Jefferson National Accelerator Facility, Newport News, VA 23606}
\author{A.V.~Pochinsky}
   \affiliation{Center for Theoretical Physics, Massachusetts Institute of Technology, Cambridge, MA 02139}
\author{D.B.~Renner\footnote{Present address: DESY Zeuthen, Theory Group, Platanenallee 6, D-15738 Zeuthen, Germany}} 
  \affiliation{ Department of Physics, University of Arizona, 1118 E 4th Street, Tucson, AZ 85721}
\author{D.G.~Richards}
  \affiliation{Thomas Jefferson National Accelerator Facility, Newport News, VA 23606}
\author{(LHPC Collaboration)}
\noaffiliation
   \date{\today}

\begin{abstract}
  We present a comprehensive study of the lowest moments of nucleon generalized
  parton distributions in $N_f=2+1$ lattice QCD 
using  domain  wall valence quarks and improved staggered sea quarks.
  Our investigation includes helicity dependent and independent generalized parton distributions for pion masses
  as low as 350 MeV and volumes as large as (3.5 fm)$^3$, for a lattice spacing of $0.124$~fm.
  We use perturbative renormalization at one-loop level with an improvement based on the non-perturbative 
  renormalization factor for the axial vector current,  and only connected diagrams are included in the isosinglet channel. 
\end{abstract}



\maketitle

% --------------------------------------------------------------------------
%
%  Introduction
%
% --------------------------------------------------------------------------
%{ \bf  Version 2.5f}

\section{\label{sec:Introduction}Introduction}

 Generalized parton distributions (GPDs)
  \cite{Mueller:1998fv,Ji:1996nm,Radyushkin:1997ki,Diehl:2003ny} 
 play a  vital role 
 in our understanding of  the structure of the nucleon in terms
  of the fundamental building blocks of QCD, the quarks and gluons.
  Before the advent of GPDs, fundamental questions as to the origin of the spin of
  the nucleon, the decomposition of the nucleon total momentum, and the distribution and density
  of the nucleon constituents in 
  position and momentum space seemed to be largely unrelated.
  In some cases, it was even unclear how to formulate these questions in a theoretically
  sound way and how to measure the underlying observables 
  experimentally.
  With the introduction of GPDs, it is possible not only to give  
  precise definitions to quantities, such as the quark and gluon angular momentum contributions 
  to the nucleon spin \cite{Ji:1996ek}
  and 
  the probability densities of quarks and gluons in impact parameter space \cite{Burkardt:2000za}, but
  also to unify and extend the successful concepts of parton distribution functions (PDFs) and
  form factors. Nucleon generalized parton distributions are experimentally accessible
  in deeply virtual Compton scattering of virtual photons off a nucleon and
  a range of other related processes 
  \cite{Stepanyan:2001sm,Airapetian:2001yk,Chekanov:2003ya,Aktas:2005ty}  . 
  Since these reactions involve in
  general convolutions of GPDs over the longitudinal momentum fraction $x$,
  which makes it difficult if not impossible to map them over the whole
  parameter space, the most stringent quantitative information on GPDs
  currently comes from quark PDFs and nucleon form factors \cite{Diehl:2004cx}. 
  
  Complementary to 
  experimental efforts, lattice QCD offers 
  a unique opportunity to calculate
  $x$-moments of GPDs from first principles.
  The first investigations of GPDs including
  studies of the quark angular momentum contributions to the nucleon spin have
  been presented by the QCDSF collaboration in quenched QCD \cite{Gockeler:2003jf} and by LHPC/SESAM
  in $N_f=2$ lattice QCD \cite{Hagler:2003jd}. Lattice results on nucleon GPDs 
  published since then   
  have provided important insights into the transverse structure of
  unpolarized nucleons \cite{LHPC:2003is}, 
  the lowest moments of polarized \cite{Schroers:2003mf}
  and tensor GPDs \cite{Gockeler:2005cj}, and
  transverse spin densities of quarks in the nucleon \cite{Diehl:2005ev,Gockeler:2006zu}. With the exception of   
  several initial studies \cite{Renner:2004ck,Edwards:2005kw}, all 
  previously published lattice results on GPDs have
  been obtained from 
  calculations in a two-flavor "heavy pion world" with pion
  masses in the range of $550$ to over $1000$ MeV. In this work, we improve on
  previous studies by presenting a comprehensive analysis of the lowest three
  moments of unpolarized and polarized GPDs in $N_f=2+1$ lattice QCD with pion
  masses as low as $350$ MeV and volumes as large as $(3.5 \text{ fm})^3$.

  The paper is organized as follows.
  We begin with an introduction to the calculation of moments of GPDs in lattice QCD in section II.
  Section III describes the hybrid approach of  
  using domain wall valence quarks with 2+1 flavors of improved staggered sea quarks.
  In section IV we present our numerical results for the generalized
  form factors, including a discussion and interpretation of 
  the quark orbital angular momentum and the transverse nucleon structure.   
Chiral extrapolations of selected lattice results to the physical pion mass are presented in section V.     
 Conclusions are given in the final section.

% --------------------------------------------------------------------------
%
%  Lattice techniques: 2pts and 3pts
%
% --------------------------------------------------------------------------

\section{\label{sec:lattice-techniques}Lattice Calculation of
Moments of Generalized Parton Distributions}

\begin{figure}[t]
% \vspace*{-0.75cm}
      \includegraphics[scale=1,clip=true,angle=0]{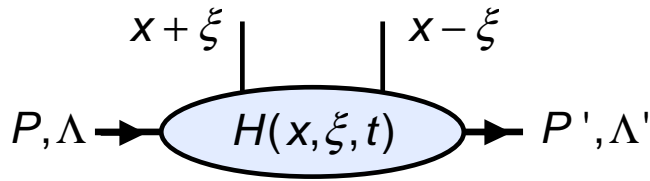}
%      \vspace*{-0.8cm}
  \caption{GPDs as part of a scattering amplitude.}
  \label{handbag1}
   \vspace*{-.5cm}
\end{figure}

 Generalized parton distributions determine off-forward matrix elements of
 gauge invariant light cone operators
\begin{equation}
 {\cal O}_\Gamma(x) =\int \!\frac{d \lambda}{4 \pi} e^{i \lambda x} \overline
  q (\frac{-\lambda n}{2})
  \Gamma \,{\cal P} e^{-ig \int_{-\lambda / 2}^{\lambda / 2} d \alpha \, n
    \cdot A(\alpha n)}\!
  q(\frac{\lambda n}{2}),
  \label{LCop}
\end{equation}
  where $x$ is the momentum fraction, $n$ is a light cone vector and $\Gamma = \fslash{n}$
  or $\Gamma =\fslash{n} \gamma_5$. The twist-2 tensor GPDs \cite{Diehl:2001pm} related to
  $\Gamma = n_\mu \sigma^{\mu j},\,j=1,2$ are not studied in this work.
  The four independent twist-2 unpolarized and polarized generalized parton distributions
  $H$, $E$, $\widetilde H$ and $\widetilde E$ are defined by the parametrizations
\bea
  \left\langle P',\Lambda ' \right|
  {\cal O}_{\not n}(x)
  \left| P,\Lambda \right \rangle
  = \dlangle\fslash{n}  \drangle H(x, \xi, t)
 + \frac{ n_\mu\Delta_\nu} {2 m}\dlangle i \sigma^{\mu \nu}\drangle E(x, \xi, t) \,,
\label{VectorOp}
\eea
and
\bea
  \left\langle P',\Lambda ' \right|
  {\cal O}_{\not n\gamma_5}(x)
  \left| P,\Lambda \right \rangle
  = \dlangle\fslash{n}\gamma_5 \drangle \widetilde H(x, \xi, t)
 + \frac{ n\cdot\Delta } {2 m}\dlangle \gamma_5\drangle \widetilde E(x, \xi, t) \,,
\label{AxialVectorOp}
\eea
where we use the short-hand notation $\dlangle \Gamma \drangle=\overline U(P',\Lambda ')\Gamma U(P,\Lambda)$
for products of Dirac spinors $U$, and where $\Delta=P'-P$, $t=\Delta^2$ and $\xi=-n\cdot\Delta/2$.
In Eqs.~(\ref{VectorOp}) and (\ref{AxialVectorOp}) we suppress the dependence of the GPDs on the
resolution scale $\mu^2$. An illustration of the GPDs parametrizing the lower part of the handbag
diagram is given in Fig.~(\ref{handbag1}).
The momentum fractions $x$ and $\xi$ both have support in the interval  $[-1,+1]$.
Depending on $x$, there are three kinematic  
regions, which offer different interpretations for the GPDs.
For $x\in [\xi,1]$ and $x\in [-1,-\xi]$, the GPDs describe the emission and reabsorption of quarks and anti-quarks,
respectively. In the case that $x$ lies in the interval $[-\xi,\xi]$, they describe the emission of
a $q \overline q$-pair.

In our lattice 
calculations, we do not work directly with the bi-local operators in Eq.~(\ref{LCop})
but instead consider moments, defined by the integral $\int_{-1}^1dx\,x^{n-1}f(x)$, of the operators in Eqs.~(\ref{VectorOp},\ref{AxialVectorOp}),
leading to towers of symmetrized, traceless local operators
\begin{equation}
 {\cal O }_{[\gamma_5]}^{\{\mu_1\ldots\mu_n\}}
  =\overline q (0) \gamma^{\{\mu_1} [\gamma_5] i\!\Dlr\!\phantom{}^{\mu _{2}}\!\cdots i\!\Dlr\!\phantom{}^{\mu _{n}\}} q(0)\,,
  \label{LocalOp}
\end{equation}
where $[\gamma_5]$ denotes the possible inclusion of the corresponding matrix, the curly brackets
represent a symmetrization over the indices $\mu_i$ and subtraction of traces, and
$\Dlr=1/2 (\stackrel{\rightarrow}{D}-\stackrel{\leftarrow}{D})$.
We relate nucleon matrix elements of the tower of local operators in Eq.~(\ref{LocalOp})
to $x$-moments of the twist-2 GPDs. To this end, we parametrize off-forward matrix elements
$ \langle P',\Lambda '|{\cal O }^{\{\mu_1\ldots\mu_n\}}| P,\Lambda\rangle $ in terms of the generalized form
factors $A_{ni}(t)$, $B_{ni}(t)$, $C_{n0}(t)$, $\widetilde A_{ni}(t)$, and $\widetilde B_{ni}(t)$.
Apart from potential difficulties related to lattice operator mixing for higher moments $n$,
lattice measurements of the operators in Eq.~(\ref{LocalOp}) become 
increasingly noisy as the number of derivatives 
increases, and we therefore restrict our calculations to $n\le 3$.
The parametrization of nucleon matrix elements of Eq.~(\ref{LocalOp}) in terms of 
generalized form factors (GFFs) for $n=1$, $2$ and $3$ reads
\cite{Diehl:2003ny,Hagler:2004yt}
\bea
 \langle P' | {\cal O}^{\mu_1} | P \rangle
 &=&\dlangle \gamma^{\mu_1 }\drangle A_{10}(t)
  + \frac{\imag}{2 m} \dlangle \sigma^{\mu_1
    \alpha} \drangle
  \Delta_{\alpha} B_{10}(t)\,, \nonumber \\ [.5cm]
 \langle P' | {\cal O}^{\lbrace \mu_1 \mu_2\rbrace} | P \rangle
 &=& \bar P^{\lbrace\mu_1}\dlangle
  \gamma^{\mu_2\rbrace}\drangle A_{20}(t)
  + \frac{\imag}{2 m} \bar P^{\lbrace\mu_1} \dlangle
  \sigma^{\mu_2\rbrace\alpha}\drangle \Delta_{\alpha} B_{20}(t)
  +\frac{1}{m}\Delta^{\{ \mu_1}   \Delta^{ \mu_2 \} }
  C_{20}(t)\,, \nonumber \\[.5cm]
  \langle P' | {\cal O}^{\lbrace\mu_1 \mu_2 \mu_3\rbrace} | P \rangle
  &=& \bar P^{\lbrace\mu_1}\bar P^{\mu_2} \dlangle
  \gamma^{\mu_3\rbrace}
  \drangle A_{30}(t)
  + \frac{\imag}{2 m} \bar P^{\lbrace \mu_1}\bar P^{\mu_2}
  \dlangle \sigma^{\mu_3\rbrace\alpha} \drangle
  \Delta_{\alpha} B_{30}(t)
  \nonumber \\
  &+& \Delta^{\lbrace \mu_1}\Delta^{\mu_2} \dlangle
  \gamma^{\mu_3\rbrace}\drangle A_{32}(t)
  + \frac{\imag}{2 m} \Delta^{\lbrace\mu_1}\Delta^{\mu_2}
  \dlangle \sigma^{\mu_3\rbrace\alpha}\drangle
  \Delta_{\alpha} B_{32}(t),
  \label{Para1}
\eea
for the unpolarized 
case, and
\bea
 \langle P' | {\cal O}_{\gamma_5}^{\mu_1} | P \rangle
 &=&\dlangle \gamma^{\mu_1 }\gamma_5\drangle \widetilde A_{10}(t)
  + \frac{1 } {2 m} \Delta^{\mu_1}\dlangle \gamma_5 \drangle
  \widetilde B_{10}(t)\,, \nonumber \\ [.5cm]
 \langle P' | {\cal O}_{\gamma_5}^{\lbrace \mu_1 \mu_2\rbrace} | P \rangle
 &=& \bar P^{\lbrace\mu_1}\dlangle
  \gamma^{\mu_2\rbrace}\gamma_5\drangle \widetilde A_{20}(t)
  + \frac{ 1 } {2 m} \Delta^{\lbrace\mu_1} \bar P^{\mu_2\rbrace} \dlangle
  \gamma_5 \drangle  \widetilde B_{20}(t) \,,
  \nonumber \\[.5cm]
  \langle P' | {\cal O}_{\gamma_5}^{\lbrace\mu_1 \mu_2 \mu_3\rbrace} | P \rangle
  &=& \bar P^{\lbrace\mu_1}\bar P^{\mu_2} \dlangle
  \gamma^{\mu_3\rbrace}\gamma_5
  \drangle \widetilde A_{30}(t)
  + \frac{1} {2 m} \Delta^{\lbrace\mu_1} \bar P^{\mu_2} \bar P^{\mu_3\rbrace}
  \dlangle \gamma_5\drangle
  \widetilde B_{30}(t)
  \nonumber \\
  &+& \Delta^{\lbrace \mu_1}\Delta^{\mu_2} \dlangle
  \gamma^{\mu_3\rbrace}\gamma_5\drangle \widetilde A_{32}(t)
  + \frac{ 1} {2 m} \Delta^{\lbrace\mu_1}\Delta^{\mu_2} \Delta^{\mu_3\rbrace}
  \dlangle\gamma_5\drangle \widetilde B_{32}(t),
  \label{Para2}
\eea
for the polarized case. Here and in the 
following we set $\bar P=(P'+P)/2$.

Using Eqs.~(\ref{VectorOp},\ref{AxialVectorOp},\ref{Para1},\ref{Para2}) it is easy
to show that Mellin-moments of the GPDs,
%\bea
\begin{align}
 H^n(\xi, t) &\equiv \int_{-1}^{1} dx\, x^{n-1} H(x, \xi, t),\,\,\,
 & E^n(\xi, t) &\equiv \int_{-1}^{1} dx\, x^{n-1} E(x, \xi, t)\, ,  \nonumber \\
 \widetilde H^n(\xi, t) &\equiv \int_{-1}^{1} dx\, x^{n-1} \widetilde H(x, \xi, t),\,\,\,
 & \widetilde E^n(\xi, t) &\equiv \int_{-1}^{1} dx\, x^{n-1} \widetilde E(x, \xi, t)\,,
  \label{moments1}
\end{align}
%\eea
are given by polynomials in the longitudinal momentum transfer $\xi$ and the GFFs.
For the lowest three moments, the corresponding relations read
%\bea
\begin{align}
 H^{n=1}(\xi, t) &= A_{10}(t),\quad & H^{n=2}(\xi, t) &= A_{20}(t)+(2\xi)^2 C_{20}(t),\quad
 & H^{n=3}(\xi, t)  &= A_{30}(t)+(2\xi)^2 A_{32}(t)\, , \nonumber \\
 E^{n=1}(\xi, t) &= B_{10}(t),\quad  & E^{n=2}(\xi, t) &= B_{20}(t)-(2\xi)^2 C_{20}(t),\quad
 & E^{n=3}(\xi, t) &= B_{30}(t)+(2\xi)^2 B_{32}(t)\, , \nonumber \\
 \widetilde H^{n=1}(\xi, t) &= \widetilde A_{10}(t),\quad & \widetilde H^{n=2}(\xi, t) &= \widetilde A_{20}(t),\quad
 & \widetilde H^{n=3}(\xi, t) &= \widetilde A_{30}(t)+(2\xi)^2 \widetilde A_{32}(t)\, , \nonumber \\
 \widetilde E^{n=1}(\xi, t) &=\widetilde B_{10}(t),\quad & \widetilde E^{n=2}(\xi, t) &= \widetilde B_{20}(t),\quad
 & \widetilde E^{n=3}(\xi, t) &= \widetilde B_{30}(t)+(2\xi)^2 \widetilde B_{32}(t)\, .
  \label{moments2}
\end{align}
%\eea
The aim of our calculation is to extract the GFFs as functions of the momentum transfer squared,
$t$, from nucleon two- and three-point-functions as described below.
Once the GFFs have been obtained, the complete $\xi$-dependence of the moments of
the GPDs is directly given by Eqs.~(\ref{moments2}).
Let us note that the Mellin-moments in Eq.~(\ref{moments1})
are taken with respect to the entire interval from $x=-1$ to $+1$. Following our discussion
below 
Eq.~(\ref{AxialVectorOp}), we find that the moments of the GPDs at $\xi=0$
correspond to sums and differences of contributions from quarks $q$ and anti-quarks
$\overline q$.  For example,
\bea
 H^n_q(\xi=0, t) \!&=&\! \int_{0}^{1} \!dx\, x^{n-1} \big ( H_q(x, 0, t)
 + (-1)^n H_{\overline q}(x,0, t) \big )\,,\nonumber\\
 E^n_q(0, t) \!&=&\! \int_{0}^{1} \!dx\, x^{n-1} \big ( E_q(x,0, t)
 + (-1)^n E_{\overline q}(x, 0, t) \big )\, ,  \nonumber \\
 \widetilde H^n_q(0, t) \!&=&\! \int_{0}^{1} \!dx\, x^{n-1} \big ( \widetilde H_q(x, 0, t)
 + (-1)^{(n-1)} \widetilde H_{\overline q}(x, 0, t) \big )\,,\nonumber\\
 \widetilde E^n_q(0, t) \!&=&\! \int_{0}^{1} \!dx\, x^{n-1} \big ( \widetilde E_q(x, 0, t)
 + (-1)^{(n-1)} \widetilde E_{\overline q}(x, 0, t) \big )\, .
  \label{moments3}
\eea
Such a simple decomposition is not possible for non-zero longitudinal momentum transfer $\xi\neq 0$.
We denote the forward limit values of the moments of $H$ and $\widetilde H$ in Eq.~(\ref{moments3})
by $\langle x^{n-1}\rangle_q=H_q^n(0, 0)=A^q_{n0}(0)$ and 
$\langle x^{n-1}\rangle_{\Delta q}=\widetilde H_q^n(0, 0)=\widetilde A^q_{n0}(0)$, where 
$\langle x^{n-1}\rangle_q$ and $\langle x^{n-1}\rangle_{\Delta q}$ correspond to the moments of unpolarized
and polarized quark parton distributions.

Below, we give a brief  
summary of the methods and techniques used to extract moments
of generalized parton distributions in lattice QCD. For 
details,  we refer the reader to \cite{Dolgov:2002zm, Hagler:2003jd}.
As usual,  the matrix elements, Eqs.~(\ref{Para1},\ref{Para2}), are calculated from the ratio of nucleon three-point and two-point functions:

\begin{figure}[t]
% \vspace*{-0.75cm}
        \includegraphics[scale=0.83,clip=false,angle=0]{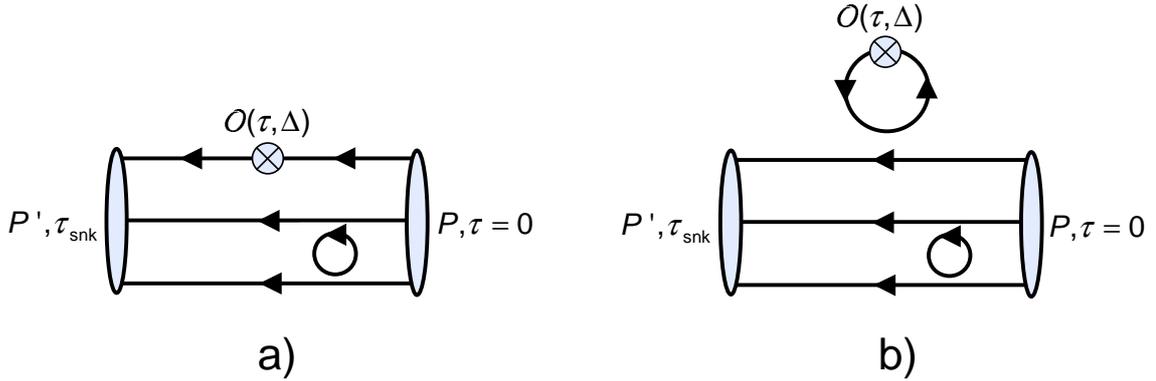}
%      \vspace*{-0.8cm}
  \caption{Connected (a)  and disconnected (b) diagrams in  unquenched lattice QCD with an operator insertion at $\tau$ and finite momentum transfer $\Delta$.}
  \label{latticediag1}
   \vspace*{-.5cm}
\end{figure}

\begin{eqnarray}
  \label{eq:n-point-def}
  C^{\text{2pt}}(\ts,P) &=& \sum_{j,k}\left(
    \Gamma_{\text{unpol}}\right)_{jk}\langle
  N_{k}(\ts,P)\overline{N}_{j}(\ts_\text{src},P) \rangle\,,
  \nonumber\\
  C_{\cal O}^{\text{3pt}}(\ts,P',P) &=& \sum_{j,k}\left(
    \Gamma_{\text{pol}}\right)_{jk}\langle N_{k}(\ts_{\text{snk}},P')
    {\cal O}(\ts,\Delta)\overline{N}_{j}(\ts_\text{src},P)\rangle\,,
\end{eqnarray}
where $\Gamma_{\text{unpol}}=(1+\gamma_4)/4$ and 
$\Gamma_{\text{pol}}=(1+\gamma_4)(1+i\gamma_5\gamma_3)/2$.
The nucleon source, $\bar{N}(\ts,P)$, and sink, $N(\ts,P)$, create and annihilate
states with the quantum numbers of the nucleon. To maximize the
overlap with the ground state, we used the smeared sources given in \cite{Dolgov:2002zm}. 
The three-point-function $C_{\cal O}^{\text{3pt}}(\ts,P',P)$
with the operator insertion at $\tau$ is illustrated in Fig.~(\ref{latticediag1})
in terms of quark propagators, showing examples of connected and 
disconnected contributions in an unquenched lattice calculation.

Using the transfer matrix formalism, we can rewrite Eqs.~(\ref{eq:n-point-def}) to obtain
\begin{eqnarray}
  \label{eq:expand-2pt}
  C^{\text{2pt}}(\ts,P) &=& e^{-E_0(P)(\ts-\ts_{\text{src}})} \left(
      Z(P)\overline{Z}(P)\right)^{1/2}\frac{E_0(P)+m}{E_0(P)}+ \mbox{higher states} \,, \\
  \label{eq:expand-3pt}
  C_{\cal O}^{\text{3pt}}(\ts,P',P) &=&
  e^{-E_0(P)(\ts-\ts_{\text{src}})-E_0(P')(\ts_{\text{snk}}-\ts)}
  \frac{\left(Z(P)\overline{Z}(P')\right)^{1/2}}{4 E_0(P')E_0(P)}\mbox{Tr}\left\{
    \Gamma_{\text{pol}}(i\fslash{P}'-m)\big(a A(t) +b B(t)+\cdots \big)(i\fslash{P}-m)
    \right\} \nonumber\\
    &+& \mbox{higher states} \, ,
\end{eqnarray}
where the factors $a,b,\ldots$ represent the prefactors (including Dirac-matrices) of the
corresponding GFFs $A(t),B(t),\ldots$ in the parametrizations in Eqs.~(\ref{Para1},\ref{Para2}),
transformed to Euclidean space. Higher states with energies $E_1>E_0$ in Eqs.~(\ref{eq:expand-2pt})
and (\ref{eq:expand-3pt}) are suppressed when $\ts_{\text{snk}}-\ts \gg 1/(E_1-E_0)$ and $\ts-\ts_\text{src}\gg 1/(E_1-E_0)$.

In order to cancel the exponentials and $Z$-factors in Eq.~(\ref{eq:expand-3pt}) 
for zero and non-zero momentum transfer $\Delta$,
we construct the ratio of two- and three-point-functions
\begin{equation}
  \label{eq:ratio}
  R_{\cal O}(\ts,P',P) = \frac{C_{\cal O}^{\text{3pt}}(\ts,P^{\prime
    },P)}
  {C^{\text{2pt}}(\ts_{\text{snk}},P^{\prime })}\left[
    \frac{C^{\text{2pt}}(\ts_{\text{snk}}-\ts+\ts_{\text{src}},P)
      \;C^{\text{2pt}}(\ts,P')\;C^{\text{2pt} }(\ts_{\text{snk}},P')}%
    {C^{\text{2pt}}(\ts_{\text{snk}}-\ts+\ts_{\text{src}},P')%
      \;C^{\text{2pt}}(\ts,P)\;C^{\text{2pt}}(\ts_{\text{snk}},P)}
  \right]^{1/2}\,.
\end{equation}
For an operator-insertion sufficiently far away from
the source and the sink in  
the Euclidean time 
direction, higher states are negligible, and the ratio $R_{\cal O}(\ts,P',P)$
exhibits a plateau in $\tau$. 
We finally average over the plateau region from
$\ts_{\text{min}}$ to $\ts_{\text{max}}$ to obtain
an averaged ratio $\overline {R}_{\cal O}(P',P)$. On a finite periodic lattice with spatial extent $L_s$,
three-momenta are given by $\vec P = 2\pi/(a L_s)\vec n$ with integer components $n_i = -L_s/2,\ldots,L_s/2$,
and for the nucleon energy we use the continuum dispersion relation $P_4=\sqrt{m^2+\vec P^2}$.
Therefore, the discrete lattice momenta result in a finite set of values for the momentum transfer squared $t$
which can be realized in our calculation.

In order to obtain symmetric and traceless operators ${\cal O}^{\{ \mu_1\mu_2\ldots\}}$,
we have to choose specific linear combinations of the indices. For the diagonal operators typical examples
are  ${\cal O }_{i=1}^{n=2}=({\cal O }^{11}+{\cal O }^{22}-{\cal O }^{33}-{\cal O }^{44})/2$  and
${\cal O }_{i=2}^{n=3}=({\cal O }^{122}+{\cal O }^{133}-2{\cal O }^{144})/\sqrt{2}$, where  ${\cal O }_1^{n=2}$
belongs to the $3$-dimensional irreducible $H(4)$-representation $\tau_1^{(3)}$ for $n=2$ and ${\cal O }_{i=2}^{n=3}$
is a member of the $8$-dimensional representation $\tau_1^{(8)}$ for $n=3$ \cite{Gockeler:1996mu}. 
The sets of operators ${\cal O }_{i}^{n}$ we are using are the same as in \cite{Hagler:2003jd}.
Altogether, there are $9$ linearly independent index combinations for $n=2$ and $12$ for $n=3$.
In order to be able to compare our results with experiment, the operators have to be renormalized
and transformed to the \msbar-scheme at a renormalization scale $\mu^2$. In general, operators mix
under renormalization, and the renormalization matrix $Z^{\cal O}$ is non-diagonal. 
We will denote the renormalized operators in the \msbar-scheme by
${\cal O }_{i}^{n,\overline{\text{MS}}}=Z_{ij}^{\cal O}{\cal O }_{j}^{n}$.
Some details concerning the renormalization procedure and numerical results for the renormalization constants
will be discussed at the end of the next section.

Based on the renormalized operators, we compute the averaged ratio $\overline{R}_{\cal O}(P',P)$
and equate it with the continuum parametrization in terms of the GFFs
given in Eq.~(\ref{eq:expand-3pt}). This is done simultaneously for all momentum
combinations $P$ and $P'$ corresponding to the same momentum transfer squared $t$
and all contributing symmetric and traceless operators ${\cal O }_{i}^{n,\overline{\text{MS}}}$,
giving a finite set of linear equations
\bea
\overline{R}_{{\cal O},k}(P_1',P_1)&=&c_{11} A(t) + c_{12} B(t) + \ldots\,, \nonumber \\
\overline{R}_{{\cal O},l}(P_2',P_2)&=&c_{21} A(t) + c_{22} B(t) + \ldots\,,  \nonumber \\
\overline{R}_{{\cal O},m}(P_3',P_3)&=&c_{31} A(t) + c_{32} B(t) + \ldots\,, \nonumber \\
&\cdots&\,,
\label{setofeqs1}
\eea
where $(P_j'-P_j)^2=t$ for all $j=1,2,3\ldots$.
The coefficients $c_{ij}$ in Eqs.~(\ref{setofeqs1}) only depend on the nucleon mass $m$ and the momenta $P,P'$
and are calculated from the traces in Eq.~(\ref{eq:expand-3pt}).
Finally, the set of equations (\ref{setofeqs1}), which in general is overdetermined, is solved numerically to extract the GFFs.
The statistical errors for the GFFs are obtained from a jackknife analysis.
%
% --------------------------------------------------------------------------
%
%  Lattice techniques: Hybrid approach
%
% --------------------------------------------------------------------------

\section{\label{sec:hybrid-approach}Lattice simulation using Domain
  wall valence quarks with staggered sea quarks}
Since calculations at physical quark masses  are
prohibitively expensive with current algorithms and machines, we have used dynamical quark configurations at the lightest masses available, and where feasible, have used chiral perturbation theory to extrapolate to the physical mass.  Staggered sea quarks with the Asqtad improved action were chosen because the computational economy of staggered quarks enabled the MILC collaboration to generate large samples of configurations at low masses on large spatial volumes~\cite{Orginos:1998ue,Orginos:1999cr}, which they freely made available to the lattice community. 

Chiral symmetry is crucial for avoiding some operator mixing, convenient for operator renormalization, and valuable for chiral extrapolation. Furthermore, the four tastes associated with staggered fermions immensely complicate calculating operator matrix elements in  nucleon states.  Hence, we chose a hybrid action utilizing chirally symmetric valence domain wall fermions (DWF) on an improved staggered fermion sea.   Although this hybrid scheme breaks unitarity at finite lattice spacing, given the arguments that 
the valence and sea actions separately approach the physical continuum limit\cite{Sharpe:2006re}, we expect that the hybrid action also approaches the physical continuum limit.
Furthermore,  partially quenched mixed action chiral perturbation theory calculations are now becoming available for quantitative control of the continuum limit.  We also note that hybrid actions have been successfully used in other contexts where, for example,  the NRQCD action for  valence quarks was combined with improved staggered sea quarks\cite{Davies:2003ik} and was successful in predicting  mass splitting in heavy quark systems.

In our calculation, we used MILC configurations~\cite{Bernard:2001av}
both from the NERSC archive and provided directly by the
collaboration. We then applied HYP-smearing~\cite{Hasenfratz:2001hp}
and bisected the lattice in the time direction. 
We have chosen gauge fields separated by 6 trajectories. 
Furthermore, we alternate between the first temporal half (time slices 0 to 31) 
and the second temporal half (time slices 32 to 63) on successive gauge configurations.
In these samples we did not find residual autocorrelations. 
The scale is set by the lattice spacing $a = 0.1241$ fm determined from heavy quark spectroscopy \cite{Aubin:2004wf} 
with an uncertainty of 2$\%$.

Domain-wall fermions~\cite{Kaplan:1992bt,Shamir:1993zy,Narayanan:1992wx} introduce
an additional fifth dimension, $L_5$. They preserve the Ward-Takahashi
identity \cite{Furman:1994ky} even at finite lattice spacing in the
limit $L_5\to\infty$. At finite values of $L_5$ a residual
explicit breaking of chiral symmetry is still present which can be
parameterized by an additional mass term in the Ward-Takahashi
identity \cite{Blum:2000kn,Blum:1998ud}. 
In our calculations, we have kept this additional mass,
$(am)_{\text{res}}$, 
at least an order of magnitude smaller
than the quark mass, $(am)_q^{\text{DWF}}$ \cite{Renner:2004ck}. 
To the extent that $(am)_{\text{res}}$ is negligible, perturbative renormalization of ${\cal O}_{[\gamma_5]}$  is independent of $\gamma_5$ in the chiral limit and 
the non-perturbative
renormalization of quark bilinear currents yields the 
same renormalization coefficients 
for the axial and the vector currents 
in the chiral limit \cite{Blum:2001sr}.

We now consider the parameters entering the DWF action. The domain wall action realizes chiral symmetry by producing right-handed states on one domain wall that decay exponentially away from the wall and exponentially decaying left handed states on the other wall.  To the extent that no low eigenmodes associated with dislocations (or rough fields) destroy the exponential decay and that the fifth dimension  $L_5$ is large enough, chiral symmetry will be nearly exact and $(am)_{\text{res}}$ will be small.  HYP smearing was essential to reduce the effect of low eigenmodes, but it is still necessary to use spectral flow to determine a value of the domain wall mass,  $M$, for which the density of low eigenmodes is as small as possible.  This was done on an ensemble of test configurations, with the result that we use $M = 1.7$. As discussed below, $L_5$ was then tuned to keep $(am)_{\text{res}}$ below 10\% of the quark mass and to have negligible effect on our lattice observables.  Finally, the quark mass was tuned to produce a pion mass equal to the Goldstone pion mass for the corresponding MILC configurations.

\subsection{\label{sec:tuning-fifth-dimens}Tuning the fifth dimension}
The extent of the fifth dimension, $L_5$, 
has been adjusted such that the residual mass, $(am)_{\text{res}}$ is at least an order of
magnitude smaller than the quark mass itself. This tuning is most relevant
at the lightest quark mass since in that case the computational cost
is largest and thus our tuning should be optimal. In addition, the
breaking of chiral symmetry is also expected to be the largest and the
resulting $L_5$ provides a minimum value needed for our calculations at the higher masses.
The residual explicit chiral symmetry breaking characterized by
$(am)_{\text{res}}$ is obtained from~\cite{Blum:2000kn}
\begin{equation}
  \label{eq:ward-taka}
  \Delta^\mu {\cal A}_\mu^a = 2m_q J_q^a(x) + 2J_{5q}^a(x)\,,
\end{equation}
where
\begin{equation}
  \label{eq:mres-def}
  J_{5q}^a(x) \approx m_{\text{res}} J_5^a(x)\,,
\end{equation}
which holds up to ${\cal O}(a^2)$.

We have run simulations using two samples of $25$ configurations with
volume $\Omega=20^3\times 32$: three degenerate dynamical Asqtad
quarks with  
bare masses $(am)_q^{\text{Asqtad,sea}}=0.050$ (denoted as
``heavy'' 
and corresponding to $m_\pi \sim 760$ MeV) and two plus one quarks with masses
$(am)_q^{\text{Asqtad,sea}}= 0.010 , 0.050$ (termed ``light'' 
and corresponding to $m_\pi \sim 350$ MeV). The
corresponding bare DWF masses have been adjusted to
$(am)_q^{\text{DWF}}=0.0810$ and $0.0138$ for the heavy and light
cases, respectively, cf.~Sec.~\ref{sec:tuning-quark-mass}.
\begin{figure}[th]
%  \label{fig:mres}
     \includegraphics[scale=0.25,clip=true]{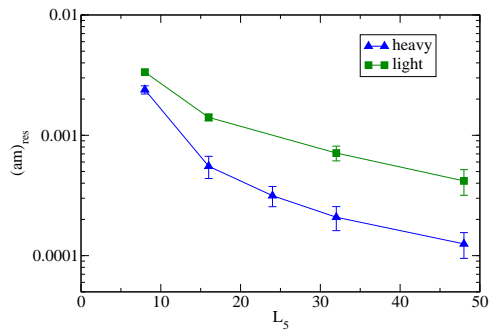}
  \caption{Residual quark mass as a function of $L_5$ for the two
    samples (heavy and light) of $25$ configurations each.}
    \label{fig:mres}
\end{figure}

The resulting residual masses obtained from Eqs.~(\ref{eq:ward-taka})
and (\ref{eq:mres-def}) are plotted in Fig.~(\ref{fig:mres}). 
In the
light quark case, $L_5=16$ just fulfills our requirement, while in the
heavy quark case $L_5=16$ more than satisfies it. This confirms our
expectation that the value of $L_5$ chosen at the lightest quark mass
sets the lower limit for the other masses as well.

\begin{figure}[htbp]
     \begin{minipage}{0.4\textwidth}
      \centering
        \includegraphics[scale=0.3,clip=true]{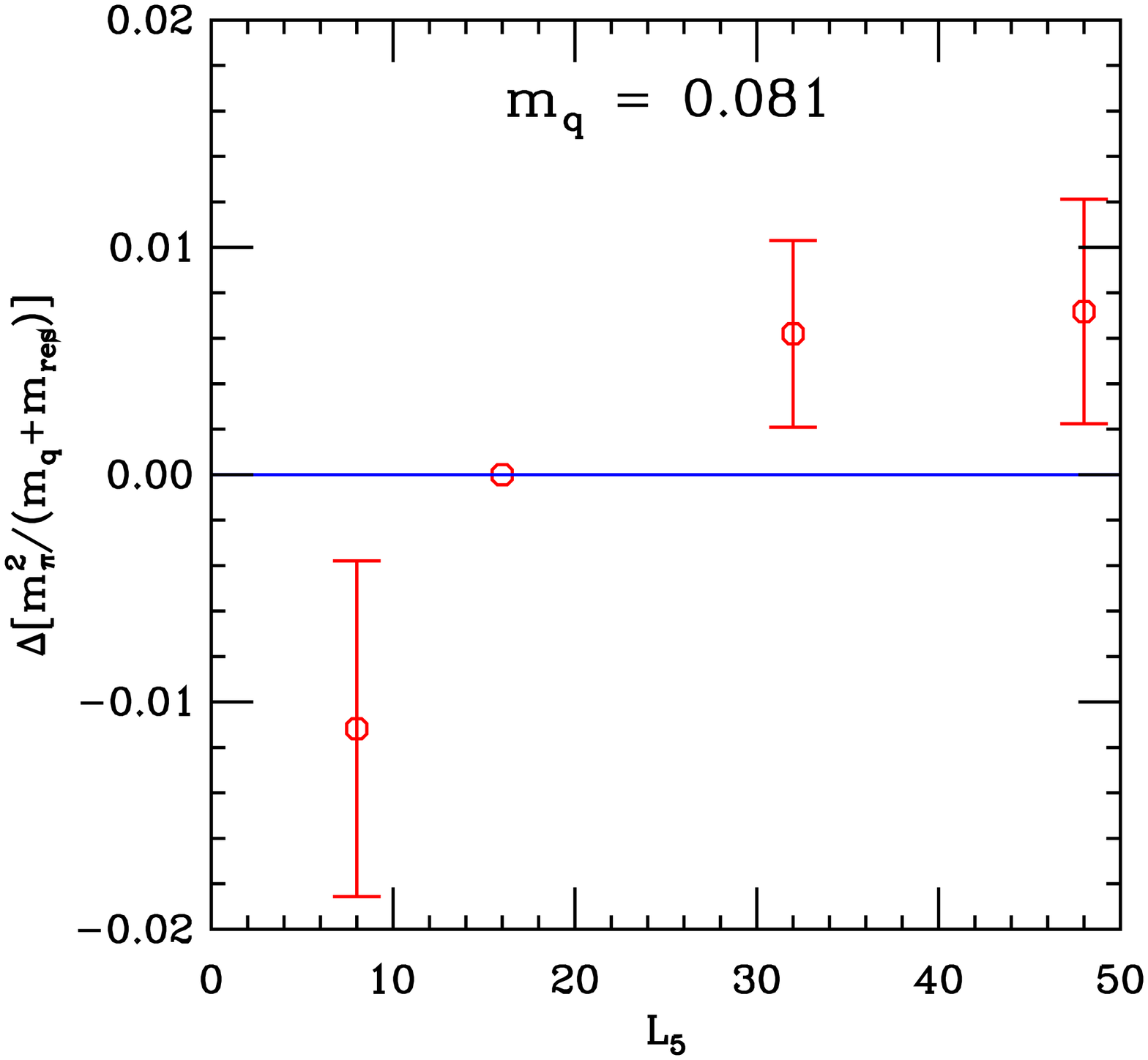}
  \caption{  Dependence of the pion mass on the extent of the fifth dimension $L_5$ for heavy quarks.}
     \label{fig:massesL5v1}
     \end{minipage}
     %\hfill
     \hspace{1cm}
     \begin{minipage}{0.4\textwidth}
      \centering
       \includegraphics[scale=0.3,clip=true]{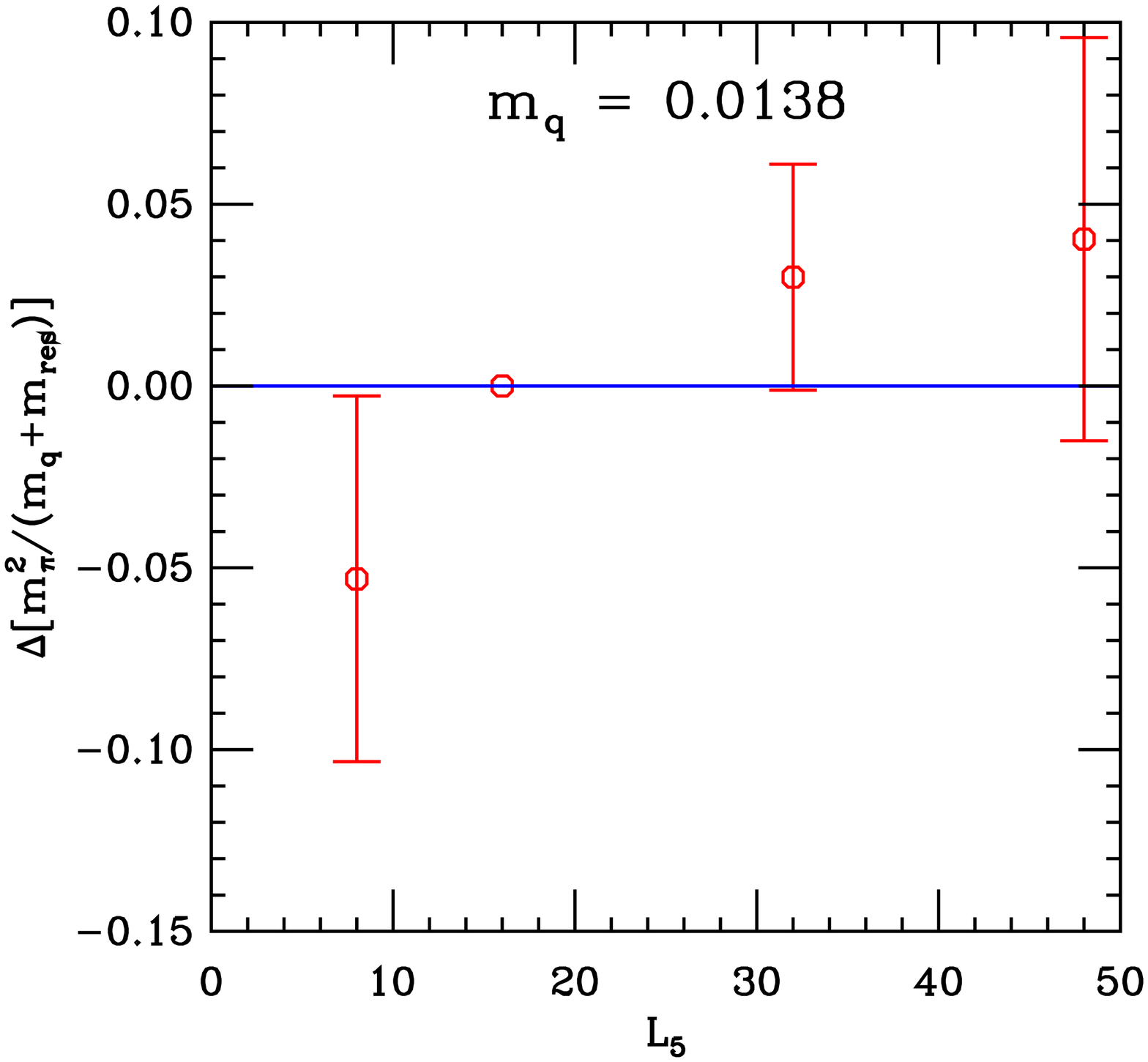}
  \caption{  Dependence of the pion mass on the extent of the fifth dimension $L_5$ for light quarks.}
      \label{fig:massesL5v2}
     \end{minipage}
   \end{figure}

\begin{figure}[htbp]
     \begin{minipage}{0.4\textwidth}
      \centering
         \includegraphics[scale=0.3,clip=true]{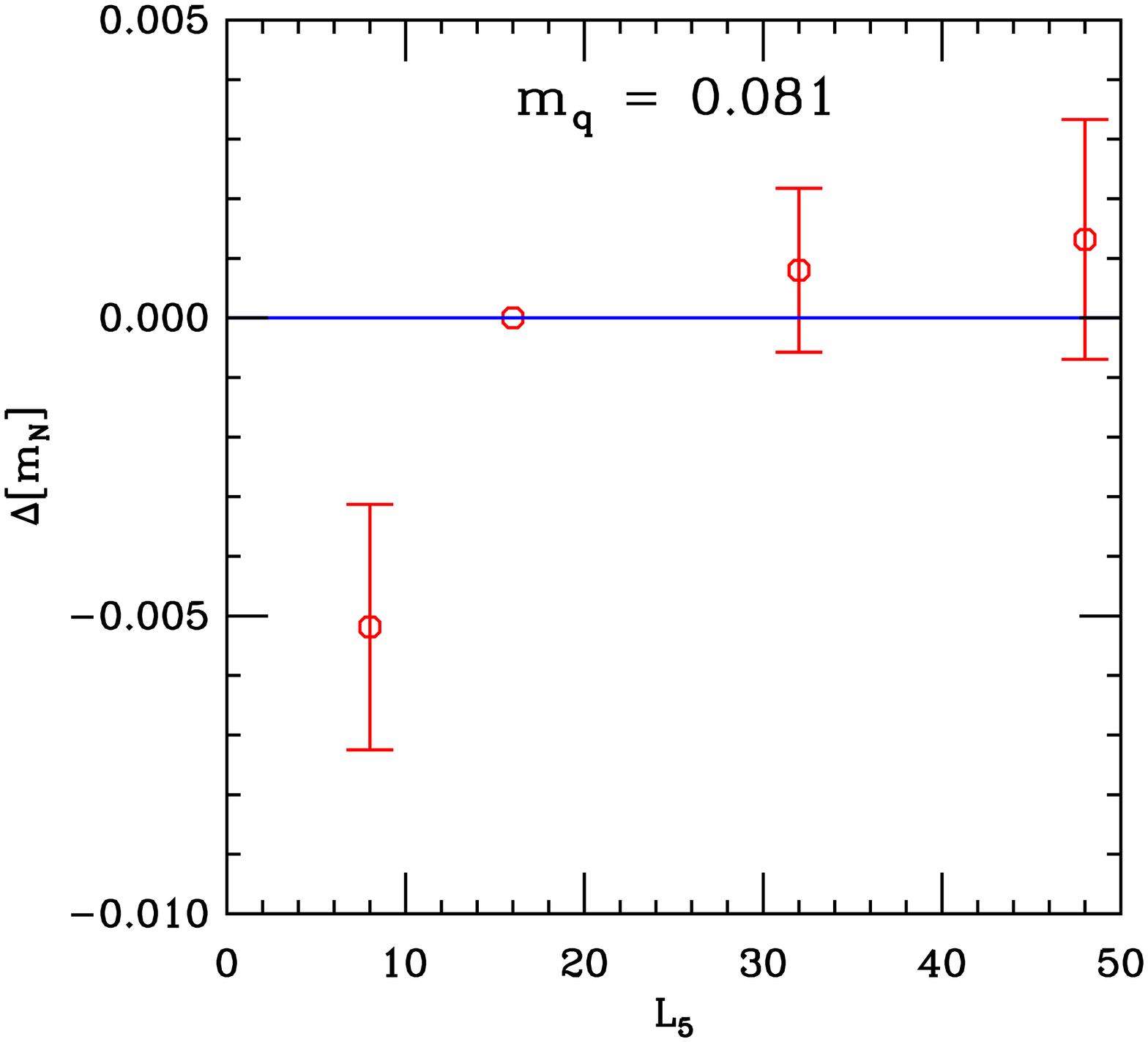}
  \caption{  Dependence of the nucleon mass on the extent of the fifth dimension $L_5$ for heavy quarks.}
     \label{fig:massesL5v3}
     \end{minipage}
     %\hfill
     \hspace{1cm}
     \begin{minipage}{0.4\textwidth}
      \centering
       \includegraphics[scale=0.3,clip=true]{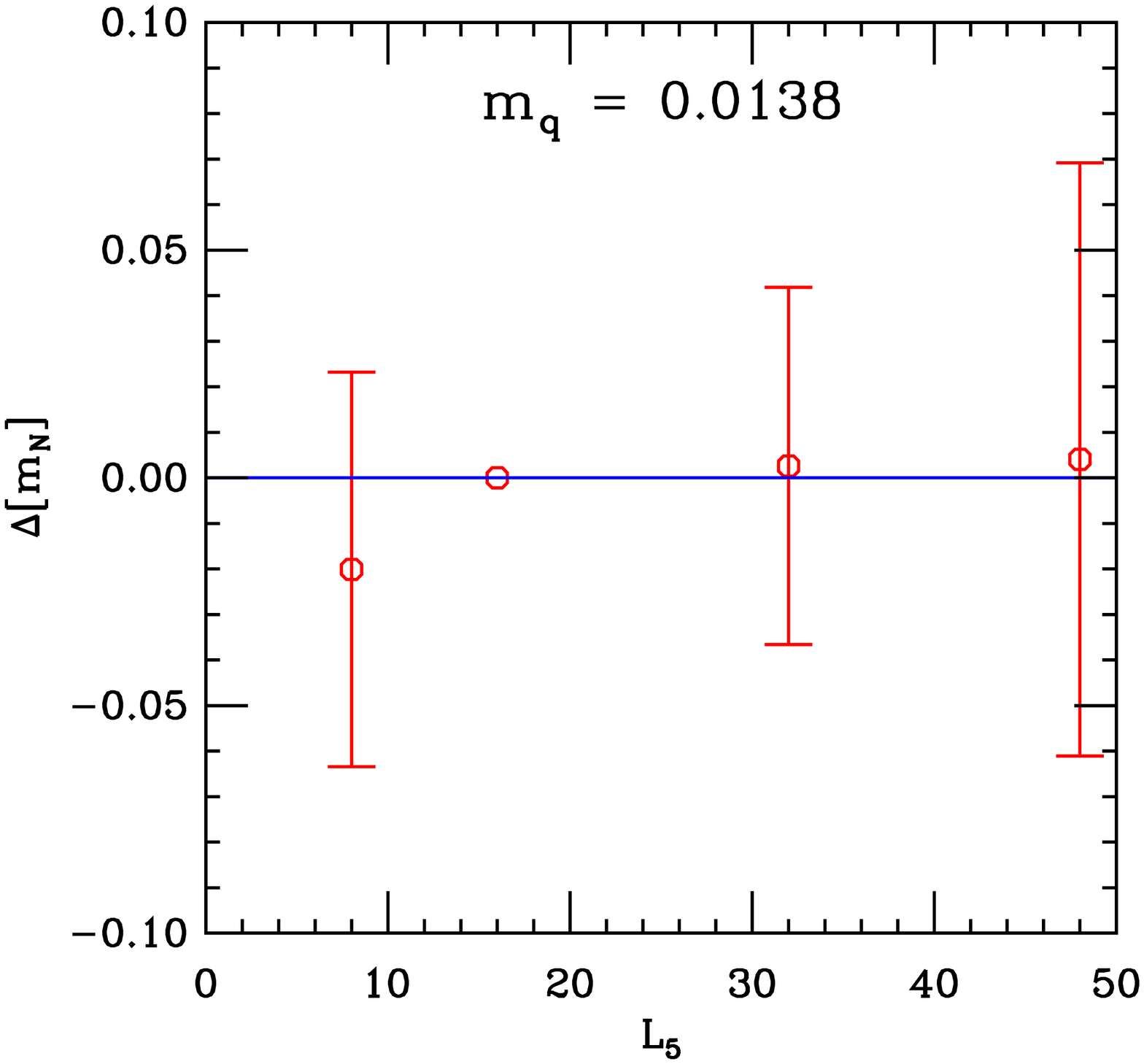}
  \caption{Dependence of the nucleon mass on the extent of the fifth dimension $L_5$ for light quarks.}
        \label{fig:massesL5v4}
     \end{minipage}
   \end{figure}

One quantitative check that $L_5 = 16$ is adequate is provided by the dependence of masses on $L_5$ as shown in  Figs.~(\ref{fig:massesL5v1}), (\ref{fig:massesL5v2}),  (\ref{fig:massesL5v3})
and  (\ref{fig:massesL5v4}). The leading effect of $m_\text{res}$ is to shift the quark mass, so that when $L_5$ is sufficiently large that this is the only effect, $m_\pi^2 \propto (m_q + m_\text{res})$.  Figs.~(\ref{fig:massesL5v1}) and (\ref{fig:massesL5v2}) show the difference in the ratio $m_\pi^2/(m_q+m_\text{res})$ at a general value of $L_5$ and at $L_5 = 16$, and indicate that the difference is essentially consistent with zero for $L_5 > 16$.  We expect the shifts in the 
 nucleon mass induced by these small shifts in the pion mass to be negligible, and indeed, Figs.~(\ref{fig:massesL5v3}) and  (\ref{fig:massesL5v4}) show that the differences between the nucleon mass at a general value of $L_5$ and at $L_5 = 16$ are consistent with zero for $L_5 > 16$.  
Hence, we choose $L_5=16$ to be a good compromise between
accuracy and performance.

\subsection{\label{sec:tuning-quark-mass}Tuning the quark mass}

We define the light quark masses in our hybrid theory by matching the pion mass in two
calculations: (i) using two plus one flavors of dynamical Asqtad
sea fermions and Asqtad valence fermions\cite{Bernard:2001av} and (ii) using 
the pion mass in our hybrid calculation with Asqtad dynamical sea fermions and valence domain-wall fermions with
$L_5=16$. Because of the four tastes and correspondingly sixteen light pseudoscalar mesons in the staggered theory, it is necessary to choose between matching the lightest pseudoscalar mass, corresponding to the Goldstone pion of the theory, or some appropriately defined average. In this work, we have chosen to match the Goldstone pion, and the results of tuning the domain wall quark mass such that the domain  wall pion mass agrees within  one percent with the Asqtad Goldstone pion mass are shown in Table~\ref{tab:parameters-summary}. The substantial difference between the bare quark masses for
Asqtad and DWF valence quarks reflects the significant difference in renormalization  for the two actions.
An observable physical difference is the fact that once the DWF quark masses have been adjusted to  fit the Asqtad Goldstone pion masses, the DWF nucleon masses are approximately 6\% lower than the corresponding Asqtad nucleon masses.  We attribute this to the range of pseudoscalar masses in the staggered theory and note that had we used a heavier quark mass so that the DWF pion fit some appropriately weighted average of the staggered pion masses, 
then the DWF nucleon would have been heavier.
\begin{table}[t]
\centering
\begin{tabular}{|c|c|c|c|c|c|c|c|c|c|}
  \hline
 dataset$\phantom{I^{\vec I}}$ & $\Omega$ & $\#$ &
    $(am)_q^{\text{Asqtad}}$ & 
    $(am)_q^{\text{DWF}}$ &
    $(am)_{\pi}^{\text{Asqtad}}$ & 
    $(am)_{\pi}^{\text{DWF}}$ &
    $(am)_N^{\text{Asqtad}}$ &
    $(am)_N^{\text{DWF}}$ & 
%    $m_N^{\text{DWF}} [MeV]$ & &
    $m_{\pi}^{\text{DWF}}$  [MeV] 
    \\ \hline
 1 & $20^3\times 32$ & $425$ &  $0.050/0.050$ & $0.0810$ & $0.4836(2)$ & $0.4773(9)$ & $1.057(5)$ & $0.986(5)$ & $758.9(1.4)$   \\
        2 &  & $350$ & $0.040/0.050$ & $0.0478$ & $0.4340(3)$ & $0.4293(10)$  & $1.003(3)$ & $0.938(8)$ & $682.6(1.6)$   \\
         3 & & $564$ & $0.030/0.050$ &  $0.0644$ & $0.3774(2)$ & $0.3747(10)$ & $0.930(3)$ & $0.869(6)$ & $595.8(1.6)$   \\
        4 &  & $486$ & $0.020/0.050$ &  $0.0313$ & $0.3109(2)$ & $0.3121(11)$ & $0.854(3)$ & $0.814(7)$  & $496.2(1.7)$  \\
         5 & & $655$ & $0.010/0.050$ &  $0.0138$ & $0.2242(2)$ & $0.2243(10)$ & $0.779(6)$ & $0.730(12)$ & $356.6(1.6)$  \\  6 & $28^3\times 32$ & $270$ & $0.010/0.050$ &  $0.0138$ &      $$ & $0.2220(9)$ &          & $0.766(15)$  & $352.3(1.4)$ \\ \hline
\end{tabular}
\caption{ The lattice volume $\Omega$ and number of configurations used for the DWF calculations and a comparison of the quark, pion, and nucleon masses in the DWF and Asqtad calculations as described in the text.}
\label{tab:parameters-summary}
\end{table}
\begin{figure}[t]
% \vspace*{-0.75cm}
        \includegraphics[scale=0.8,clip=true,angle=0]{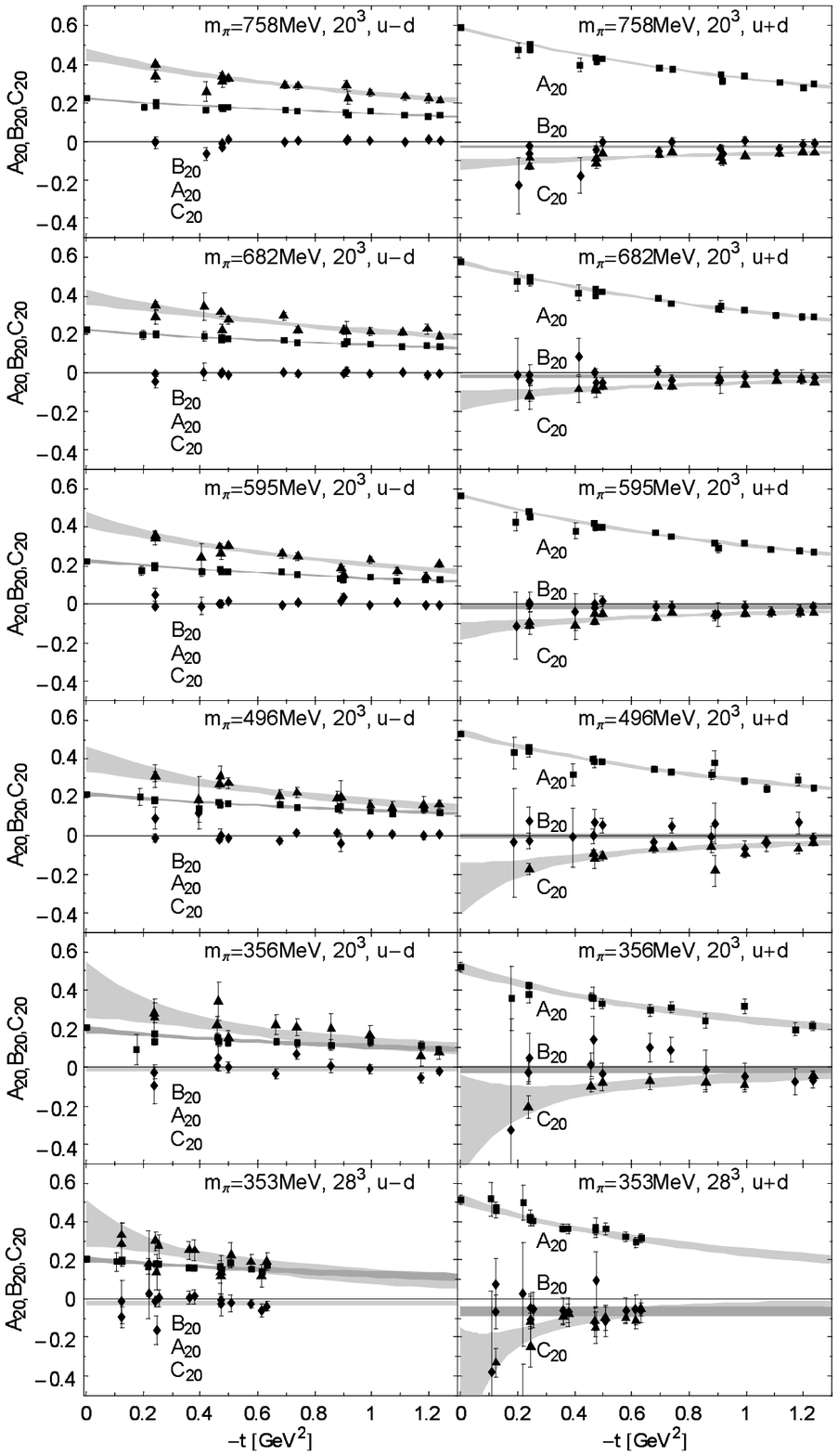}
%      \vspace*{-0.8cm}
  \caption{ Unpolarized (vector) generalized $n=2$ form factors for the flavor combinations 
  $u-d$ (left) and $u+d$ (right). Disconnected contributions are not included.}
  \label{ABC1st}
\end{figure}
\begin{figure}[t]
% \vspace*{-0.75cm}
        \includegraphics[scale=0.8,clip=true,angle=0]{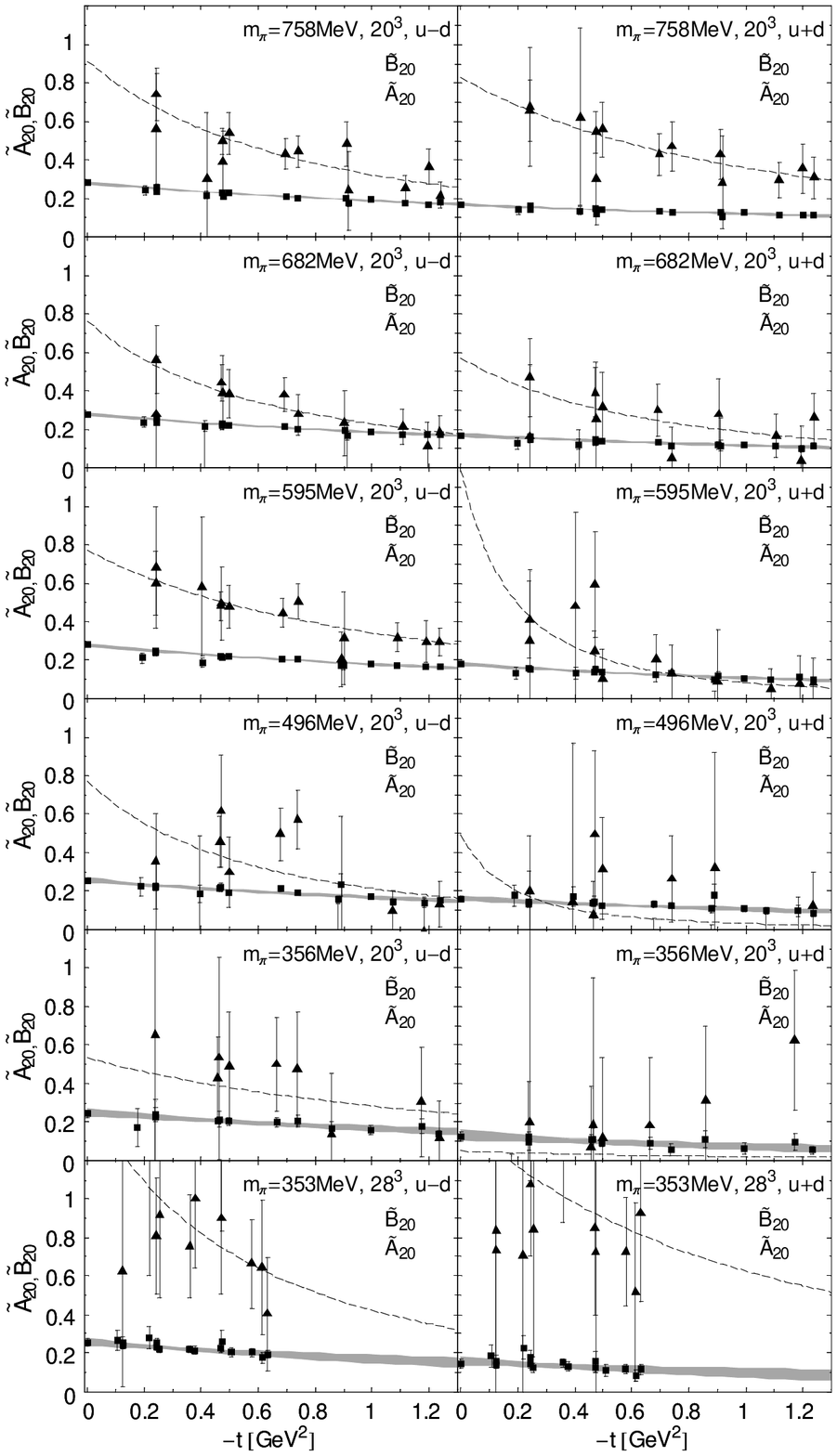}
%      \vspace*{-0.8cm}
  \caption{Polarized (axial vector) generalized $n=2$ form factors for the flavor combinations 
  $u-d$ (left) and $u+d$ (right). Disconnected contributions are not included.}
  \label{ABpol1st}
\end{figure}

\subsection{\label{sec:renormalization}Operator renormalization}

The quark bilinear operators in Eq.~[\ref{LocalOp}] are renormalized using a combination of one-loop perturbation theory and non-perturbative renormalization of the axial vector current.  

Our lattice calculations using lattice regularization with cutoff $1/a$ are related to physical observables at scale $\mu^2$ in the
$\overline{\text{MS}}$  renormalization scheme in 1-loop perturbation theory by
\begin{eqnarray}
\mathcal{O}^{\overline{\text{MS}}}_i(\mu^2) &=&\sum_j\left(\delta_{ij}+\frac{g_0^2}{16\pi^2}\,
\frac{N_c^2-1}{2N_c}
\left(\gamma^{\overline{\text{MS}}}_{ij}\log(\mu^2a^2)-(B^{LATT}_{ij}
-B^{\overline{\text{MS}}}_{ij})\right)\right)\cdot \mathcal{O}^{LATT}_j(a^2) \nonumber \\
&=& Z^\mathcal{O}_{ij} \cdot \mathcal{O}^{LATT}_j(a^2),
\label{app-renorm1}
\end{eqnarray}
where the  anomalous dimensions $\gamma_{ij}$ and the finite constants $B_{ij}$ have been calculated for domain wall fermions with HYP smearing in Refs.~\cite{Bistrovic:2005aa,Bistrovic:2007ab}. 
Because the renormalization factors for operators with and without $\gamma_5$ are identical at quark mass zero, we use mass independent renormalization with all renormalization constants defined at quark mass zero.
The renormalization factors $Z^\mathcal{O}_{ij}$ for domain wall mass $M=1.7$ used in this work are tabulated in Table \ref{tab:Zs}, using the results for the one-loop coupling constant $g^2/(12 \pi^2)=1/53.64$ 
from Refs.~\cite{Bistrovic:2005aa,Bistrovic:2007ab}. 
\begin{table}[ht]
  \begin{tabular}{|l|l|l|}
    \hline 
    operator$\phantom{I^{\vec I}}$ & $H(4)$ & $Z^{\cal O,\text{pert}}$ \\ \hline
    $\bar q[\gamma_5]\gamma_{\{\mu} D_{\nu\}} q$ \,\,& $\tau^{(3)}_1$ & $0.962$ \\
    $\bar q[\gamma_5]\gamma_{\{\mu} D_{\nu\}} q$ \,\,& $\tau^{(6)}_1$ & $0.968$ \\
    $\bar q[\gamma_5]\gamma_{\{\mu} D_{\nu} D_{\rho\}} q$ \,\,& $\tau^{(4)}_2$ & $0.980$ \\
    $\bar q[\gamma_5]\gamma_{\{\mu} D_{\nu} D_{\rho\}} q$ \,\,& $\tau^{(8)}_1$ & $0.982$ \\
     \hline
  \end{tabular}
  \caption{Perturbative 1-loop lattice renormalization constants for the $\overline{\text{MS}}$ scheme at a scale $\mu^2=1/a^2$.}
      \label{tab:Zs}
\end{table}
By virtue of the suppression of loop
integrals by HYP smearing, the ratio of the one-loop perturbative
renormalization factor for a general bilinear operator to the
renormalization factor for the axial current is within a few percent
of unity, suggesting adequate convergence for this ratio at one-loop level. Since one element in the calculation common to all operators arising from the wave function renormalization in the fifth dimension is not small, it is desirable to determine this one common factor non-perturbatively. This is accomplished using the fact that the
 renormalization factor, $Z_A$, for the four dimensional axial current operator $A_\mu =   \bar q \gamma_\mu \gamma_5 q $ may be calculated using the five dimensional conserved axial current for domain wall fermions ${\cal A}_\mu$ by the relation\cite{Blum:2000kn} $ \langle {\cal A}_\mu(t) A_\mu(0)\rangle = Z_A  \langle A_\mu(t) A_\mu(0)\rangle$. 
 Hence the
complete renormalization factor is written as the exact axial current
renormalization factor times the ratio of the perturbative
renormalization factor for the desired operator divided by the
perturbative renormalization factor for the axial current.  That is, 
\begin{equation}
Z^{\cal O} = \frac{Z^{\cal O,\text{pert}}}{Z^{\text{pert}}_{A}} \cdot Z^{\text{nonpert}}_{A}\,.
\end{equation}
In the continuum, because of Lorentz invariance, the totally symmetric operator $\bar q[\gamma_5]\gamma_{\{\mu} D_{\nu} D_{\rho\}} q$ cannot mix with the mixed symmetry operator
$\bar q[\gamma_5]\gamma_{ [ \mu} D_{\{\nu]} D_{\rho \} } q$, where the square brackets denote antisymmetrization.  
In contrast,  on the lattice, both operators appear in the same representation,  $\tau^{(8)}_1$, so that they can and do mix.  However, the mixing coefficient\cite{Bistrovic:2005aa,Bistrovic:2007ab}, $Z^{\cal O}_{ij} = 2.88 \times 10^{-3}$, is very small, so that we have ignored the contribution of the mixed symmetry operator in this present work.

All results below have been transformed to a scale of $\mu^2=4$ GeV$^2$.

% --------------------------------------------------------------------------
%
%  Results for lowest moments
%
% --------------------------------------------------------------------------

\section{\label{sec:results}Numerical Results for the Generalized Form Factors}

Since two point functions taken at the sink $\ts= \ts_{\text{snk}}$,
$C^{\text{2pt}}(\ts_{\text{snk}},P^\prime)$ and
$C^{\text{2pt}}(\ts_{\text{snk}},P)$ in the ratio Eq.~(\ref{eq:ratio})
decay exponentially for the full
Euclidean distance $\ts_{\text{snk}}-\ts_{\text{src }}$, they are
particularly subject to statistical noise.
In the worst case, they may become negative, which we observe
for three values of the momentum transfer for the dataset $m=0.01, 20^3$.
%and $m=0.01, 28^3$. 
The corresponding datapoints are excluded from our analysis.
Our numerical results for the complete set of unpolarized 
and polarized $n=1,2,3$ isovector and isosinglet GFFs 
as functions of the momentum transfer squared are provided in appendix \ref{appendix1}.
\begin{figure}[htbp]
% \vspace*{-0.75cm}
        \includegraphics[scale=0.8,clip=true,angle=0]{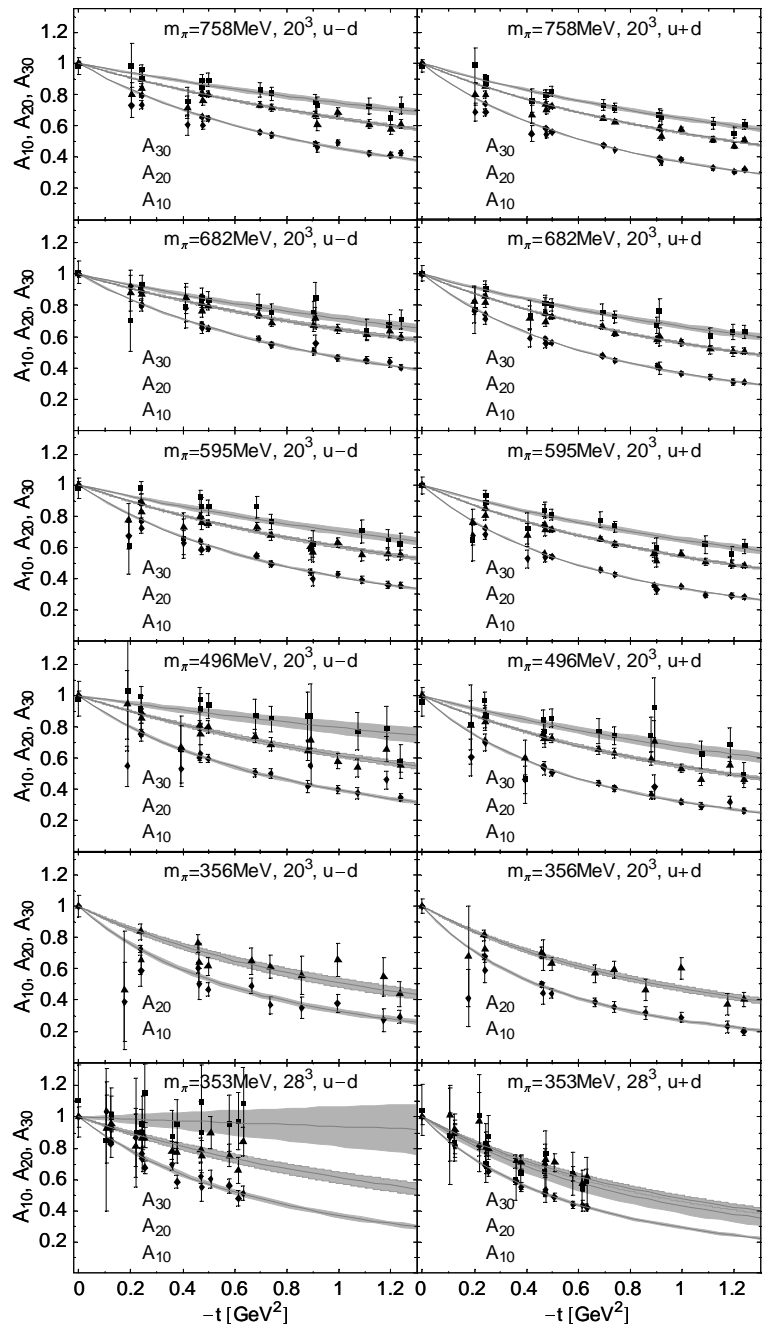}
%      \vspace*{-0.8cm}
  \caption{Flattening of the slope of the $A_{n0}$ GFFs with increasing $n$ for flavor combinations
  $u-d$ (left) and $u+d$ (right). The solid curves and error bands correspond to dipole fits described in the text. Disconnected contributions are not included.}
  \label{A123}
\end{figure}

In Figs.~(\ref{ABC1st}) and (\ref{ABpol1st}) we show results for the vector generalized form factors
$A_{20}$, $B_{20}$, $C_{20}$ and axial vector GFFs $\widetilde A_{20}$, $\widetilde B_{20}$
as functions of the momentum transfer squared $t$.
We observe that the absolute values in the isovector and isosinglet channels
are in qualitative agreement with the predictions from large $N_c$ counting rules, see e.g. \cite{Goeke:2001tz},
for the unpolarized GFFs
\begin{equation}
|A_{20}^{u+d}|\sim N_c^2 \gg |A_{20}^{u-d}|\sim N_c, \qquad |B_{20}^{u-d}|\sim N_c^3 \gg |B_{20}^{u+d}|\sim N_c^2
, \qquad |C_{20}^{u+d}|\sim N_c^2 \gg |C_{20}^{u-d}|\sim N_c\,.
\label{ChPTE20}
\end{equation}
In the polarized case, the inequalities from the counting rules are not  satisfied nearly as strongly.  Whereas the counting rules predict:
\begin{equation}
|\widetilde A_{20}^{u-d}|\sim N_c^2 \gg |\widetilde A_{20}^{u+d}|\sim N_c,
\qquad |\widetilde B_{20}^{u-d}|\sim N_c^4 \gg |\widetilde B_{20}^{u+d}|\sim N_c^3\,,
\label{ChPTE20}
\end{equation}
our results for $\widetilde A_{20}^{u-d}$ are only slightly larger than $\widetilde A_{20}^{u+d}$, and although the errors are large, $\widetilde B_{20}^{u-d}$ appears to be comparable to $\widetilde B_{20}^{u+d}$ rather than dominating it.
Finally, our results disagree with the predicted hierarchy
between different types of GFFs:
\begin{equation}
|B_{20}^{u-d}|\sim N_c^3 \gg |A_{20}^{u+d}|\sim N_c^2\,,
\label{ChPTE20}
\end{equation}
since the lattice results (at non-zero $t$) clearly give $A_{20}^{u+d} > B_{20}^{u-d}$.
It would be valuable to understand why these counting rules are only partially satisfied.

For future reference, it is important to note that the GFF $C_{20}$ , which gives rise
to the $\xi$-dependence of the $n=2$ moment of the GPDs $H(x,\xi,t)$ and $E(x,\xi,t)$,
is compatible with zero for $u-d$, over the full range of
momentum transfer squared $t\approx -0.12\ldots -1.2$ GeV$^2$.
Similarly, the isosinglet GFF $B_{20}^{u+d}$, which is one of the terms in the contribution of the total angular momentum to the nucleon spin,  is compatible with zero within errors. 
We will study both these GFFs in  detail in section V.

We now consider the behavior of the slopes of the GFFs $A_{n0}$ and their relation to the transverse size of the nucleon.
Since $\int_{-1}^{1} dx\,x^{n-1} H(x,\xi=0,t)=A_{n0}(t)$, it is evident
that the GFFs for increasing $n$ correspond to increasing average momentum fractions $\overline{\langle x\rangle}$.
As the average momentum fraction gets larger, or equivalently
as $n\to \infty$, we expect the $t$-slope of the GFFs $A_{n0}$ to flatten. 
This may be understood in terms of the 
light cone Fock representation \cite{Brodsky:2000xy,Diehl:2000xz}
by the fact that 
the final state nucleon wavefunction for a struck quark with 
momentum fraction $x$ and initial transverse momentum $k_\perp^{\text{in}}$ depends on the 
transverse momentum $k_\perp = k_\perp^{\text{in}} - (1-x)\Delta_\perp$.   
Hence, a large transverse momentum transfer $t = -\Delta^2_\perp$  
can be better absorbed without causing breakup of the bound state by quarks with large momentum fraction $x$.
 \begin{figure}[t]
     \begin{minipage}{0.4\textwidth}
      \centering
          \includegraphics[scale=0.4,clip=true,angle=0]{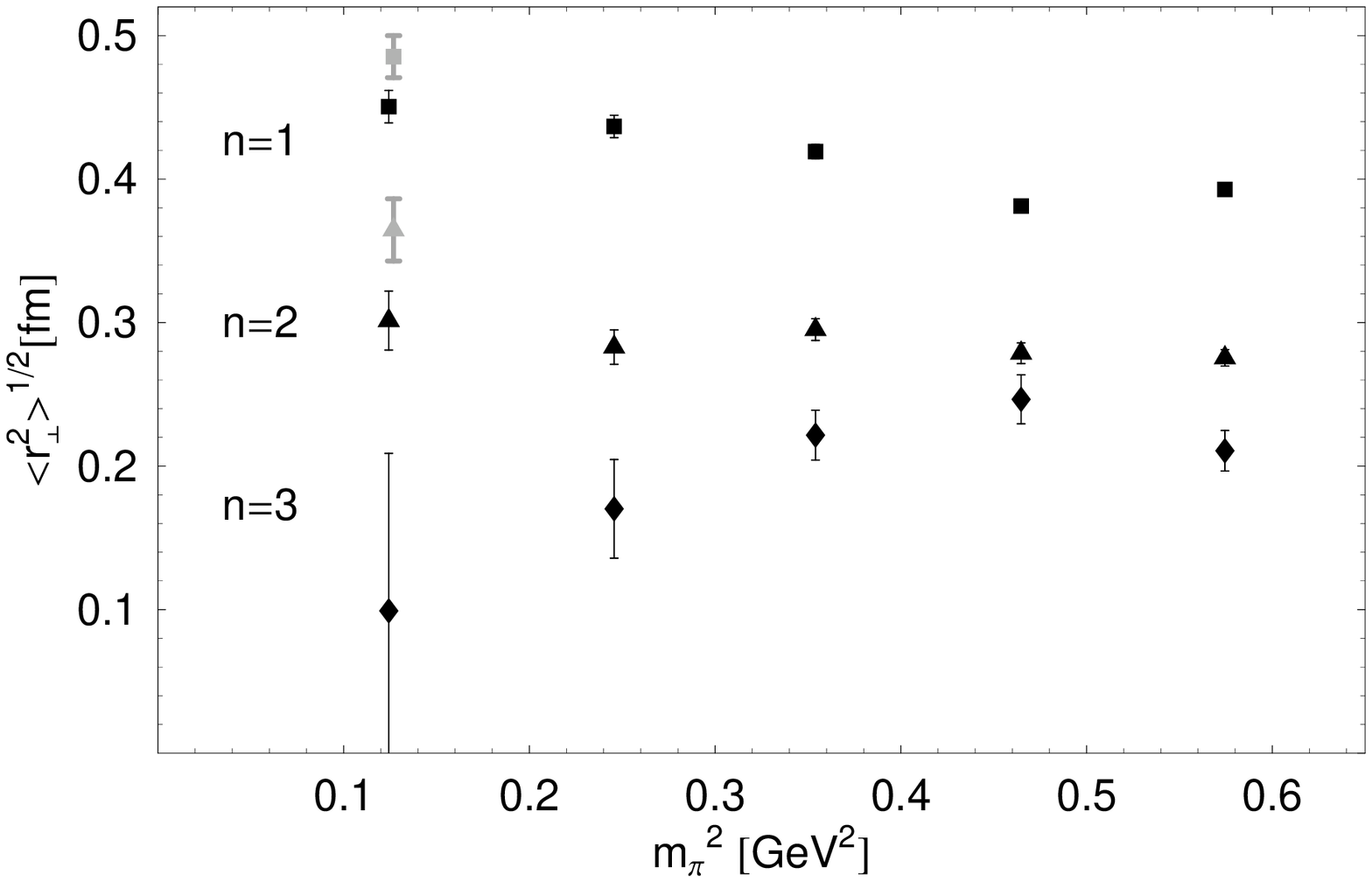}
       \caption{Two dimensional rms radii of the vector GFFs versus $m_\pi^2$ for the flavor combination $u-d$. The results  for $m_\pi=354$ MeV, $L^3=20^3$ are displayed in gray.  }\label{rmsvsmpi}
     \end{minipage}
     %\hfill
     \hspace{1cm}
     \begin{minipage}{0.4\textwidth}
      \centering
          \includegraphics[scale=0.4,clip=true,angle=0]{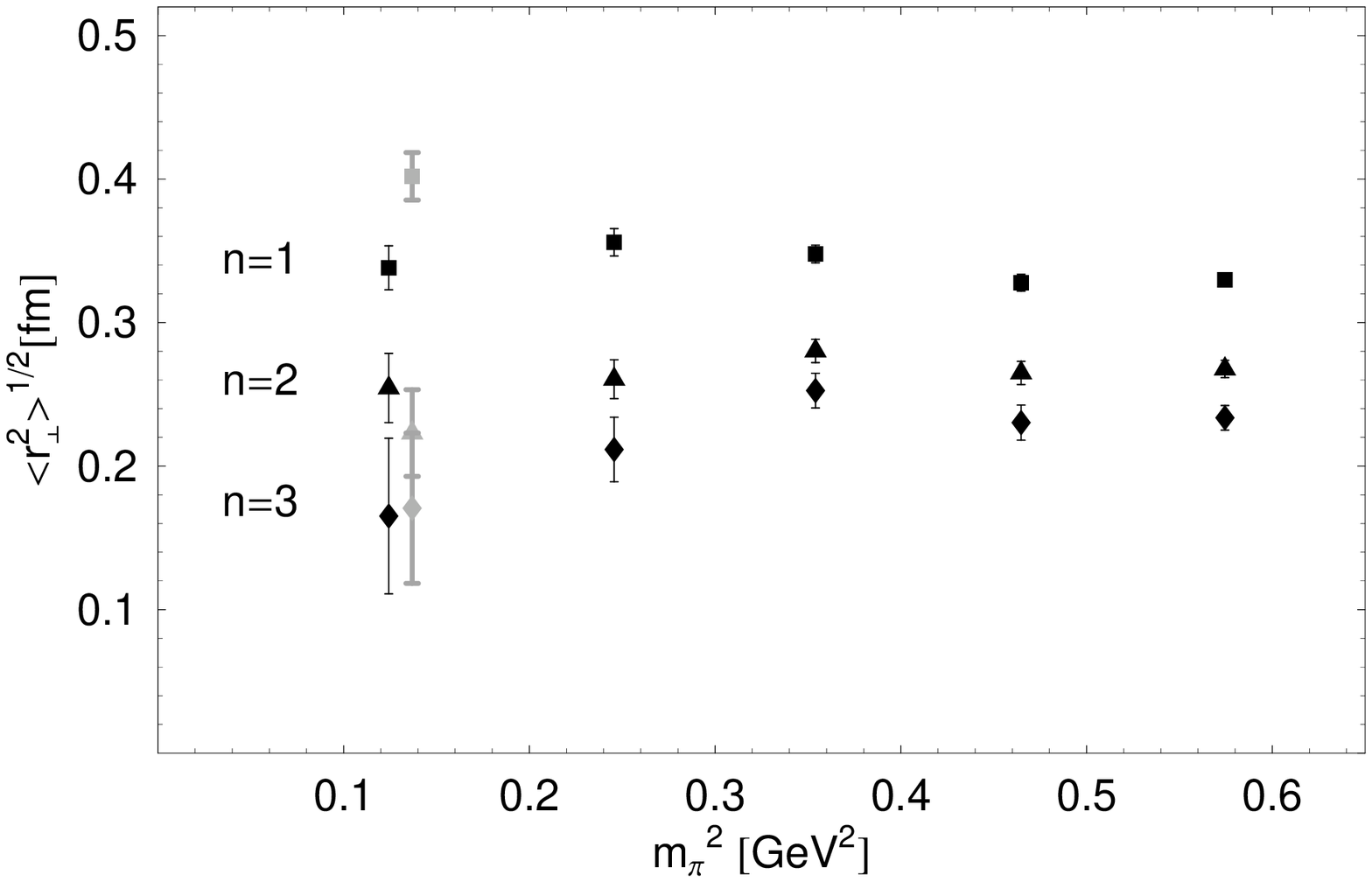}
       \caption{Two dimensional rms radii of the axial vector GFFs versus $m_\pi^2$ for the flavor combination $u-d$. The results  for $m_\pi=354$ MeV, $L^3=20^3$ are displayed in gray and slightly shifted to
       the right for clarity.  }\label{Axialrmsvsmpi}
     \end{minipage}
   \end{figure} 
\begin{figure}[htbp]
% \vspace*{-0.75cm}
       \includegraphics[scale=0.7,clip=true,angle=0]{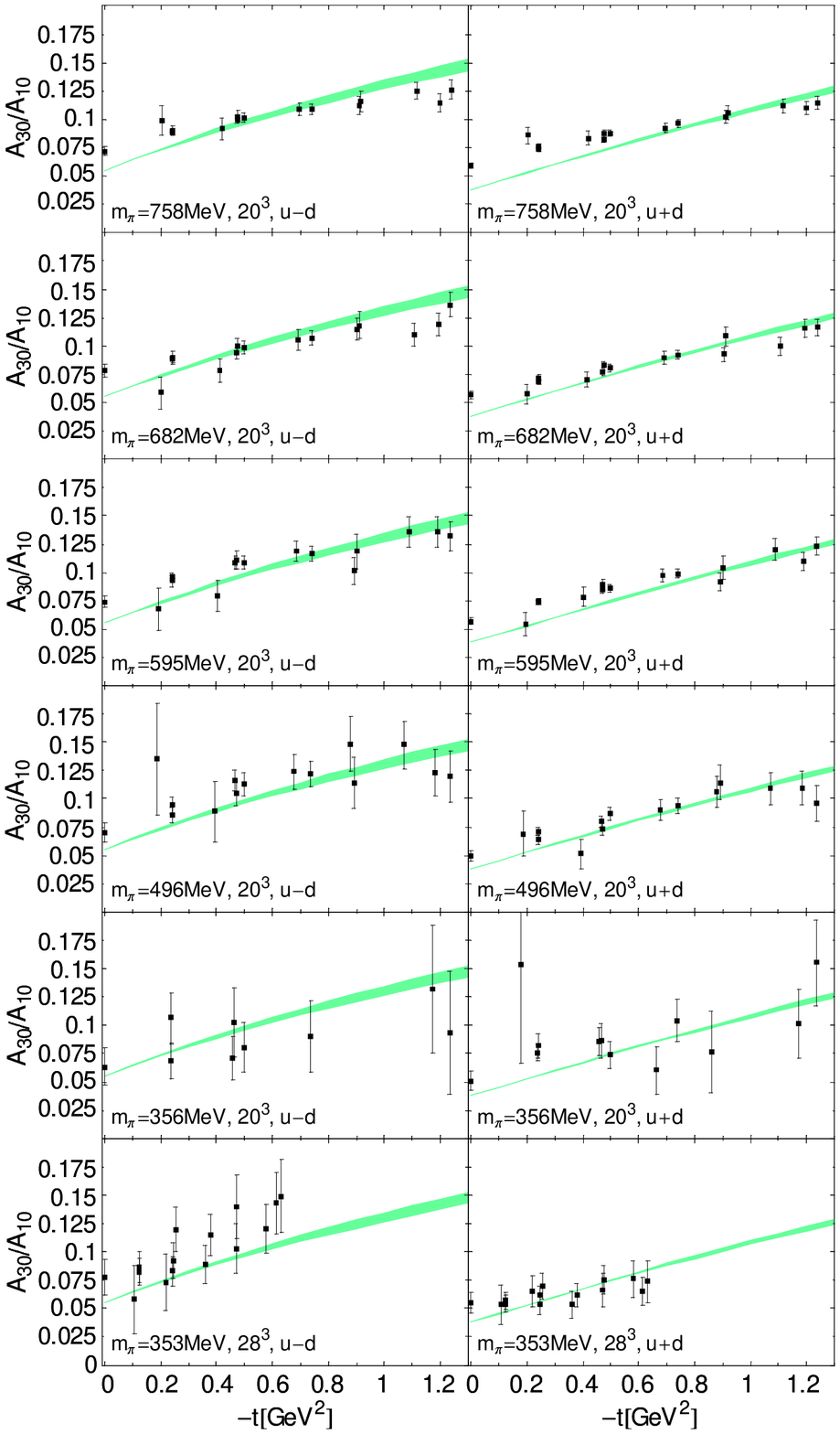}
%      \vspace*{-0.8cm}
  \caption{Ratio of generalized form factors $A_{30}(t)/A_{10}(t)$ for the flavor combinations
  $u-d$ (left) and $u+d$ (right) compared  with the parametrization in Ref. \cite{Diehl:2004cx}.
  Disconnected contributions are not included.}
  \label{A30overA10}
\end{figure}
Additional insight is obtained by considering the impact parameter dependent GPD,  
$H(x,b^2_\perp)$, which has a probability interpretation and is the Fourier transform \cite{Burkardt:2000za} with respect to transverse momentum transfer of $H(x,\xi=0,t=-\Delta_\perp^2)$:
\begin{equation}
H(x,b^2_\perp)=\int \frac{d^2\Delta_\perp}{(2\pi)^2} e^{-ib_\perp\cdot \Delta_\perp}H(x,\xi=0,t=-\Delta_\perp^2)\,,
\label{ChPTE20}
\end{equation}
where $\Delta_\perp$ is the transverse momentum transfer. The impact
parameter $b_\perp$ corresponds to the distance of the active quark 
from the center of momentum
of the nucleon. As $x\rightarrow 1$, a single quark will carry all the longitudinal
momentum of the nucleon and therefore represent its center of momentum, so that
the impact parameter distribution in this limit is strongly peaked around the origin, 
$H(x\rightarrow 1,b^2_\perp)\propto \delta^2(b^2_\perp)$.
The corresponding flattening of the GFFs in the momentum transfer $t$ is clearly visible in 
Fig.~(\ref{A123}), where we compare the slopes of the 
GFFs $A_{(n=1,2,3)0}$ which have been normalized to unity at $t=0$. 

Dipole fits to the GFFs in Fig.~(\ref{A123}), denoted by the solid lines and statistical error bands, enable us to determine the slopes of the form factors and thus express the
 two- and three-dimensional rms radii $ \langle r^2\rangle_\perp^{1/2}$ and
$ \langle r^2\rangle^{1/2}$ in terms of the dipole masses $m_D$
\begin{equation}
 \langle r^2_\perp\rangle=\frac{2}{3}\langle r^2\rangle=\frac{8}{m_D^2}\,.
%  \label{LCop}
\end{equation}
Since the range of values for the momentum transfer $t$ is much smaller for the
large volume ($L^3=28^3$) dataset, we have restricted the dipole fits for all datasets to
the overlapping region of $t=0\ldots -0.8$ GeV$^2$. Our results for the 2d rms radii versus
the pion mass squared are presented in Figs.~(\ref{rmsvsmpi}) and (\ref{Axialrmsvsmpi}).
These results confirm the dramatic dependence of the transverse rms radius on the  moment and thus the average momentum fraction as first observed\cite{LHPC:2003is} for pion masses 750 MeV and higher, and show that this dependence increases as the pion mass decreases. Indeed, considering the ratio of the $n=3$ moment to the $n=1$ moment, which both correspond to the same sum or difference of quarks and antiquarks, we observe that for vector GFFs this ratio decreases from approximately $0.58$ to $0.22$ as the pion mass decreases from 750 MeV to 350 MeV, and for axial vector GFFs, it decreases from roughly $0.71$ to $0.43$.

\begin{figure}[htbp]
% \vspace*{-0.75cm}
       \includegraphics[scale=0.7,clip=true,angle=0]{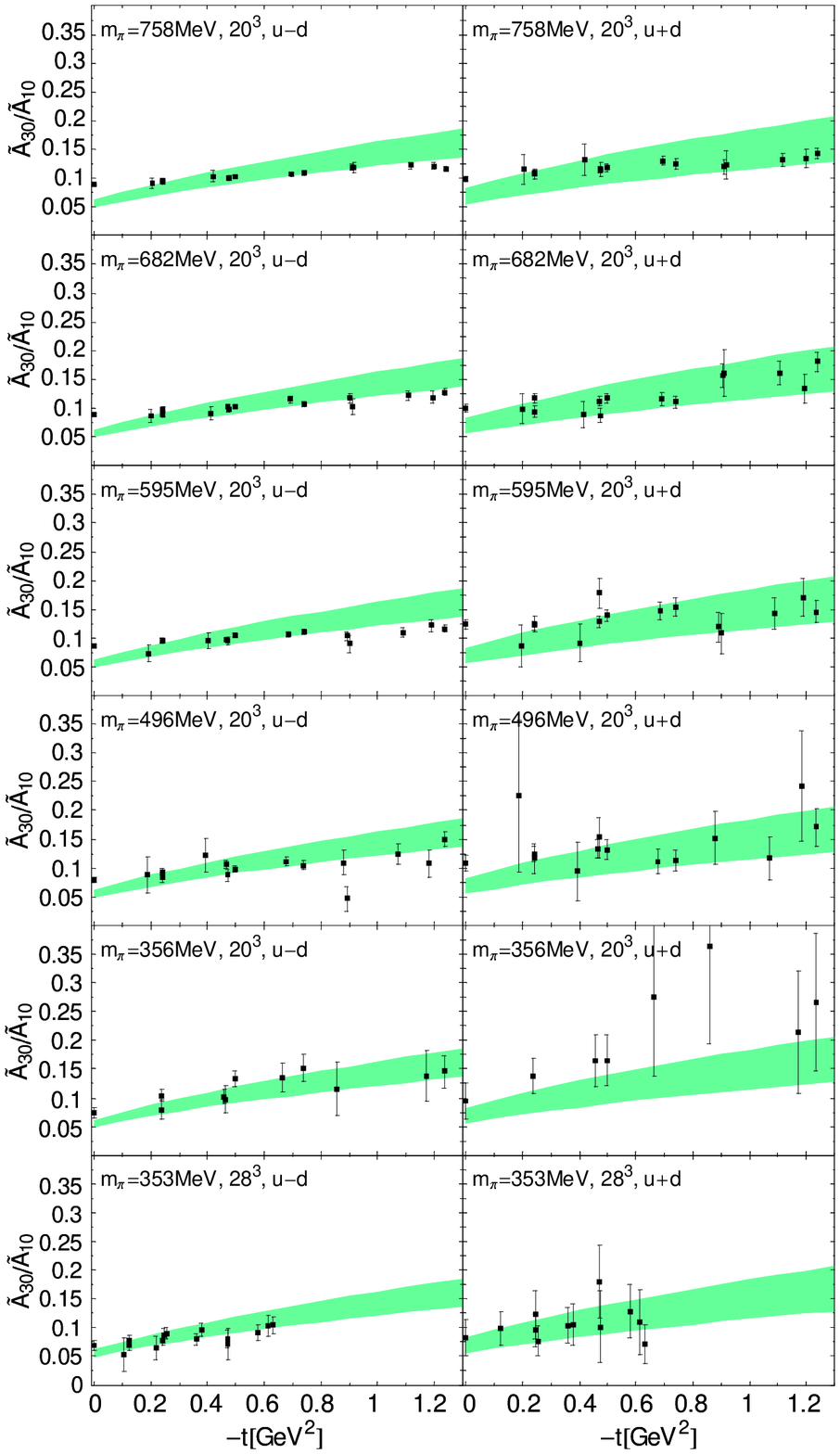}
%      \vspace*{-0.8cm}
  \caption{Ratio of polarized generalized form factors $\widetilde A_{30}(t)/\widetilde A_{10}(t)$ for the flavor combinations $u-d$ (left) and $u+d$ (right) compared with the parametrization in Ref. \cite{Diehl:2004cx}.
  Disconnected contributions are not included.}
  \label{A30poloverA10pol}
\end{figure}

In Figs.~(\ref{A30overA10}) and (\ref{A30poloverA10pol}) 
we present a first comparison of our results for ratios of generalized form factors
$A_{30}(t)/A_{10}(t)$ and $\widetilde A_{30}(t)/\widetilde A_{10}(t)$ 
to the parametrization by Diehl et al.\cite{Diehl:2004cx}
as function of the momentum transfer squared $t$. As the pion mass decreases,
the slope of our results approaches that of the phenomenological 
parametrization. Our results clearly indicate that a factorized ansatz 
for the GPDs in $x$ and $t$, which would lead
to constant ratios in Figs.~(\ref{A30overA10}) and (\ref{A30poloverA10pol})
breaks down already for small values of the momentum transfer squared $|t|\ll 1$ GeV$^2$.

\begin{figure}[th]
     \begin{minipage}{0.48\textwidth}
      \centering
          \includegraphics[scale=0.41,clip=true,angle=0]{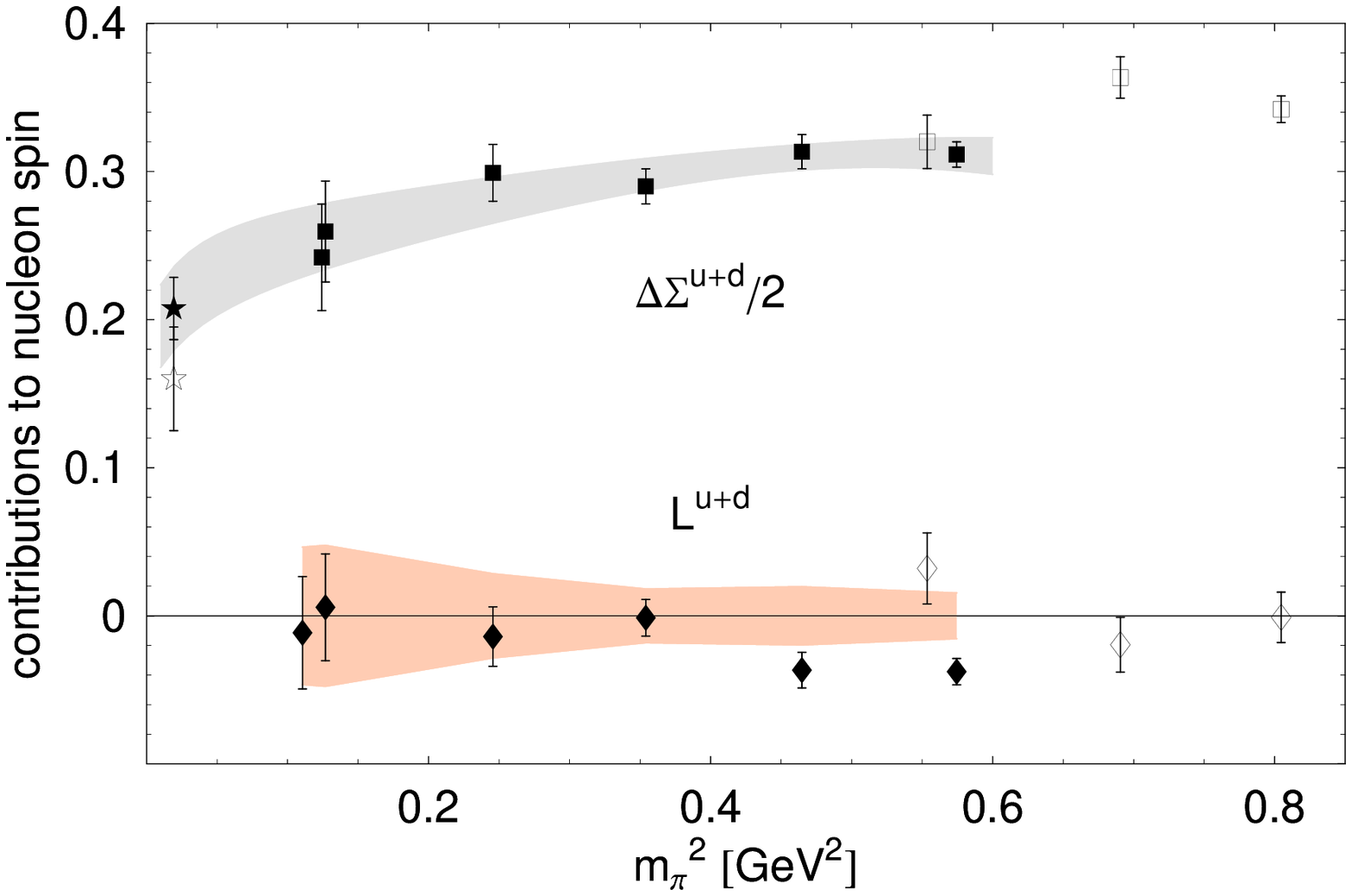}
       \caption{Total quark spin and orbital angular momentum contributions to the spin of the nucleon. The filled and open stars represent values given in HERMES 2007 \cite{Airapetian:2007aa} and 1999 \cite{Ackerstaff:1999ey} respectively and open symbols represent earlier LHPC/SESAM calculations. The error bands are explained in the text. Disconnected contributions are not included.\newline\newline\newline\label{Lq1}}
     \end{minipage}
     %\hfill
     \hspace{0.5cm}
     \begin{minipage}{0.48\textwidth}
      \centering
          \includegraphics[scale=0.4,clip=true,angle=0]{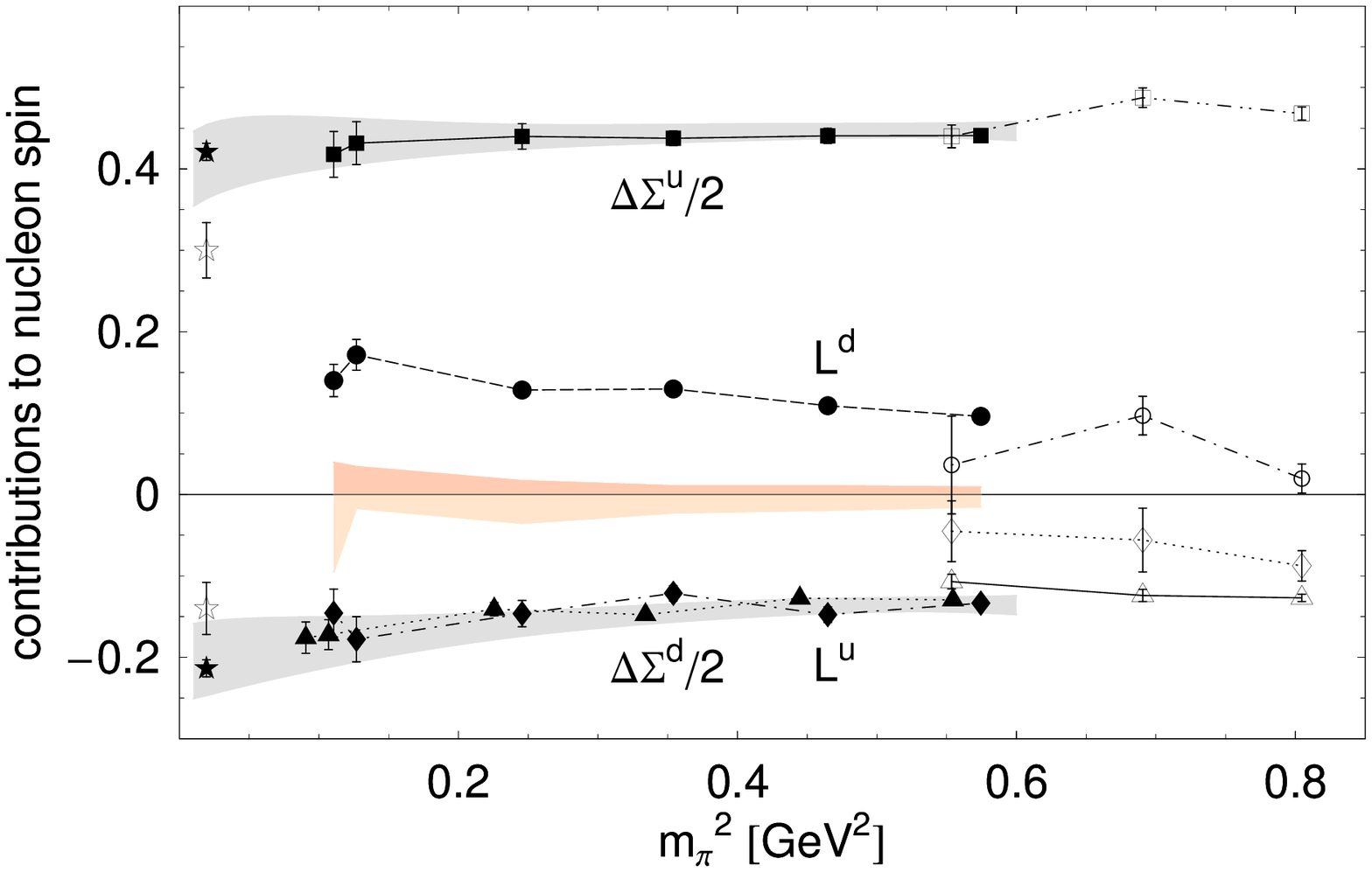}
       \caption{Quark spin and orbital angular momentum contributions to the spin of the nucleon for up and down quarks.
       Squares and triangles denote $\Delta \Sigma^u$ and $\Delta \Sigma^d$ respectively, 
       and diamonds and circles denote $L^u$ and $L^d$ respectively. The filled and open stars represent values given in 
       HERMES 2007 \cite{Airapetian:2007aa} and 1999 \cite{Ackerstaff:1999ey} respectively and open symbols represent earlier
       LHPC/SESAM calculations. The error bands are explained in the text. Disconnected contributions are not included.\label{Lq2}}
     \end{minipage}
   \end{figure}

The GFFs $A^q_{20}(t=0)=\langle x\rangle_q$ and $B^q_{20}(t=0)$ enable us to compute
the total quark angular momentum contribution to the nucleon spin \cite{Ji:1996ek}, 
$J^q=1/2( A^q_{20}(0) + B^q_{20}(0) )$. 
Figures~(\ref{Lq1}) and (\ref{Lq2})
show results for the quark spin $\widetilde A^q_{10}(t=0)/2=\Delta\Sigma^q /2$ 
and the orbital angular momentum $L^q=J^q-\Delta\Sigma^q/2$ contributions to the nucleon spin 
$S=1/2$ versus the pion mass squared. Preliminary chiral extrapolations of $\Delta\Sigma^{q}$
based on self-consistently improved one-loop ChPT 
\cite{Beane:2005rj,Beane:2006gj,Beane:2006kx,Edwards:2006qx,DrusPDFs} for $\Delta\Sigma^{u+d}$ and  
ChPT including the $\Delta$ resonance \cite{Hemmert:2003cb,Beane:2004rf} for $g_A=\Delta\Sigma^{u-d}$
and are shown as shaded bands.
The values for $B^q_{20}(t=0)$ have been obtained from a
linear extrapolation of $B^{u+d}_{20}(t)$ and a dipole extrapolation for $B^{u,d}_{20}(t)$ in $t$.
The resulting uncertainty in $B^q_{20}(t=0)$, which contributes to the uncertainty in $L_q$, 
depends on the details of the corresponding fit, such as the functional form and range of $t$, and is therefore partially systematic. To allow the reader
to assess the absolute statistical errors, we represent the errors for $L_q$ coming from the extrapolation in $t$ by 
error bands around the $m_\pi^2$-axis in Figs.~(\ref{Lq1}) and (\ref{Lq2}). 
Experimental results for the quark spin fractions $\Delta\Sigma^{u+d}$ and $\Delta(u,d)=\Delta\Sigma^{u,d}$
are represented by open stars for the prediction given in the HERMES publication from 1999 \cite{Ackerstaff:1999ey} and filled stars for the 2007 HERMES results \cite{Airapetian:2007aa}. The significant difference between the new HERMES results, which are consistent with recent COMPASS results\cite{Alexakhin:2006vx}, 
and the values given in \cite{Ackerstaff:1999ey} is probably 
to a large extent due to the simple Regge-parametrization which has been used in \cite{Ackerstaff:1999ey} to compute
the contribution to $\Delta\Sigma$ coming from the low $x$-region. It is gratifying that
the new values are much closer to our lattice results.

These results reveal two remarkable features of the quark contributions to the nucleon spin.  The first is that the magnitude of the orbital angular momentum contributions of the up and down quarks, $L^u$ and $L^d$, are separately quite substantial, starting at  0.15 at $m_\pi = 750$ MeV and increasing to nearly 0.20 at 350 MeV, and yet they cancel nearly completely at all pion masses. The second is the close cancellation between the orbital and spin contributions of the d quarks, $L^d$ and $\Delta \Sigma^d /2$ for all pion masses. It would be valuable to understand the physical origin of both features.

%
%\newpage
\section{Chiral extrapolations}
Our ultimate goal is to use the combination of full QCD lattice calculations in the chiral regime and chiral perturbation theory to  extrapolate to the physical pion mass, to extrapolate to infinite volume, to extrapolate in momentum transfer, and to correct for lattice artifacts,  with all the relevant low energy constants being determined solely from lattice data.  
Significant progress has been made in many aspects of chiral perturbation theory
(ChPT) relevant to the nucleon observables addressed in this work
\cite{Chen:2001eg,Chen:2001pv,Arndt:2001ye,Detmold:2001jb,Belitsky:2002jp,Young:2002ib,Beane:2004rf,Detmold:2005pt,Chen:2005ab,Edwards:2006qx,
Diehl:2006js,Ando:2006sk,Diehl:2006ya,Dorati:2007bk,Wang:2007iw}.
Although important developments have been made  in correcting for our hybrid action
\cite{Chen:2005ab,Tiburzi:2005is,Bar:2005tu,Aubin:2006hg,Chen:2006wf} and finite volume \cite{Beane:2004rf,Detmold:2005pt}, results for the relevant GFFs are not yet available. In this work we will focus on ChPT treatment of the pion mass and momentum dependence.

The basic problem is that currently, there is not yet unambiguous evidence supporting a particular counting scheme and its convergence criteria, leading to a range of alternative re-summations, and 
there is similar ambiguity concerning the choice of degrees of freedom,  such as when and if it is essential to include the $\Delta$ resonance. When complete results for the observables of interest are available, it will be interesting to compare four approaches: heavy baryon ChPT (HBChPT)\cite{Diehl:2006js,Ando:2006sk,Diehl:2006ya}, covariant ChPT in the baryon sector (BChPT)\cite{Dorati:2007bk}, self-consistently improved one-loop ChPT
\cite{Beane:2005rj,Beane:2006gj,Beane:2006kx,Edwards:2006qx,DrusPDFs}, and ChPT with finite-range regulators\cite{Detmold:2001jb,Young:2002ib,Wang:2007iw}. Although self-consistent improvement by utilizing values of parameters like $f_\pi$ and $g_A$ calculated on the lattice at the relevant pion mass and finite-range regulators appear to improve the behavior of ChPT at larger values of the pion mass, based on the results available in the literature, we will focus on the two formulations HBChPT and BChPT.

Heavy baryon ChPT, which we will subsequently always refer to as HBChPT,  assumes that $m_\pi$ and the magnitude of the spatial three-momentum, $p$, are much smaller than the nucleon mass in the chiral limit, $m^0_N \sim 890$ MeV, and the chiral scale $\Lambda_\chi = 4 \pi f_\pi = 1.17$ GeV,  and simultaneously expands in powers of the four quantities $\epsilon =\{ \frac{p}{\Lambda_\chi}, \frac{m_\pi}{\Lambda_\chi},\frac{p}{m^0_N},\frac{m_\pi}{m^0_N}\}$.  In contrast, covariant baryon ChPT, which, slightly changing the notation of Ref.~\cite{Dorati:2007bk}, we will subsequently always refer to as CBChPT, does not treat $m^0_N$ and  $\Lambda_\chi$ as comparable scales, but rather keeps all powers $(\frac{1}{m^0_N})^n$ generated by the couplings included in the ChPT Lagrangian. Thus, it is a resummation that includes terms that would contribute in higher order to HBChPT and may be thought of as recoil corrections. The HBChPT results of refs.~\cite{Diehl:2006js,Ando:2006sk,Diehl:2006ya} and the CBChPT results of ref.~\cite{Dorati:2007bk} have the desirable property that they use the same regularization  scheme, so that truncation of the higher order terms in CBChPT yields the corresponding  result in HBChPT, a feature that we will utilize below. One of our primary objectives will be to assess the regimes of applicability of these two alternative formulations.
For notational convenience, we will refer to the generic momentum in both theories as $p$, and order both theories in powers of $p^n$. We would like to note, however, that the counting scheme of the HBChPT-approaches in
\cite{Diehl:2006js,Ando:2006sk,Diehl:2006ya} differs from the one used in the CBChPT-approach of \cite{Dorati:2007bk}.

The HBChPT\cite{Diehl:2006js,Ando:2006sk,Diehl:2006ya} and CBChPT\cite{Dorati:2007bk} results for GFFs, including the dependence on the momentum transfer squared, $t$,
enable us to investigate for the first time possible non-trivial correlations in 
the $m_\pi$- and $t$-dependence of GFFs.
It is  interesting to note that to order $\mathcal{O}(p^2)$ in HBChPT,  unpolarized and polarized GFFs are independent of each other and depend on separate  chiral limit values and counter-terms.
In contrast, in CBChPT, the pion-mass dependence of the isovector momentum fraction of quarks $\langle x\rangle_{u-d}$ 
is {\em simultaneously} controlled by both the chiral limit values $\langle x\rangle_{u-d}^0$ and 
$\langle x\rangle_{\Delta u-\Delta d}^0$, as is also the case for  $\langle x\rangle_{\Delta u-\Delta d}$. To  $\mathcal{O}(p^2)$, however, CBChPT 
does not include insertions from pion operators, and it turns out that the $t$-dependence 
for the isosinglet (and isovector) case to this order 
is therefore essentially linear and decouples from the pion mass dependence. Once CBChPT 
calculations have been pushed to higher orders, it will be interesting to study the combined non-analytic $(t,m_\pi)$-dependence of the full set of polarized and unpolarized GFFs based on this approach.
For the time being, we will investigate possible non-trivial correlations in $m_\pi$ and $t$ in the framework
of covariant CBChPT by including partial $\mathcal{O}(p^3)$-corrections as discussed below.

\begin{table}[t]
  \begin{tabular}{|c|c|c|c|c|c|c|c|c|c|c|}
    \hline 
    $f^0_\pi$ [GeV] & $m^0_N$ [GeV] & $g^0_A$ &$\langle x\rangle^{\pi,0}_{u+d}$& $c_1$[GeV$^{-1}$] & $c_2$[GeV$^{-1}$] & $c_3$[GeV$^{-1}$] & $g^0_{\pi N\Delta}$ & $\Delta=m^0_\Delta -m^0_N$ [GeV] \\ \hline
    $0.092$ & $0.89$ & $1.2$ & 0.5 & -0.9 & 3.2 & -3.4 & $3/(2^{3/2}) g^0_A$ & 0.3\\
     \hline
  \end{tabular}
  \caption{Low energy constants used in the chiral extrapolations.}
      \label{tab:LECs}
\end{table}
The low energy constants  used for the chiral extrapolations are summarized in Table \ref{tab:LECs}.   Ultimately, all these values will be determined simultaneously by a global fit to a full set of lattice calculations of all the relevant observables using the same lattice action, but at present they  are chosen as follows.
We will use $f^0_\pi=0.092$ GeV, $m^0_N=0.89$ GeV and $g^0_A=1.2$ as the chiral limit values 
of the pion decay constant, nucleon mass and the axial vector coupling constant, respectively, and for notational simplicity, will subsequently omit  the upper index $0$ for these quantities. We note that these values are compatible within statistical errors with hybrid lattice calculations of  $f_\pi$, $m_N$, and $g_A$.
In addition, we need the low energy constants $c_1$, $c_2$
and $c_3$ for the chiral extrapolation of $C_{20}^{u+d}$ in the framework
of HBChPT in section \ref{subsecHBC20}. Here, $c_2=3.2$~GeV$^{-1}$ and $c_3=-3.4$~GeV$^{-1}$ are taken from 
Refs.~\cite{Procura:2006bj,Dorati:2007bk} and 
$c_1=-0.90$~GeV$^{-1}$ has been obtained from a CBChPT fit to our nucleon mass lattice data,
which provides us with a parametrization for the pion mass dependence of the nucleon
mass in our simulation needed for some of the chiral extrapolations below. 
Depending on the order of ChPT, diagrams with insertions of pion operators contribute 
for the isosinglet GFFs, which introduces the momentum fraction of quarks in the pion
in the chiral limit, $\langle x\rangle^{\pi,0}_{u+d}$, as an additional low energy constant.
Ultimately, we will calculate this quantity  from chiral fits to hybrid
lattice results for the pion, but for now we use $\langle x\rangle^{\pi,0}_{u+d}=0.5$ 
\cite{Best:1997qp,Brommel:2005ee,Capitani:2005jp}, which is obtained from lattice calculations that are in reasonable agreement with phenomenology \cite{Sutton:1991ay,Gluck:1999xe,Wijesooriya:2005ir}.
Finally, for the chiral extrapolation of the total quark angular momentum in the framework
of HBChPT including the $\Delta$-resonance, we use the nucleon-$\Delta$ mass splitting $\Delta=0.3$~GeV, which is consistent with lattice calculations 
and  the large-$N_c$ relation $g^0_{\pi N\Delta}=3/(2^{3/2}) g^0_A$ for the pion-nucleon-$\Delta$ coupling 
$g_{\pi N\Delta}$. This latter result will soon be superseded by extrapolation of lattice calculations of the $N-\Delta$ transition form factor\cite{Alexandrou:2003ea,Alexandrou:2004xn,Alexandrou:2006mc}.

\begin{table}[ht]
\centering
\begin{tabular}{|c|c|c|c|c|c|c|c|c}
  \hline
section&  GFF & HBChPT & CBChPT & expected dependence on $m_\pi,t$\\ \hline\hline
 A & $A_{20}^{u-d}$ & $\mathcal{O}(p^2)$ & $\mathcal{O}(p^2)$ & non-analytic in $m_\pi$, $\approx$ linear in $t$ \\ \hline
  B & $B_{20}^{u-d}$ & & $\mathcal{O}(p^2)$  +  corr. of $\mathcal{O}(p^3)$
             & non-analytic in $m_\pi$, $\approx$ linear in $t$ \\ \hline
 C & $C_{20}^{u-d}$ & & $\mathcal{O}(p^2)$  + corr. of $\mathcal{O}(p^3)$
             & non-analytic in $m_\pi$, $\approx$ linear in $t$ \\ \hline
 D &  $A_{20}^{u+d}$ & & $\mathcal{O}(p^2)$ 
          + corr. of $\mathcal{O}(p^3)$ & non-analytic in $m_\pi$ and $t$ \\\hline
 I, E & $B_{20}^{u+d}$ & $\mathcal{O}(p^2)$ & $\mathcal{O}(p^2)$
          +  $\mathcal{O}(p^3)$-CTs & non-analytic in $m_\pi$ and $t$ \\\hline
 F & $C_{20}^{u+d}$ &  & $\mathcal{O}(p^2)$
          + corr. of $\mathcal{O}(p^{3,4})$
            & non-analytic in $m_\pi$ and $t$ \\\hline
  G & $J^{u+d}=1/2(A+B)_{20}^{u+d}$ & & $\mathcal{O}(p^2)$ 
          +  corr. of $\mathcal{O}(p^3)$ &  \\\hline
  H & $E_{20}^{u+d}=(A+t/(4m_N)^2B)_{20}^{u+d}$ & $\mathcal{O}(p^2)$ &
         &linear in $m_\pi^2$ and $t$ \\\hline
H &  $M_{20}^{u+d}=(A+B)_{20}^{u+d}$ & $\mathcal{O}(p^2)$ &
         &non-analytic in $m_\pi$ and $t$ \\\hline
 J & $C_{20}^{u+d}$ & $\mathcal{O}(p^2)$ & 
            & non-analytic in $m_\pi$ and $t$ \\\hline
 K &$J^{u+d}=1/2(A+B)_{20}^{u+d}$ & $\mathcal{O}(p^2)$ &
                 &  \\\hline
 K & $J^{u+d}=1/2(A+B)_{20}^{u+d}$  & $\mathcal{O}(p^2)$ with $\Delta$ &
                 &  \\\hline     
\end{tabular}
\caption{Overview of different approaches to the $(m_\pi,t)$-dependence of GFFs in ChPT studied in sections \ref{sec:A20IsoVecrel}-\ref{sec:OAMHB}.}
\label{tab:ChPT-approaches}
\end{table}
Our chiral extrapolations are organized as follows, and summarized in 
 Table \ref{tab:ChPT-approaches}.
 We begin in  section \ref{sec:A20IsoVecrel} with a comparison of CBChPT and HBChPT
extrapolations of the isovector GFF $A^{u-d}_{20}(t)$ and show that whereas CBChPT 
yields a satisfactory fit over the range of pion masses used in the lattice calculations, HBChPT only produces fits to the lowest few points. Hence, in sections \ref{sec:B20IsoVecrel} through \ref{sec:C20Isosinglet}, we study 
the pion mass and $t$-dependence of the isovector GFFs $B^{u-d}_{20}$ and $C^{u-d}_{20}$ and of the isoscalar GFFs $A^{u+d}_{20}$,  $B^{u+d}_{20}$ and $C^{u+d}_{20}$.  
This is followed in section \ref{AMcovariant} by a discussion of our results for the angular momentum of quarks,
based on the CBChPT extrapolations for $A_{20}(t=0)$ and $B_{20}(t=0)$.
In the counting scheme of \cite{Diehl:2006js,Ando:2006sk,Diehl:2006ya} insertions of
pion operators occur at $\mathcal{O}(p^2)$ in HBChPT, leading to a non-analytic combined 
dependence on $m_\pi$ and $t$ for the GFFs $B^{u+d}_{20}(t)$ and $C^{u+d}_{20}(t)$ , 
which we study in sections \ref{EMIsoSingletextr},  \ref{subsecB20}, and \ref{subsecHBC20}. 
Finally, in section \ref{sec:OAMHB}, we study the pion mass dependence of the 
total quark angular momentum $J_q$ in  HBChPT, both including \cite{Chen:2001pv} and excluding
explicit $\Delta$ degrees of freedom and compare
with the corresponding CBChPT results.  
 The chiral extrapolations  of $B^{u+d}_{20}(t)$ and $C^{u+d}_{20}(t)$ in sections \ref{sec:C20Isosinglet}, \ref{subsecB20}, and \ref{subsecHBC20}
are the first parametrizations of their combined $(m_\pi,t)$-dependence,
 providing valuable insights into non-trivial correlations of $m_\pi$ and $t$.
Although we have a considerable amount of lattice data available for reasonably
small values of $|t|\leq 0.25\text{GeV}^2$ and  $m_\pi \leq 500\text{MeV}$, we  use 
an extended set of results for $|t|<0.48\text{GeV}^2$, $m_\pi<700\text{MeV}$ in most of the fits 
to improve the statistics.

%%%%%%%%%%%%%%%%%%%%%%%%%%%%%%%%%%%%%%%%%%%%%%%%%%%%%%%%%%%%%%%%%%%%%%%%%%%%%%%%%%%%%%%%%%%%%%%%%%%%%%%%%%%%%%%%%%%%%%%%%%%%%%
\subsection{\label{sec:A20IsoVecrel}CBChPT extrapolation of  $A_{20}^{u-d}(t)$}
\begin{figure}[thbp]
     \begin{minipage}{0.4\textwidth}
      \centering
          \includegraphics[scale=0.4,clip=true,angle=0]{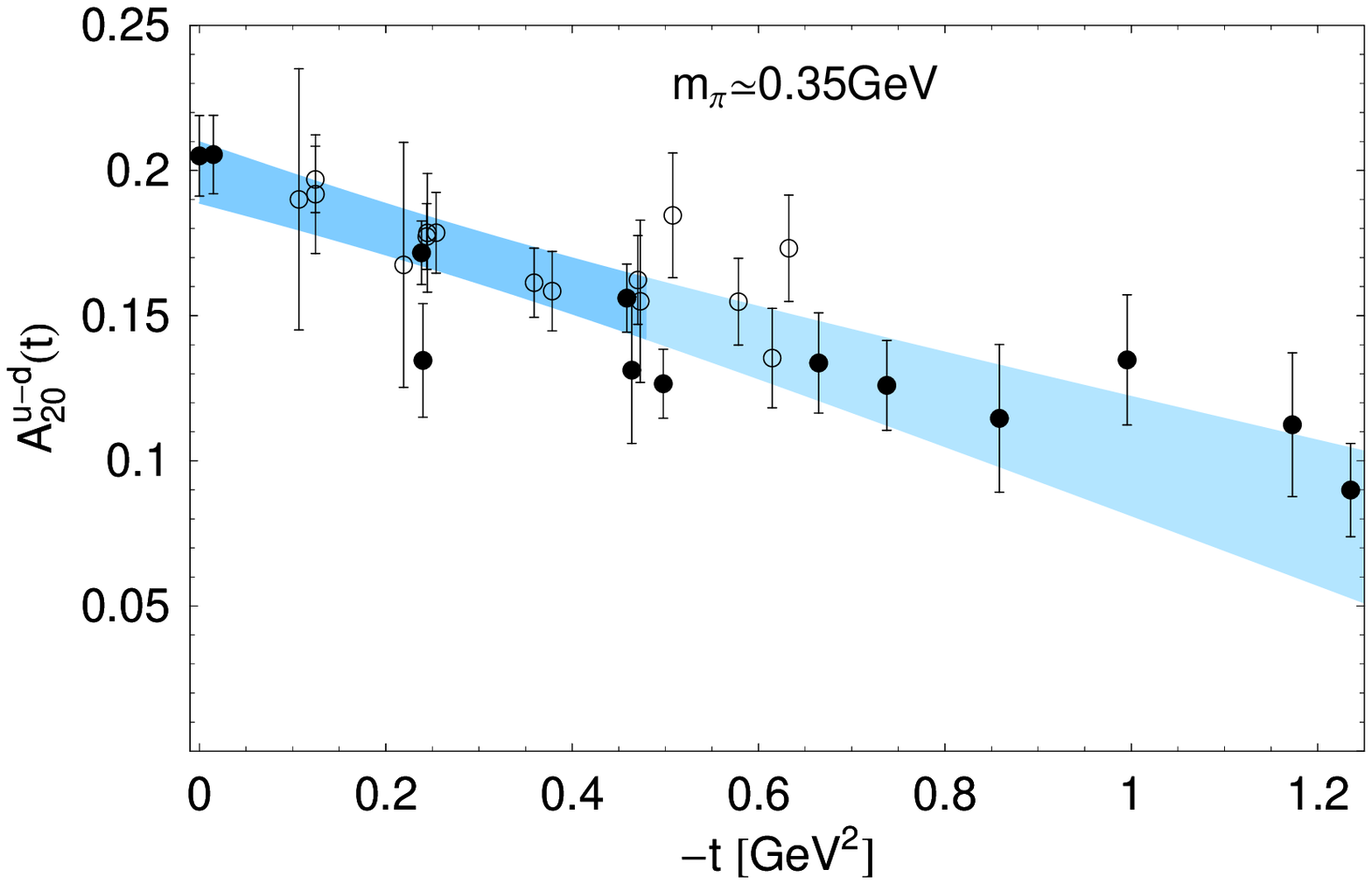}
%      \vspace*{-0.8cm}
  \caption{Lattice results for $A_{20}^{u-d}(t)$ at $m_\pi\approx 0.35$ GeV together with the result of a global simultaneous chiral fit using Eqs.~(\ref{ChPTA20umdp4}), (\ref{ChPTB20umdp4}) and (\ref{ChPTC20umdp4}).}\label{A20IsoVecv3}
     \end{minipage}
     %\hfill
     \hspace{1cm}
     \begin{minipage}{0.4\textwidth}
      \centering
          \includegraphics[scale=0.4,clip=true,angle=0]{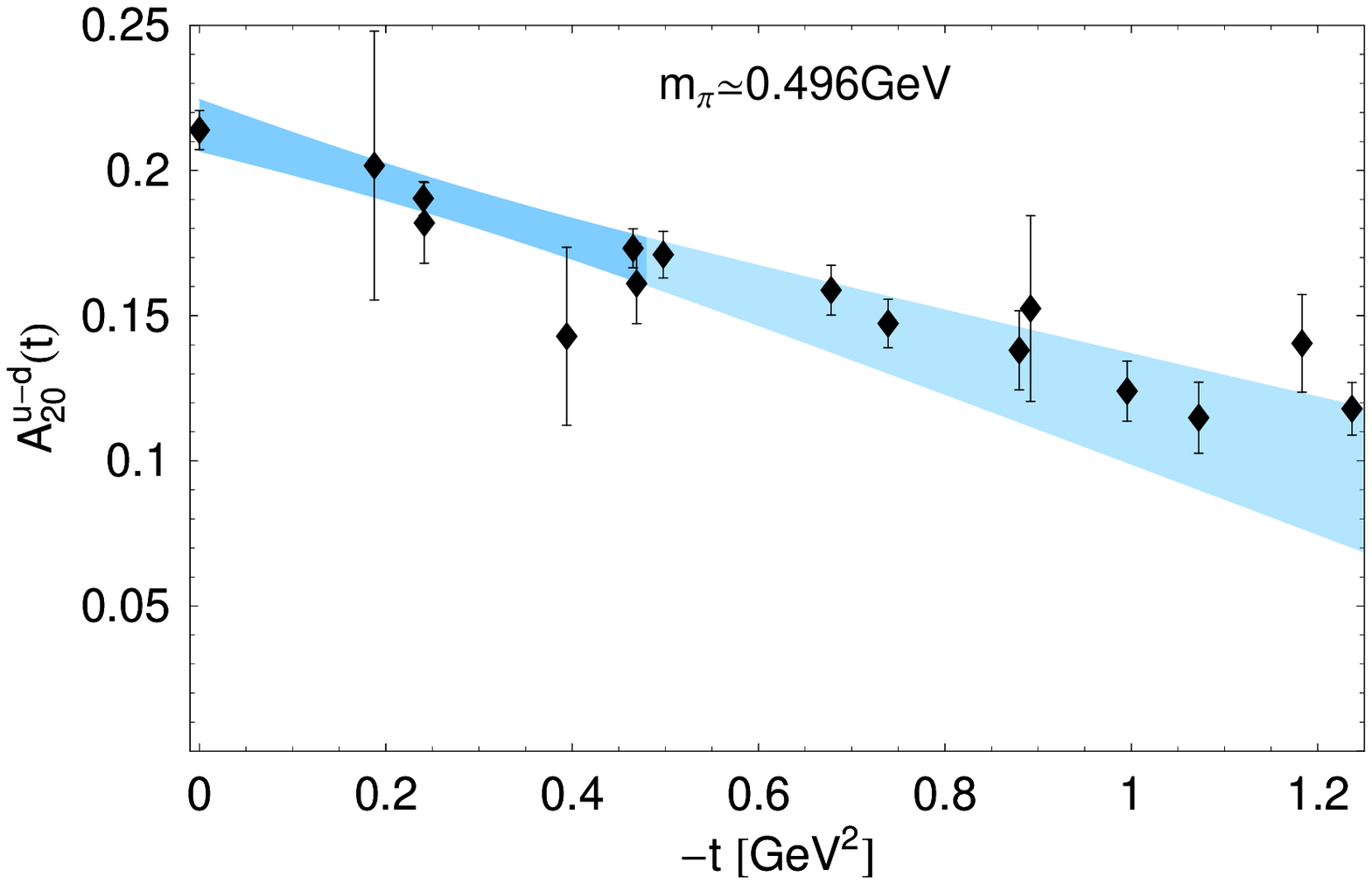}
%      \vspace*{-0.8cm}
  \caption{Lattice results for $A_{20}^{u-d}(t)$ at $m_\pi\approx 0.496$ GeV together with the result of a global simultaneous chiral fit using Eqs.~(\ref{ChPTA20umdp4}), (\ref{ChPTB20umdp4}) and (\ref{ChPTC20umdp4}).}\label{A20IsoVecv4}
     \end{minipage}
   \end{figure}

The $\mathcal{O}(p^2)$ CBChPT result\cite{Dorati:2007bk} for the isovector GFF $A_{20}^{u-d}(t)$ is
\begin{equation}
A_{20}^{u-d}(t,m_\pi)=A_{20}^{0,u-d}\bigg(f_A^{u-d}(m_\pi) + \frac{g_A^2}{192 \pi^2f_\pi^2}h_A(t,m_\pi)\bigg)  + \widetilde A_{20}^{0,u-d} j_A^{u-d}(m_\pi) + 
 A_{20}^{m_\pi,u-d} m_\pi^2 + A_{20}^{t,u-d} t\,,
\label{ChPTA20umdp4}
\end{equation}
where $f_A^{u-d}(m_\pi)$, $h_A(t,m_\pi)$ and  $j_A^{u-d}(m_\pi)$ contain the non-analytic dependence on the 
pion mass and momentum transfer squared and $A_{20}^{0,u-d} \equiv A_{20}^{u-d}(t=0,m_\pi=0)$.
Because of the small prefactor, the term $\propto h_A(t,m_\pi)$ is of 
$\mathcal{O}(10^{-3})$ for $m_\pi\le700$ MeV, $|t|<1$ GeV$^2$
and therefore numerically negligible. Thus, there are essentially no correlations of $t$ and $m_\pi$ present,
and the dependence on $t$ is only due to the counter term $(A_{20}^{t,u-d}\, t)$.
We use the value $\widetilde A_{20}^{0,u-d}=0.17$ obtained from a chiral fit to our lattice results
for $\widetilde A_{20}^{u-d}(t=0)=\langle x\rangle_{\Delta u-\Delta d}$ \cite{Edwards:2006qx}. 
Since the low energy constant $A_{20}^{0,u-d}$ is a common parameter in the CBChPT-formulae 
for the GFFs $A_{20}^{u-d}$, $B_{20}^{u-d}$ and $C_{20}^{u-d}$, we performed a simultaneous fit
based on Eq.~(\ref{ChPTA20umdp4}) (for $A_{20}^{u-d}$), Eq.~(\ref{ChPTB20umdp4}) (for $B_{20}^{u-d}$) and Eq.~(\ref{ChPTC20umdp4})
 (for $C_{20}^{u-d}$) with a total of $9$ (1 common and 8 separate) fit parameters
%From a fit with the three free parameters $A_{20}^{m_\pi,u-d}$, $A_{20}^{t,u-d}$ and $A_{20}^{0,u-d}$ 
to over $120$ lattice datapoints. The details of the CBChPT-extrapolations and the results for
the GFFs  $B_{20}^{u-d}$ and $C_{20}^{u-d}$ will be discussed below in sections 
\ref{sec:B20IsoVecrel} and \ref{sec:C20IsoVecrel}, respectively.
We find $A_{20}^{0,u-d}=0.133(9)$ and $\langle x\rangle_{u-d}=A_{20}^{u-d}(t=0,m_{\pi,\text{phys}})=0.157(10)$ at the physical point. 
This is in very good agreement with phenomenological results from CTEQ and MRST \cite{DurhamDatabase} 
PDF-parametrizations, $\langle x\rangle_{u-d}^{\text{MRST2001}}=0.157(5)$ and $\langle x\rangle_{u-d}^{\text{CTEQ6}}=0.155(5)$.
%errors estimated by comparing CTEQ, MRST and Alekhin  
%
\begin{figure}[h]
     \begin{minipage}{0.4\textwidth}
      \centering
          \includegraphics[scale=0.4,clip=true,angle=0]{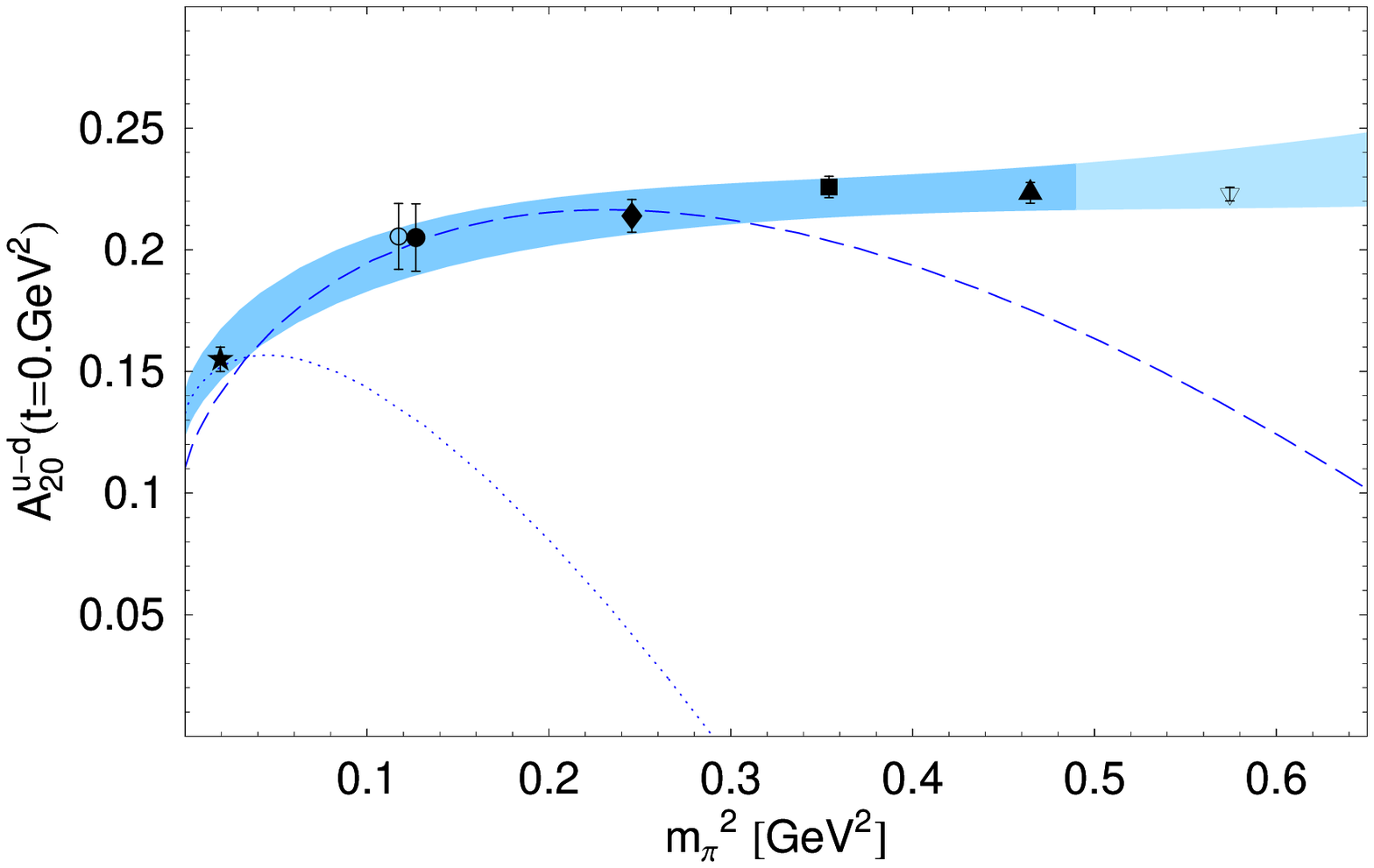}
%          \includegraphics[scale=0.4,clip=true,angle=0]{A20IsoVec_relp4_w_HBChPTcurve_tp0_v3}
%      \vspace*{-0.8cm}
  \caption{Lattice results for $A_{20}^{u-d}$ at $t=0$ GeV$^2$ versus $m_\pi^2$. The error band is the result of
  a global simultaneous chiral fit using Eqs.~(\ref{ChPTA20umdp4}), (\ref{ChPTB20umdp4}) and (\ref{ChPTC20umdp4}). The phenomenological result from CTEQ6 is indicated by the star. The heavy-baryon-limit of the CBChPT fit is shown by the dotted line, and a HBChPT fit to the lattice data for $|t|<0.3\text{GeV}^2$ and $m_\pi<0.5\text{GeV}$ is shown by the dashed line.}\label{A20IsoVecv1}
     \end{minipage}
     %\hfill
     \hspace{1cm}
     \begin{minipage}{0.4\textwidth}
      \centering
          \includegraphics[scale=0.4,clip=true,angle=0]{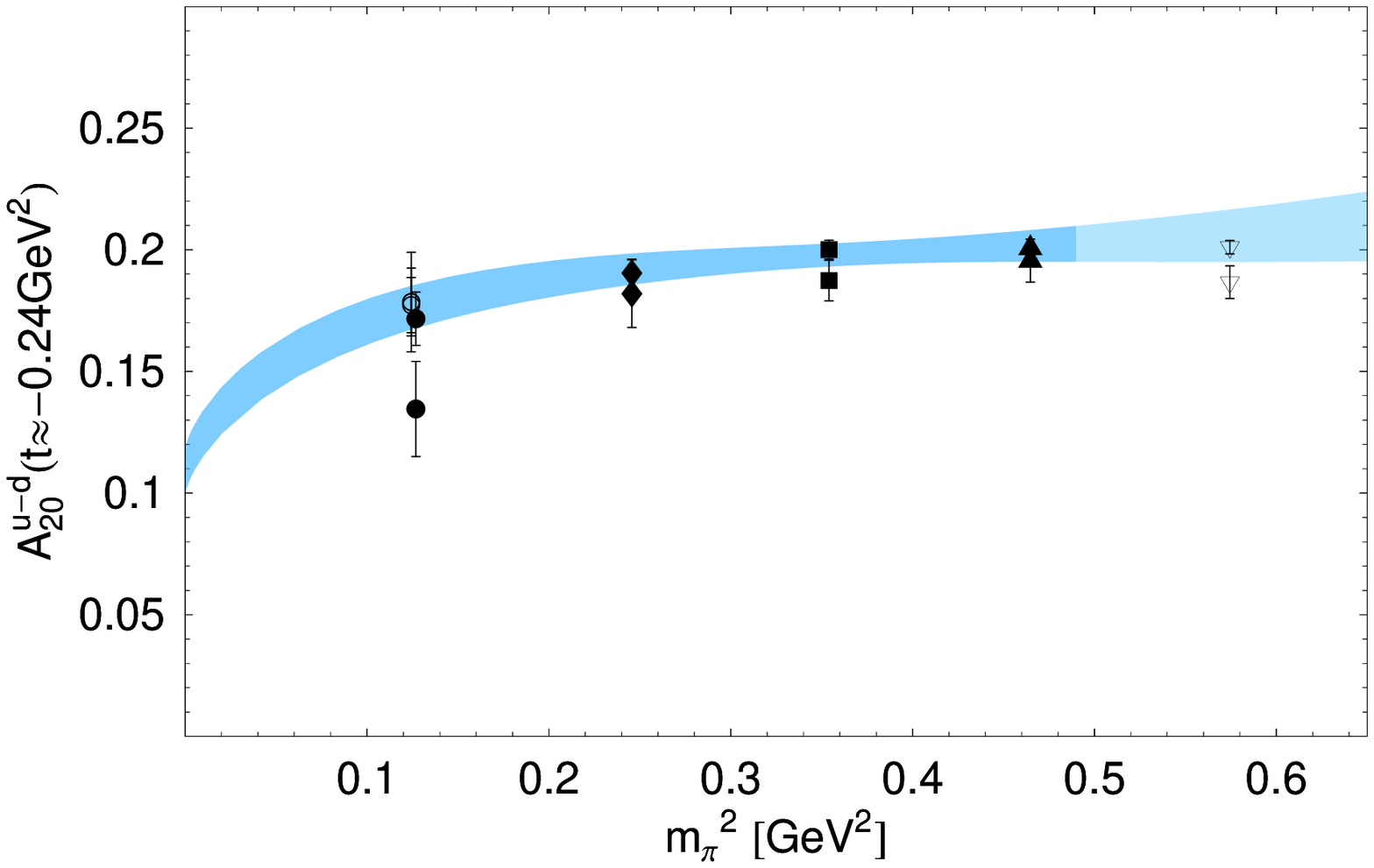}
%      \vspace*{-0.8cm}
  \caption{Lattice results for $A_{20}^{u-d}$ at $t\approx-0.24$ GeV$^2$ versus $m_\pi^2$ together with the result of a global simultaneous chiral fit using Eqs.~(\ref{ChPTA20umdp4}), (\ref{ChPTB20umdp4}) and (\ref{ChPTC20umdp4}).\newline\newline\newline\newline}\label{A20IsoVecv2}
     \end{minipage}
   \end{figure}
The results of the fit are shown in Figs.~(\ref{A20IsoVecv1}) and (\ref{A20IsoVecv2}).
The dependence of $A_{20}^{u-d}(t)$ on the momentum transfer squared is presented in Figs.~(\ref{A20IsoVecv3}) and (\ref{A20IsoVecv4}), where we again obtain a good description of the lattice data.

To study the difference between HBChPT and CBChPT, we took the heavy baryon limit of CBChPT while keeping the same values of the fit parameters, and obtained the dotted line in Fig.~(\ref{A20IsoVecv3}).  This curve only overlaps the CBChPT curve for  $m_\pi<m_{\pi,\text{phys}}$ and  drops off sharply for $m_\pi>m_{\pi,\text{phys}}$, indicating the quantitative importance of the truncated terms when using the coefficients from the CBChPT fit.  In addition, it is important to ask the separate question of how well the lattice data can be fit with the HBChPT expression when the coefficients are determined directly by a best fit to the data. The dashed curve in Fig.~(\ref{A20IsoVecv1}) shows the result of fitting  our lattice data for $|t|<0.3\text{GeV}^2$ and $m_\pi<0.5\text{GeV}$, and indicates that HBChPT describes the behavior of our lattice data over a significantly smaller range of pion masses than CBChPT.

\begin{figure}[thbp]
     \begin{minipage}{0.4\textwidth}
      \centering
          \includegraphics[scale=0.4,clip=true,angle=0]{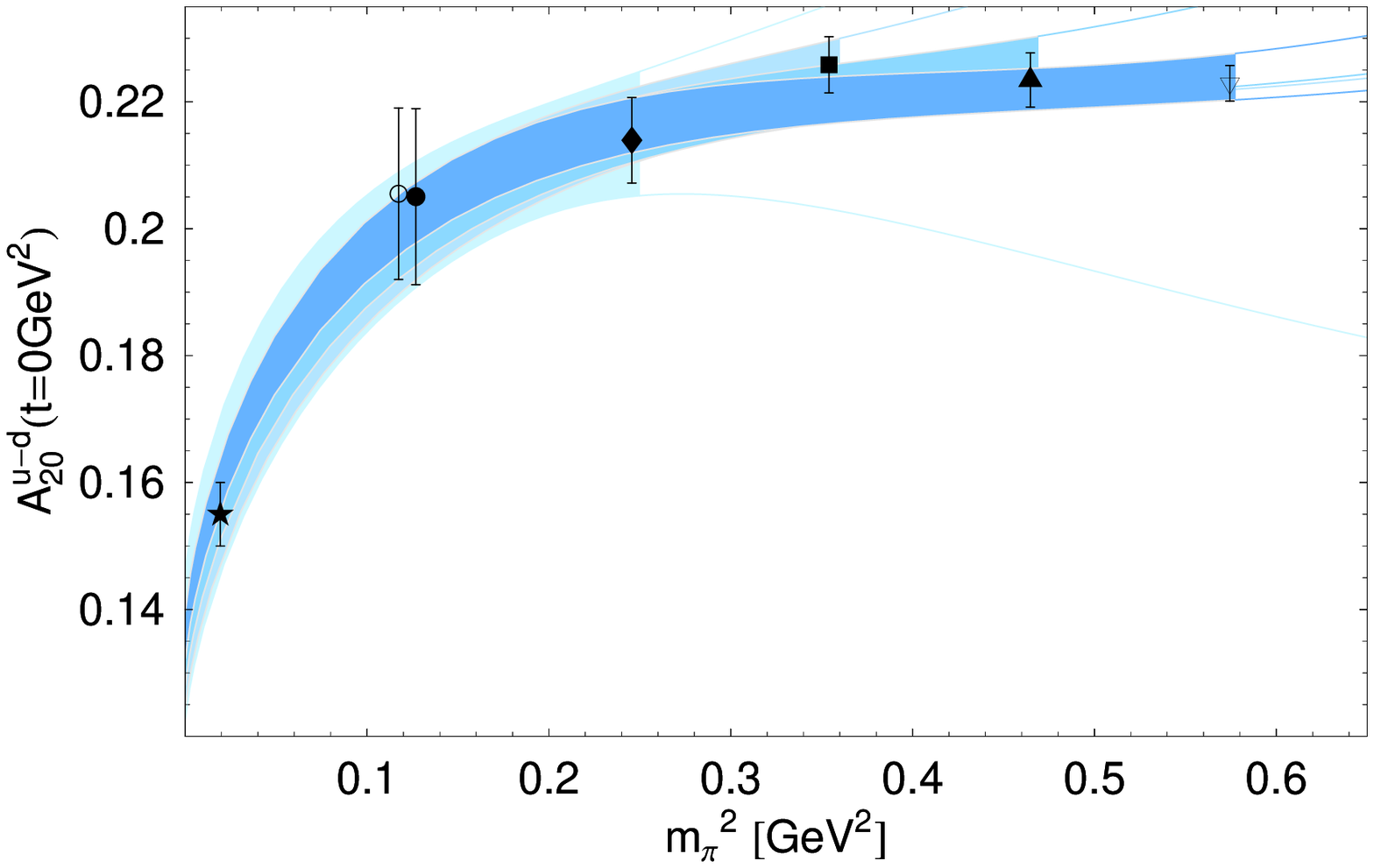}
%      \vspace*{-0.8cm}
  \caption{Lattice results for $A_{20}^{u-d}$ at $t=0$ GeV$^2$ versus $m_\pi^2$ together with chiral fits based on Eq.~(\ref{ChPTA20umdp4}). The four different error bands represent chiral fits to lattice results including pion 
masses in the regions $m_\pi<500$, $600$, $685$ and $760$ MeV. \newline}\label{A20IsoVecmPiRegions}
     \end{minipage}
     %\hfill
     \hspace{1cm}
     \begin{minipage}{0.4\textwidth}
      \centering
         \includegraphics[scale=0.4,clip=true,angle=0]{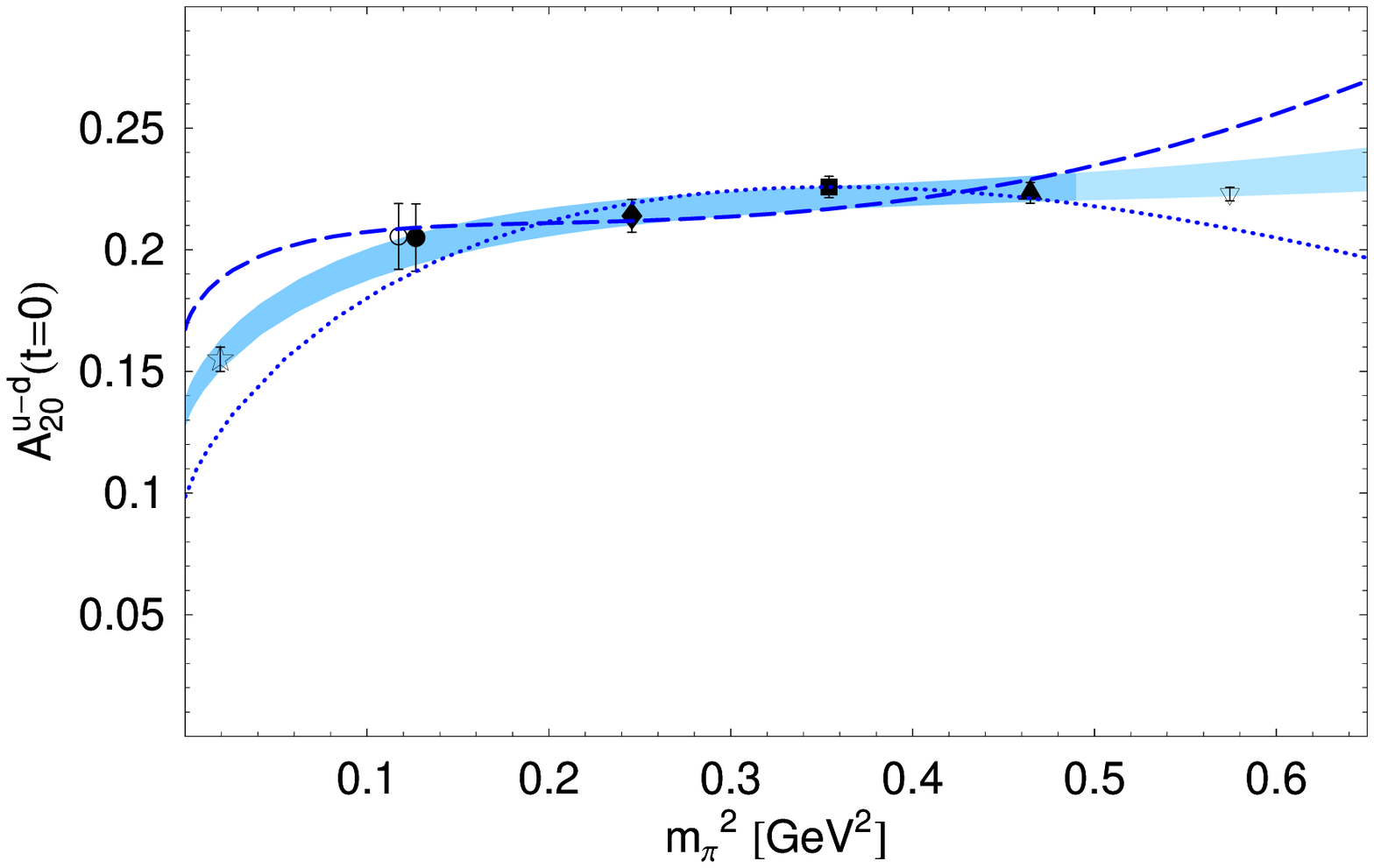}
%      \vspace*{-0.8cm}
  \caption{Chiral fits including the terms in Eq.~(\ref{ChPTA20umdp4}) plus an additional $m_\pi^3$ term as described in the text.  
  The narrow band is the original band omitting this term, the dashed line corresponds to $\delta_A^{(3),u-d}=1$, and
  the dotted line to $\delta_A^{(3),u-d}=-1$. 
 % Estimate of $\mathcal{O}(p^3)$-corrections to the chiral extrapolation of $A_{20}^{u-d}$
%  as advocated in Ref.~\cite{Dorati:2007bk}. The lines are based on fits using Eq.~(\ref{ChPTA20umdp4}) plus an additional term $\propto m_\pi^3$ as explained in the text. The dashed line corresponds to $\delta_A^{(3),u-d}=1$, and
%  the dotted line to $\delta_A^{(3),u-d}=-1$. This type of $\mathcal{O}(p^3)$-correction $\propto m_\pi^3$
% is clearly inconsistent with the behavior of the data.
}\label{A20umdOrderp3}
 \end{minipage}
\end{figure}

Because limitations in computational resources presently require us to include lattice data extending to such large pion masses, it would clearly be desirable to carry out a chiral perturbation theory analysis consistently including all terms of $\mathcal{O}(p^3)$. In the absence of the requisite full ChPT analysis, we have studied uncertainties in the chiral extrapolations by repeating the fit for different maximal values of the included pion masses. Figure~(\ref{A20IsoVecmPiRegions}) shows a comparison of the chiral extrapolations of $A_{20}^{u-d}$, based on fits to the lattice data in the regions $m_\pi<500$, $600$, $685$ and $760$ MeV. We find that the extrapolations to the chiral limit fully agree within statistical errors in all four cases. Note that the experimental point shown in Fig.~(\ref{A20IsoVecmPiRegions}) was not included in the fits, but each of the four  analyses is consistent with it.  This insensitivity to the upper mass cutoff shows that the strong bending towards the physical point is not driven by the large pion mass region, where $\mathcal{O}(p^3)$ corrections would be largest. Furthermore, all four chiral fits are, within statistical errors, in agreement with the lattice data points at large pion masses. This indicates that the present statistical error envelope is comparable to any systematic effects due to  higher order corrections.

Another prescription to estimate $\mathcal{O}(p^3)$ corrections that has been advocated in the literature Ref.~\cite{Dorati:2007bk} is simply adding a single $m^3_\pi$ term and, assuming both ``naturalness" of the coefficient and the lack of other functional forms, seeing what error band arises from varying the coefficient from -1 to +1. Thus, to explore this possibility, following Ref.~\cite{Dorati:2007bk}, we have added to the result in Eq.~(\ref{ChPTA20umdp4}) the term  $\delta_A^{(3),u-d}m_\pi^3/(\Lambda_\chi^2 m_N^0)$, where  $\Lambda_\chi\approx1.2$ GeV is the chiral symmetry breaking scale, and varied the constant  $\delta_A^{(3),u-d}$ in the range $-1,\ldots,+1$. The results of fits to the lattice data for  $A_{20}^{u-d}$ including this additional term are shown in Fig.~(\ref{A20umdOrderp3}), where the error band corresponds to  $\delta_A^{(3),u-d}=0$, the dashed line corresponds to  $\delta_A^{(3),u-d}=+1$, and the dotted line corresponds to $\delta_A^{(3),u-d}=-1$. From the figure, we note that this $m^3_\pi$ term alone with coefficients +1 and -1 is clearly inconsistent with the behavior of the data.  Theoretically, this is not unreasonable, since the foundation of the $\overline{\mbox{IR}}$ regularization scheme is a resummation of classes of terms, and here a single cubic term has been arbitrarily singled out. Analogous fits with similar qualitatively inconsistent behavior were also obtained for $A_{20}^{u+d}$, treated in a later section. 
%Hence, although this prescription may provide useful estimates in other contexts, it is not useful in this work, and hence we do not %include it in subsequent fits.
Hence, although this prescription may provide useful estimates in other contexts, we do not believe it is useful in this work, and 
hence we do not include it in subsequent fits.

   %
   %
%
%   
%%%%%%%%%%%%%%%%%%%%%%%%%%%%%%%%%%%%%%%%%%%%%%%%%%%%%%%%%%%%%%%%%%%%%%%%%%%%%%%%%%%%%%%%%%%%%%%%%%%%%%%%%%%%%%%%%%%%%%%%%%%%%%
\subsection{\label{sec:B20IsoVecrel}CBChPT extrapolation of $B_{20}^{u-d}(t)$}
The  $\mathcal{O}(p^2)$ CBChPT calculation \cite{Dorati:2007bk} for the isovector $B_{20}$ GFF gives 
\begin{equation}
B_{20}^{u-d}(t,m_\pi)=\frac{m_N(m_\pi)}{m_N} B_{20}^{0,u-d} + A_{20}^{0,u-d} h_B^{u-d}(t,m_\pi) 
+ \frac{m_N(m_\pi)}{m_N}\bigg\{\delta_{B}^{t,u-d}\,t + \delta_{B}^{m_\pi,u-d}\,m_\pi^2  \bigg\}\,,
\label{ChPTB20umdp4}
\end{equation}
where $m_N(m_\pi)$ is the pion mass dependent nucleon mass, 
$B_{20}^{0,u-d} \equiv B_{20}^{u-d}(t=0,m_\pi=0)$,
and where we have included estimates of $\mathcal{O}(p^3)$-corrections in form of $(\delta_{B}^{t,u-d}\,t)$
and $(\delta_{B}^{m_\pi,u-d}\,m_\pi^2)$. The low energy constants
$B_{20}^{0,u-d}$, $\delta_{B}^{t,u-d}$ and $\delta_{B}^{m_\pi,u-d}$ are treated as free parameters and may be 
obtained from a fit to the lattice data.
The non-analytic dependence on $m_\pi$ and $t$ is given by $h_B^{u-d}(t,m_\pi)$, but
it turns out that this function is approximately independent of $t$, $h_B^{u-d}(t,m_\pi)\approx h_B^{u-d}(m_\pi)$
for $m_\pi\le700$ MeV, $|t|<1$ GeV$^2$. The $t$-dependence is therefore in practice linear due
to the $\mathcal{O}(p^3)$-correction term.
\begin{figure}[t]
     \begin{minipage}{0.4\textwidth}
      \centering
          \includegraphics[scale=0.4,clip=true,angle=0]{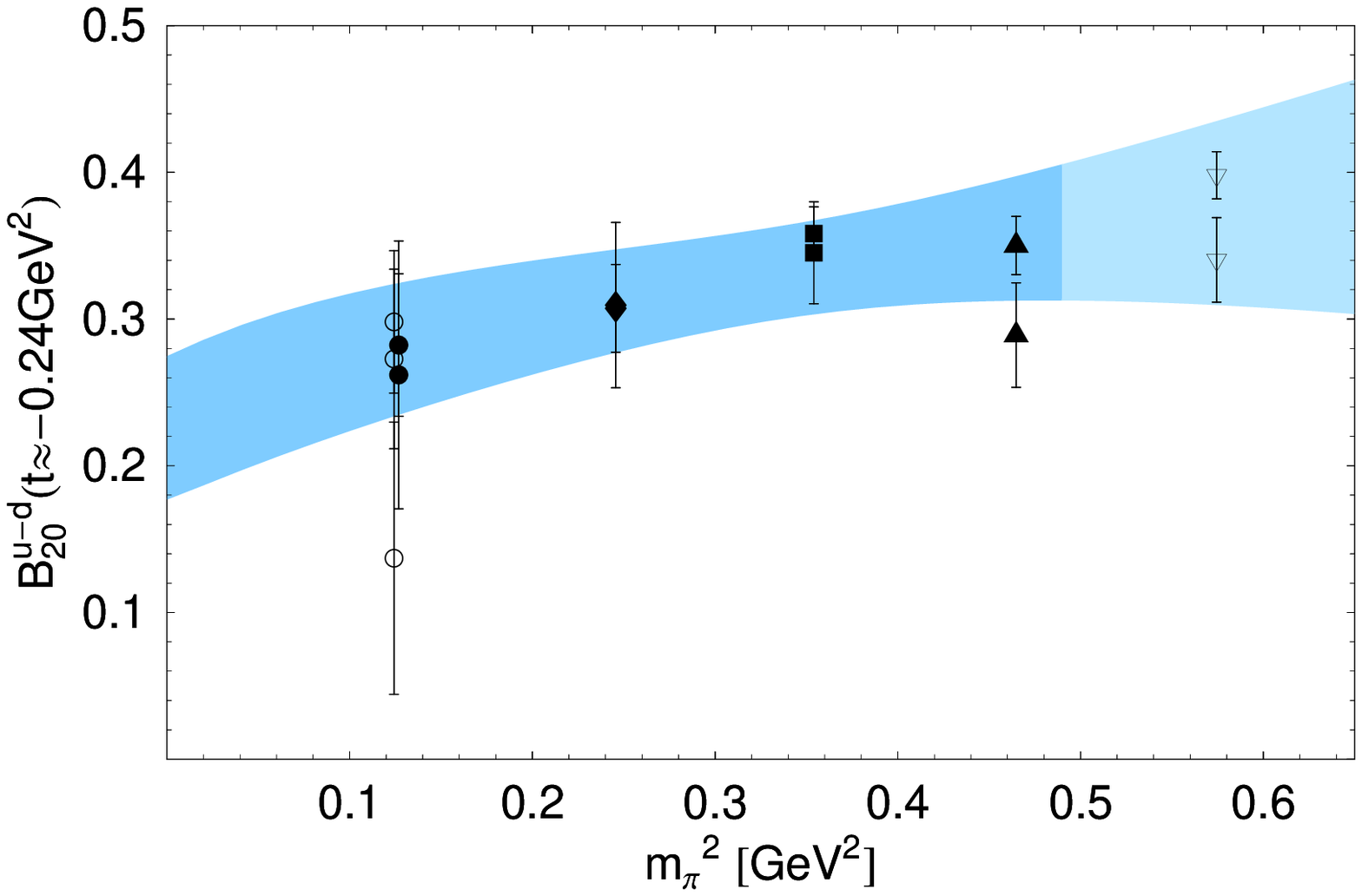}
%         \includegraphics[scale=0.4,clip=true,angle=0]{B20IsoVecExtended_relp4_tp24}
%      \vspace*{-0.8cm}
  \caption{Lattice results for $B_{20}^{u-d}$ at $t\approx-0.24$ GeV$^2$ versus $m_\pi^2$ together 
  with the result of a global simultaneous chiral fit using Eqs.~(\ref{ChPTA20umdp4}), (\ref{ChPTB20umdp4}) and (\ref{ChPTC20umdp4}).}\label{B20p4v1}
     \end{minipage}
     %\hfill
     \hspace{1cm}
     \begin{minipage}{0.4\textwidth}
      \centering
          \includegraphics[scale=0.4,clip=true,angle=0]{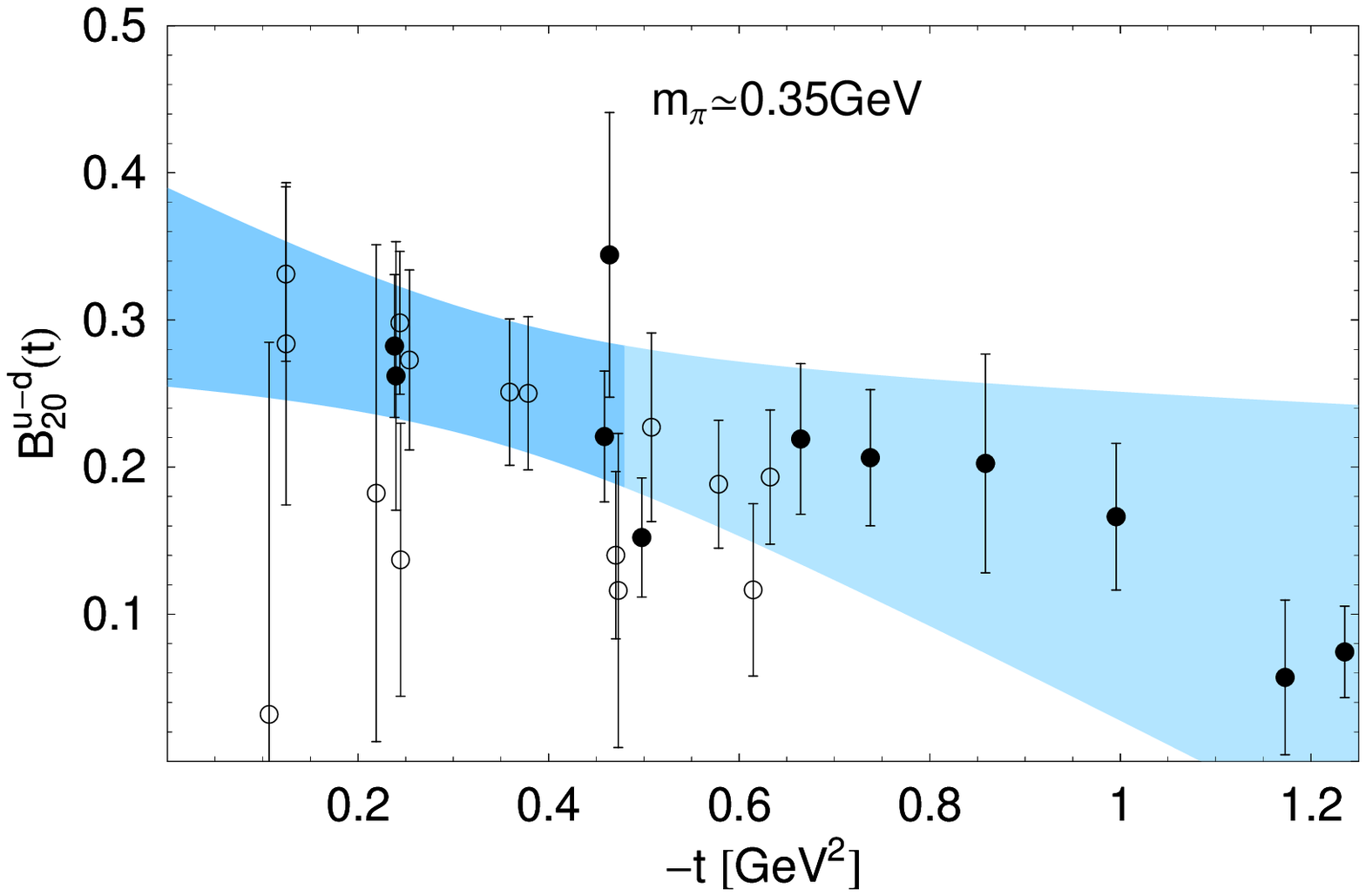}
%      \vspace*{-0.8cm}
  \caption{Lattice results for $B_{20}^{u-d}(t)$ at $m_\pi\approx350$ MeV versus $(-t)$
  together with the result of a global simultaneous chiral fit using Eqs.~(\ref{ChPTA20umdp4}), (\ref{ChPTB20umdp4}) and (\ref{ChPTC20umdp4}).}
  \label{B20p4v2}
     \end{minipage}
   \end{figure}
For the $m_\pi$-dependent nucleon mass we use 
$\mathcal{O}(p^4)$ CBChPT \cite{Gail:2006ai,Dorati:2007bk} fitted to our lattice results for $m_N$.
The chiral extrapolation of $B_{20}^{u-d}$ is based on a global simultaneous fit as discussed in the previous section
with $A_{20}^{0,u-d}$ as common fit parameter in Eqs.~(\ref{ChPTA20umdp4}), (\ref{ChPTB20umdp4}) and (\ref{ChPTC20umdp4}).  
We obtain $B_{20}^{0,u-d}=0.263(62)$ and $B_{20}^{u-d}(t=0,m_{\pi,\text{phys}})=0.273(63)$ 
at the physical pion mass. Results of the fit are shown in Figs.~(\ref{B20p4v1}) and (\ref{B20p4v2}).
\subsection{\label{sec:C20IsoVecrel}CBChPT extrapolation of $C_{20}^{u-d}(t)$}
The pion mass dependence of the isovector GFF $C_{20}$  in CBChPT to  $\mathcal{O}(p^2)$  is
very similar to that of the isovector $B_{20}$ above and given by \cite{Dorati:2007bk}
\begin{equation}
C_{20}^{u-d}(t,m_\pi)=\frac{m_N(m_\pi)}{m_N} C_{20}^{0,u-d} + A_{20}^{0,u-d} h_C^{u-d}(t,m_\pi) 
+ \frac{m_N(m_\pi)}{m_N}\bigg\{\delta_{C}^{t,u-d}\,t + \delta_{C}^{m_\pi,u-d}\,m_\pi^2 \bigg\}\,,
\label{ChPTC20umdp4}
\end{equation}
where $C_{20}^{0,u-d} \equiv C_{20}^{u-d}(t=0,m_\pi=0)$. 
\begin{figure}[htbp]
     \begin{minipage}{0.4\textwidth}
      \centering
          \includegraphics[scale=0.4,clip=true,angle=0]{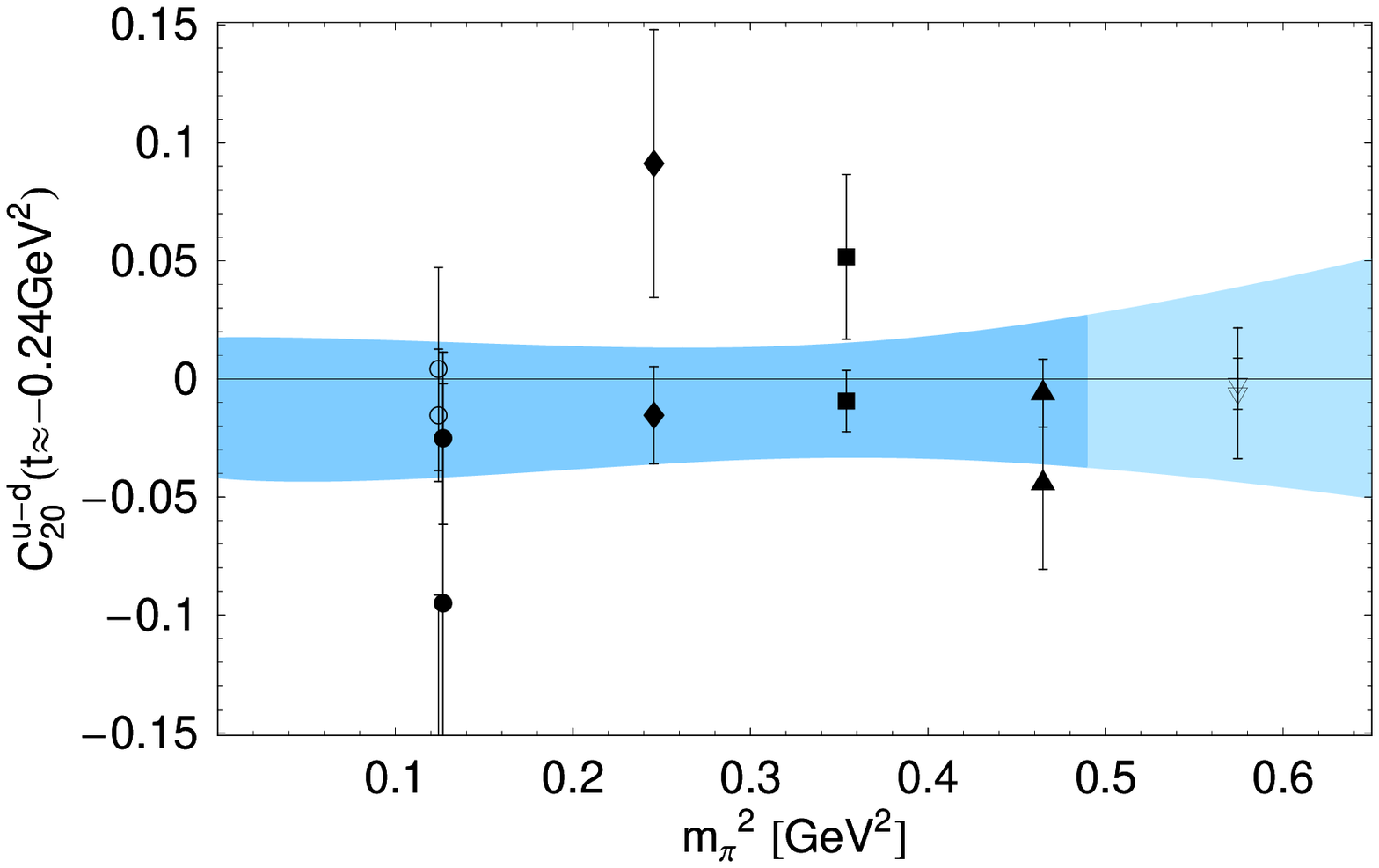}
%          \includegraphics[scale=0.4,clip=true,angle=0]{C20IsoVecExtended_relp4_tp24}
%      \vspace*{-0.8cm}
  \caption{Lattice results for $C_{20}^{u-d}$ at $t\approx-0.24$ GeV$^2$ versus $m_\pi^2$ together 
  with the result of a global simultaneous chiral fit using Eqs.~(\ref{ChPTA20umdp4}), (\ref{ChPTB20umdp4}) and (\ref{ChPTC20umdp4}).}
  \label{C20p4v1}
     \end{minipage}
     %\hfill
     \hspace{1cm}
     \begin{minipage}{0.4\textwidth}
      \centering
          \includegraphics[scale=0.4,clip=true,angle=0]{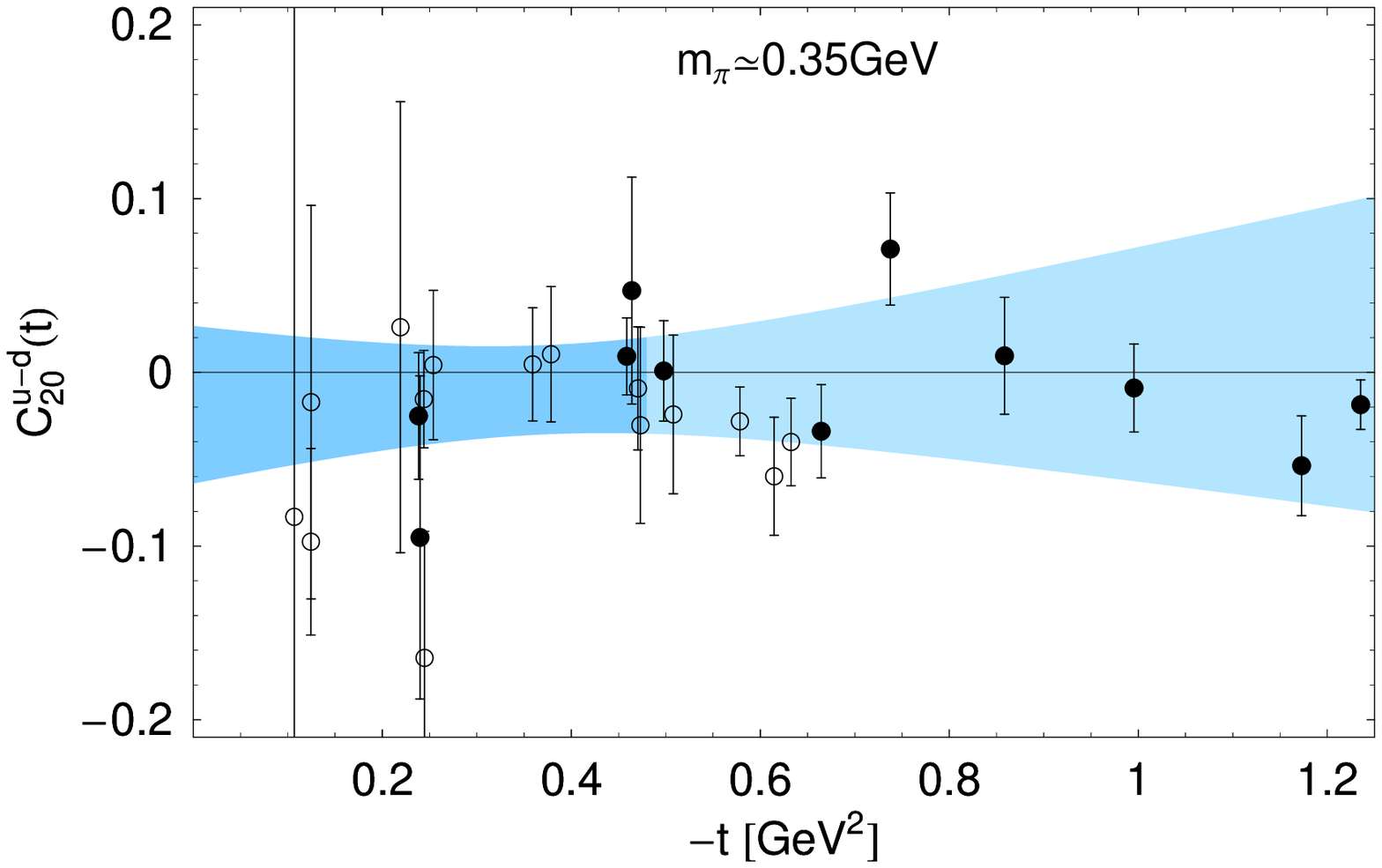}
%      \vspace*{-0.8cm}
  \caption{Lattice results for $C_{20}^{u-d}(t)$ at $m_\pi\approx350$ MeV versus $(-t)$ 
  together with the result a global simultaneous chiral fit using Eqs.~(\ref{ChPTA20umdp4}), (\ref{ChPTB20umdp4}) and (\ref{ChPTC20umdp4}).}
  \label{C20p4v2}
     \end{minipage}
   \end{figure}
As in the case of $B_{20}^{u-d}$, $(\delta_{C}^{t,u-d}\,t)$ 
and $(\delta_{C}^{m_\pi,u-d}\,m_\pi^2)$ represent 
$\mathcal{O}(p^3)$-correction terms, and it turns out that $h_C^{u-d}(t,m_\pi)$ 
is practically independent of $t$. From a global simultaneous chiral fit based on 
Eqs.~(\ref{ChPTA20umdp4}), (\ref{ChPTB20umdp4}) and (\ref{ChPTC20umdp4}) with common
fit parameter $A_{20}^{0,u-d}$ as discussed in section \ref{sec:A20IsoVecrel}, 
%With the same input parameters as for $B_{20}^{u-d}$ and the same number of free parameters, 
we obtain $C_{20}^{0,u-d}=-0.017(39)$ and $C_{20}^{u-d}(t=0,m_{\pi,\text{phys}})=-0.017(41)$ at the physical point.
The results are presented in Figs.~(\ref{C20p4v1}) and (\ref{C20p4v2}).
We note that $C^{u-d}_{20}(t)$ is roughly one order of magnitude smaller than 
$A^{u-d}_{20}(t)$ and $B^{u-d}_{20}(t)$ and fully compatible with zero within errors, $C_{20}^{u-d}(t)\approx 0$.
This implies a rather mild dependence of the $n=2$ moment of the GPDs $H^{u-d}(x,\xi,t)$ and $E^{u-d}(x,\xi,t)$ 
on the longitudinal momentum transfer $\xi$ (at least for small $\xi$), so that
$H^{n=2}_{u-d}(\xi,t) \approx A^{u-d}_{20}(t)$ and $E^{n=2}_{u-d}(\xi,t) \approx B^{u-d}_{n0}(t)$.
%\vspace{-3mm}
\subsection{\label{sec:A20Isosinglet}CBChPT extrapolation of $A_{20}^{u+d}(t)$}
The (total) isosinglet momentum fraction of quarks, $A_{20}^{u+d}(t=0)=\langle x\rangle_{u+d}$
is not only an important hadron structure observable on its own but is in addition an essential
ingredient for the computation of the total angular momentum contribution of quarks to the nucleon spin,
$J^{u+d}=1/2(A_{20}^{u+d}(0)+B_{20}^{u+d}(0))$. 
The combined $(t,m_\pi)$-dependence in CBChPT is given by \cite{Dorati:2007bk}:
\begin{equation}
A_{20}^{u+d}(t,m_\pi)=A_{20}^{0,u+d}\bigg(f^{u+d}_A(m_\pi) - \frac{g_A^2}{64 \pi^2f_\pi^2}h_A(t,m_\pi)\bigg)  + 
 A^{m_\pi,u+d}_{20} m_\pi^2 + A^{t,u+d}_{20} t + \Delta A_{20}^{u+d}(t,m_\pi) + \mathcal{O}(p^3)\,,
\label{ChPTA20umdp4new}
\end{equation}
where $A_{20}^{0,u+d}  \equiv A_{20}^{u+d}(t=0,m_\pi=0)$, $f^{u+d}_A(m_\pi)$ and 
$h_A(t,m_\pi)$ contain the non-analytic dependence on the 
pion mass and momentum transfer squared, and the constants $A^{m_\pi,u+d}_{20}$ and $A^{t,u+d}$
may be obtained from a fit to the lattice data.
In this counting scheme, contributions from operator
insertions in the pion line proportional to the momentum fraction of quarks in the pion
in the chiral limit, $\langle x\rangle^{\pi,0}_{u+d}$, are of order $\mathcal{O}(p^3)$. 
However, in order to see if such contributions could be relevant
for the pion masses and values of the momentum transfer squared accessible in our calculation,
we include the estimate of the $\mathcal{O}(p^3)$-contribution 
$\Delta A_{20}^{u+d}$ provided in \cite{Dorati:2007bk} in the fit to the lattice data points.

\begin{figure}[htbp]
     \begin{minipage}{0.4\textwidth}
      \centering
          \includegraphics[scale=0.4,clip=true,angle=0]{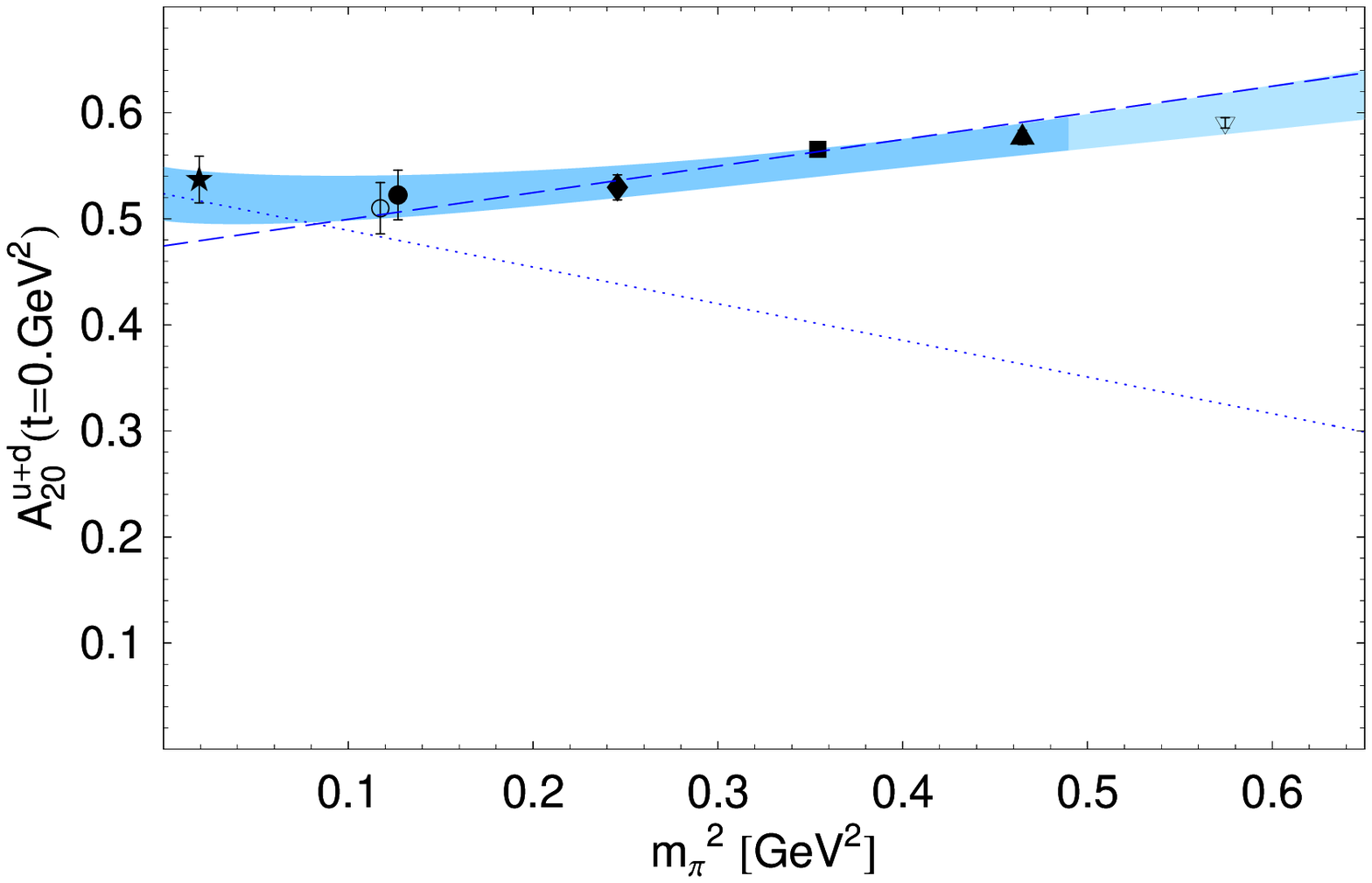}
%          \includegraphics[scale=0.4,clip=true,angle=0]{A20IsoSingl_relp4_t0_v3}
%      \vspace*{-0.8cm}
  \caption{Lattice results for $A_{20}^{u+d}$ at $t=0$ GeV$^2$ versus $m_\pi^2$. The error band is the result of 
  a global simultaneous chiral fit using Eqs.~(\ref{ChPTA20umdp4new}), (\ref{ChPTB20updp4}) and (\ref{ChPTC20updp4}). 
 The phenomenological value from CTEQ6 is denoted by a star. The heavy-baryon-limit of the CBChPT fit is 
  shown by the dotted line, and a HBChPT fit to the lattice data for $|t|<0.3\text{GeV}^2$ and $m_\pi<0.5\text{GeV}$ is shown by the dashed line.}\label{A20v3}
     \end{minipage}
     %\hfill
     \hspace{1cm}
     \begin{minipage}{0.4\textwidth}
      \centering
          \includegraphics[scale=0.4,clip=true,angle=0]{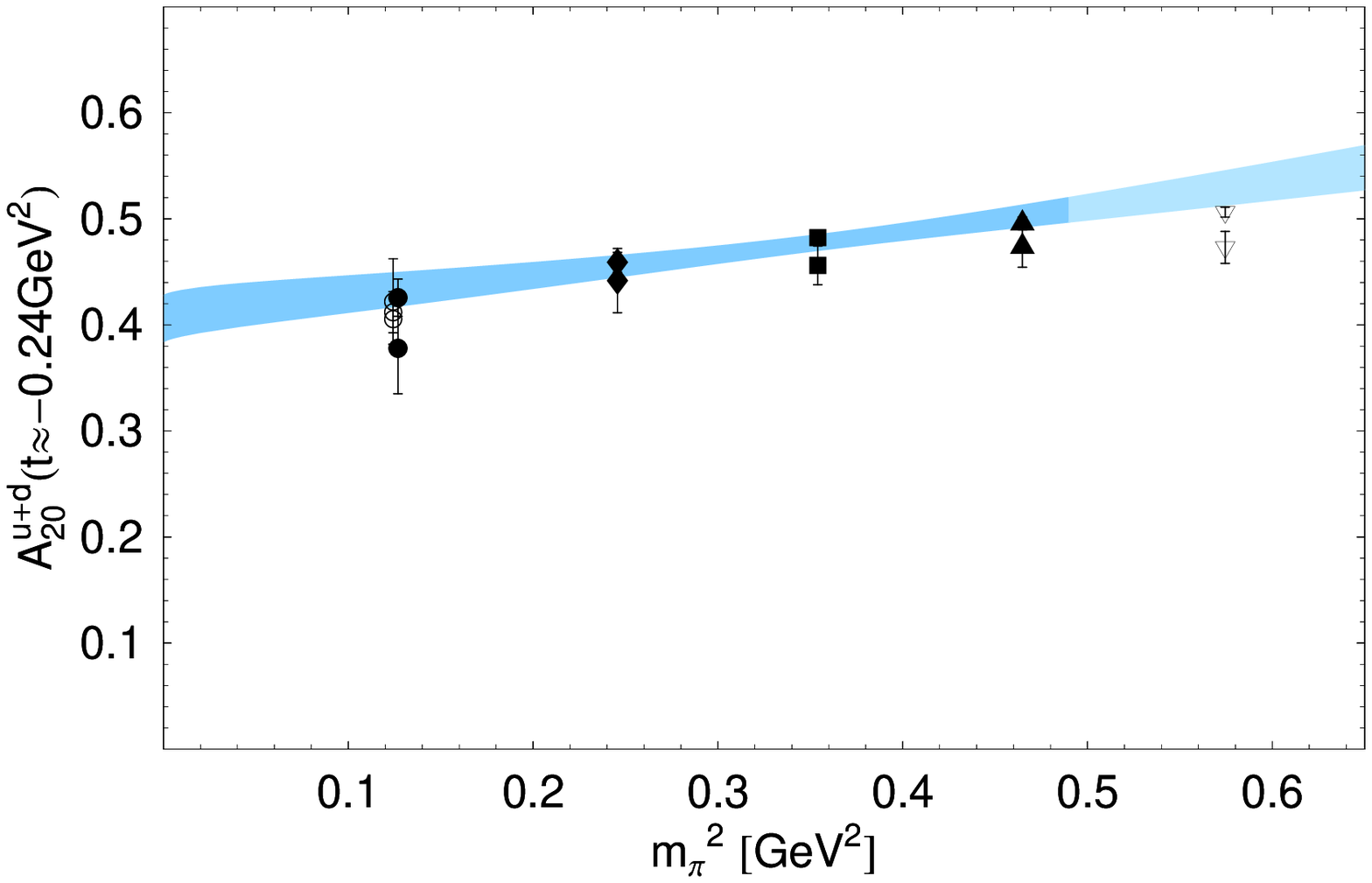}
%      \vspace*{-0.8cm}
  \caption{Lattice results for $A_{20}^{u+d}(t)$ at $t\approx-0.24$ GeV$^2$ versus $m_\pi^2$ together with the result of  a global simultaneous chiral fit using Eqs.~(\ref{ChPTA20umdp4new}), (\ref{ChPTB20updp4}) and (\ref{ChPTC20updp4}).\newline\newline\newline\newline}\label{A20v4}
     \end{minipage}
   \end{figure}

\begin{figure}[htbp]
     \begin{minipage}{0.4\textwidth}
      \centering
          \includegraphics[scale=0.4,clip=true,angle=0]{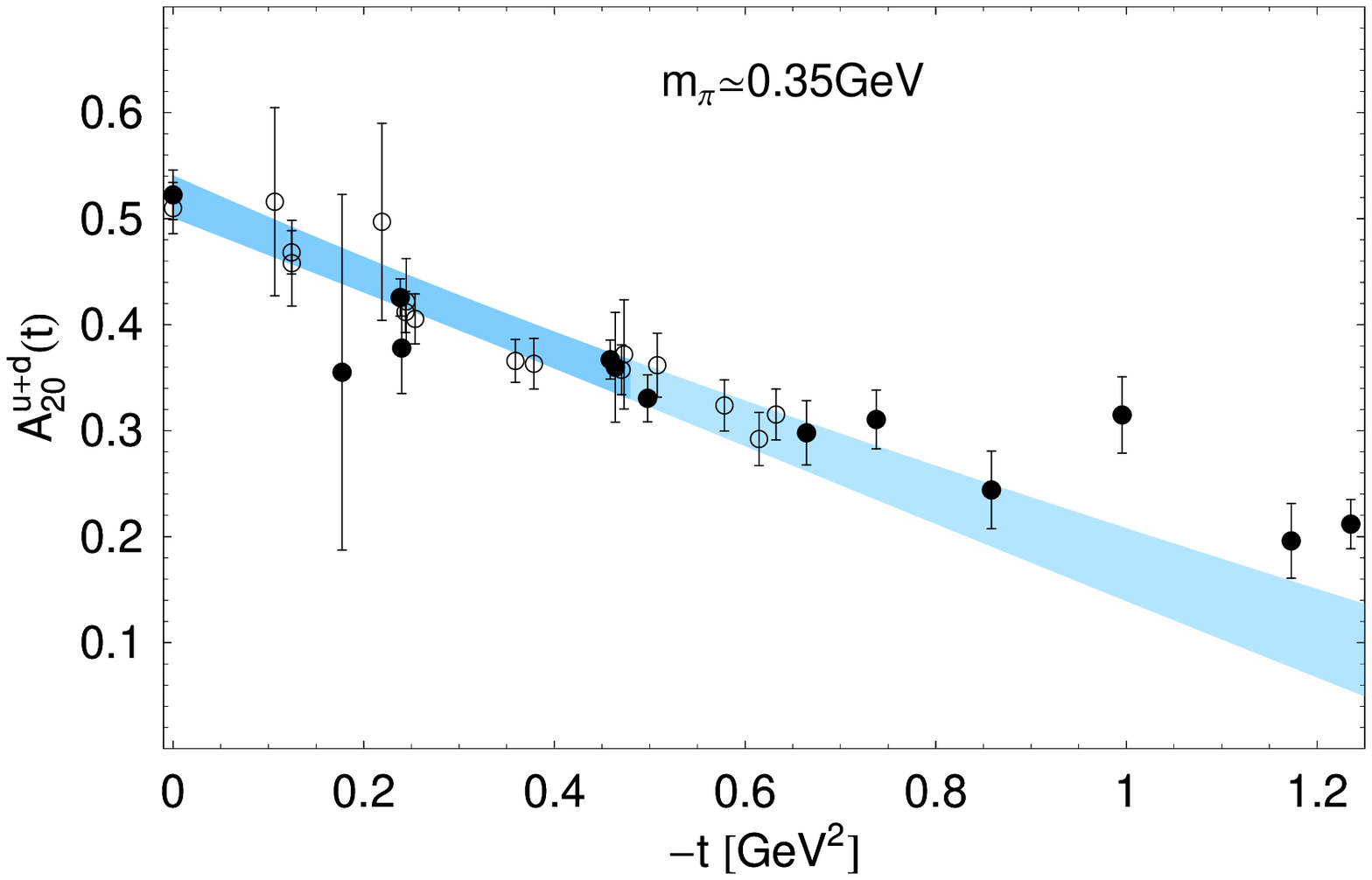}
%      \vspace*{-0.8cm}
  \caption{Lattice results for $A_{20}^{u+d}(t)$ at $m_\pi\approx 0.35$ GeV together with the result of  a global simultaneous chiral fit using Eqs.~(\ref{ChPTA20umdp4new}), (\ref{ChPTB20updp4}) and (\ref{ChPTC20updp4}).}\label{A20v5}
     \end{minipage}
     %\hfill
     \hspace{1cm}
     \begin{minipage}{0.4\textwidth}
      \centering
          \includegraphics[scale=0.4,clip=true,angle=0]{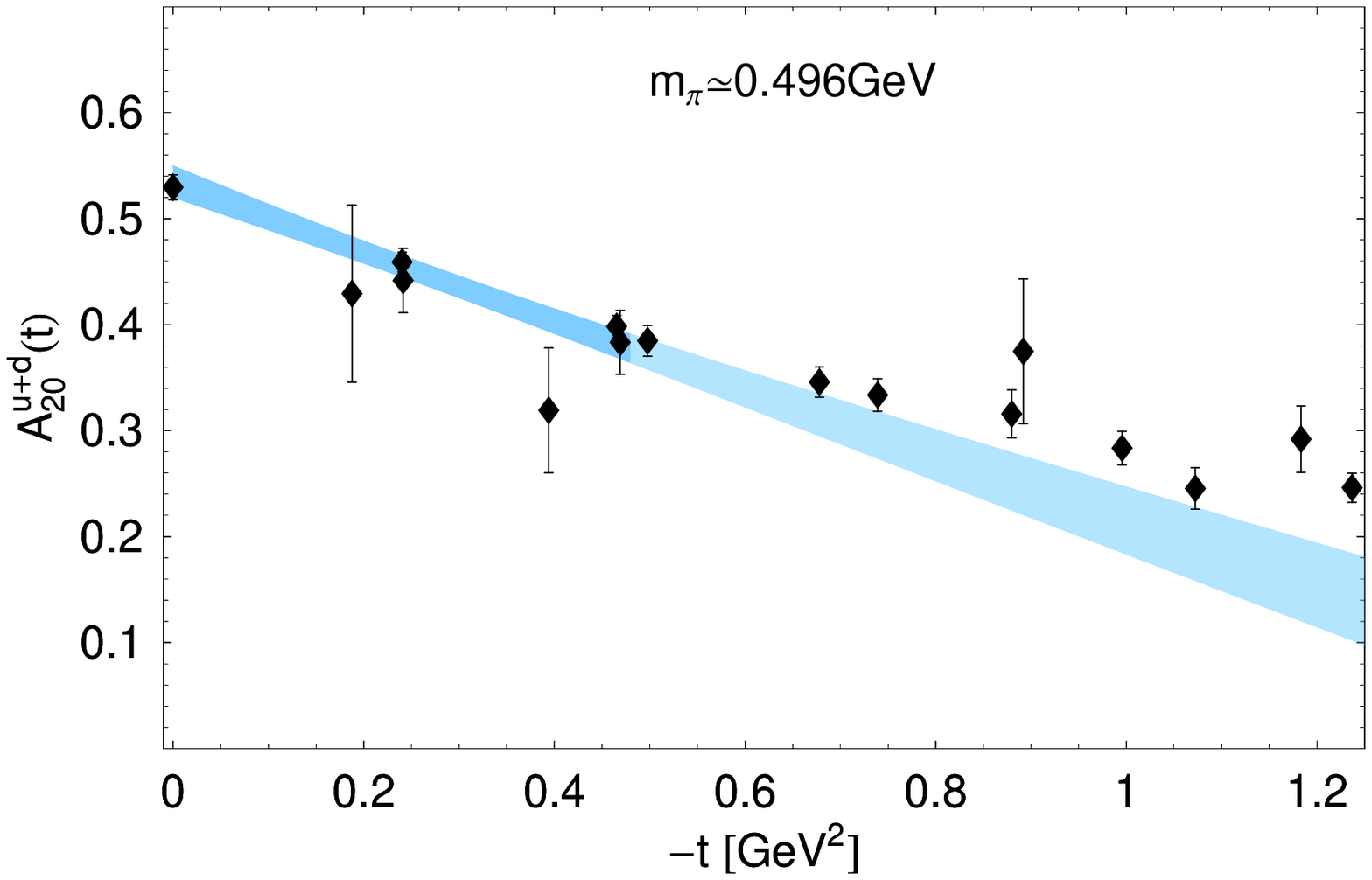}
%      \vspace*{-0.8cm}
  \caption{Lattice results for $A_{20}^{u+d}(t)$ at $m_\pi\approx 0.496$ GeV together with the result of a global simultaneous chiral fit using Eqs.~(\ref{ChPTA20umdp4new}), (\ref{ChPTB20updp4}) and (\ref{ChPTC20updp4}).}\label{A20v6}
     \end{minipage}
   \end{figure}

Similar to the isovector case discussed in the previous sections, the low energy constant
$A_{20}^{0,u+d}$ is a common parameter in the chiral 
extrapolation formulae for the isosinglet GFFs $A_{20}^{u+d}$, $B_{20}^{u+d}$ and $C_{20}^{u+d}$.
Using $\langle x\rangle^{\pi,0}_{u+d}=0.5$ from Table~\ref{tab:LECs} as an input parameter, we performed
a simultaneous fit to over 120 lattice data points for these three GFFs, based on 
Eqs.~(\ref{ChPTA20umdp4new}), (\ref{ChPTB20updp4}) and (\ref{ChPTC20updp4}), with 1 common and 8 separate
low energy constants as fit parameters. For the details of the CBChPT results for $B_{20}^{u+d}$ and $C_{20}^{u+d}$
we refer to the sections \ref{sec:B20Isosinglet} and \ref{sec:C20Isosinglet} below.
%Using $\langle x\rangle^{\pi,0}_{u+d}=0.5$ from Table~\ref{tab:LECs} as an input parameter,
%fitting the three parameters $A_{20}^{0,u+d}$, $A^{m_\pi,u+d}_{20}$ and $A^{t,u+d}$ to more than $40$ lattice data points gives $A_{20}^{0,u+d}=0.524(14)$ and
The chiral fit gives $A_{20}^{0,u+d}=0.524(25)$ and 
$\langle x\rangle_{u+d}=A_{20}^{u+d}(t=0,m_{\pi,\text{phys}})=0.520(24)$ at the physical point. 
Again, this is in very good agreement with phenomenological results from CTEQ 
and MRST \cite{DurhamDatabase} parametrizations, $\langle x\rangle_{u+d}^{\text{MRST2001}}=0.538(22)$ and $\langle x\rangle_{u+d}^{\text{CTEQ6}}=0.537(22)$. A variation of the input parameter 
$\langle x\rangle^{\pi,0}_{u+d}$ by $\pm$10\% only leads to a small change in $A_{20}^{0,u+d}(t=0)$ of $\mathcal{O}(1\%)$, which is significantly smaller than the statistical error of $\approx5\%$.
The results of the fit are shown in Figs.~(\ref{A20v3}) and (\ref{A20v4}).
We would like to note that the slight upwards bending in Fig.~(\ref{A20v3}) at low $m_\pi$, and
therefore the good agreement with the phenomenological value, is due to the 
$\mathcal{O}(p^3)$-contribution $\Delta A_{20}^{u+d}$.
It has to be seen if this somewhat unusual curvature persists once the full $\mathcal{O}(p^3)$
contribution is available and fitted to the lattice results. The inclusion of
contributions from disconnected diagrams could also require a different 
extrapolation in $m_\pi$. The dependence of $A_{20}^{u+d}(t)$ on $t$ at fixed values of 
$m_\pi$ is presented in Figs.~(\ref{A20v5}) and (\ref{A20v6}).

As in the case of $A^{u-d}$, we also consider the heavy baryon limit of the CBChPT fit, giving the result $A_{20}^{u+d}(t=0,m_\pi)=A_{20}^{0,u+d}+A_{20}^{m_\pi,u+d} m_\pi^2$
represented by the dotted line in Fig.~(\ref{A20v3}), which
agrees with the CBChPT result only over a very limited range at low pion masses. Notably,
while the lattice results for $A_{20}^{u+d}$ are rising for larger pion masses, the heavy-baryon-limit curve has 
the opposite slope with negative $A_{20}^{m_\pi,u+d}$.
However, a direct HBChPT fit with free coefficients (see also section \ref{EMIsoSingletextr})
shown by the dashed curve leads to a positive $A_{20}^{m_\pi,u+d}$ and
a reasonable description of the lattice datapoints.
%

%%
%%%%%%%%%%%%%%%%%%%%%%%%%%%%%%%%%%%%%%%%%%%%%%%%%%%%%%%%%%%%%%%%%%%%%%%%%%%%%%%%%%%%%%%%%%%%%%%%%%%%%%%%%%%%%%%%%
\subsection{\label{sec:B20Isosinglet}CBChPT extrapolation of $B_{20}^{u+d}(t)$}
The dependence on the pion mass and the momentum transfer squared
of the isosinglet $B_{20}$ GFF in  $\mathcal{O}(p^2)$ CBChPT is given by \cite{Dorati:2007bk}
\begin{equation}
B_{20}^{u+d}(t,m_\pi)=\frac{m_N(m_\pi)}{m_N} B_{20}^{0,u+d} + A_{20}^{0,u+d} h_B^{u+d}(t,m_\pi) 
+ \Delta B_{20}^{u+d}(t,m_\pi)+ \frac{m_N(m_\pi)}{m_N}\bigg\{\delta_{B}^{t,u+d}\,t 
+ \delta_{B}^{m_\pi,u+d}\,m_\pi^2\bigg\}  + \mathcal{O}(p^3) \,,
\label{ChPTB20updp4}
\end{equation}
where $B_{20}^{0,u+d} \equiv B_{20}^{u+d}(t=0,m_\pi=0)$,
and the terms $\Delta B_{20}$, $\delta_{B}^{t,u+d}\,t$ and $\delta_{B}^{m_\pi,u+d}\,m_\pi^2$ are of $\mathcal{O}(p^3)$ and represent only a part the full $\mathcal{O}(p^3)$ contribution. The a priori unknown constants $B_{20}^{0,u+d}$,
$\delta_{B}^{t,u+d}$ and $\delta_{B}^{m_\pi,u+d}$ may be obtained from a fit to the lattice data.
A fit to our lattice results based on Eq.~(\ref{ChPTB20updp4}) turns out to be unstable and produces large values for the counter term
parameter $\delta_{B}^{m_\pi,u+d}\approx 15$. This can be seen as indication that other
higher order correction terms of $\mathcal{O}(p^3)$ not yet included in Eq.~(\ref{ChPTB20updp4})
are numerically important and needed to stabilize the extrapolation. We note that
the counting scheme of \cite{Dorati:2007bk} suggests that $\Delta B_{20}$ is not a 
dominant $\mathcal{O}(p^3)$-contribution concerning the pion mass dependence, at least for $t=0$.
This can be seen to some extent from the heavy-baryon-limit of Eq.~(\ref{ChPTB20updp4}), which
does not reproduce the full coefficient, $\propto(A+B)_{20}^{0,u+d}$, of the $m_\pi^2\log(m_\pi^2)$-term  
in  HBChPT (see e.g. \cite{Chen:2001pv,Diehl:2006js,Ando:2006sk}), but rather gives a term $\propto A_{20}^{0,u+d}m_\pi^2\log(m_\pi^2)$ without the $B_{20}^{0,u+d}m_\pi^2\log(m_\pi^2)$ contribution.

\begin{figure}[htbp]
     \begin{minipage}{0.4\textwidth}
      \centering
          \includegraphics[scale=0.4,clip=true,angle=0]{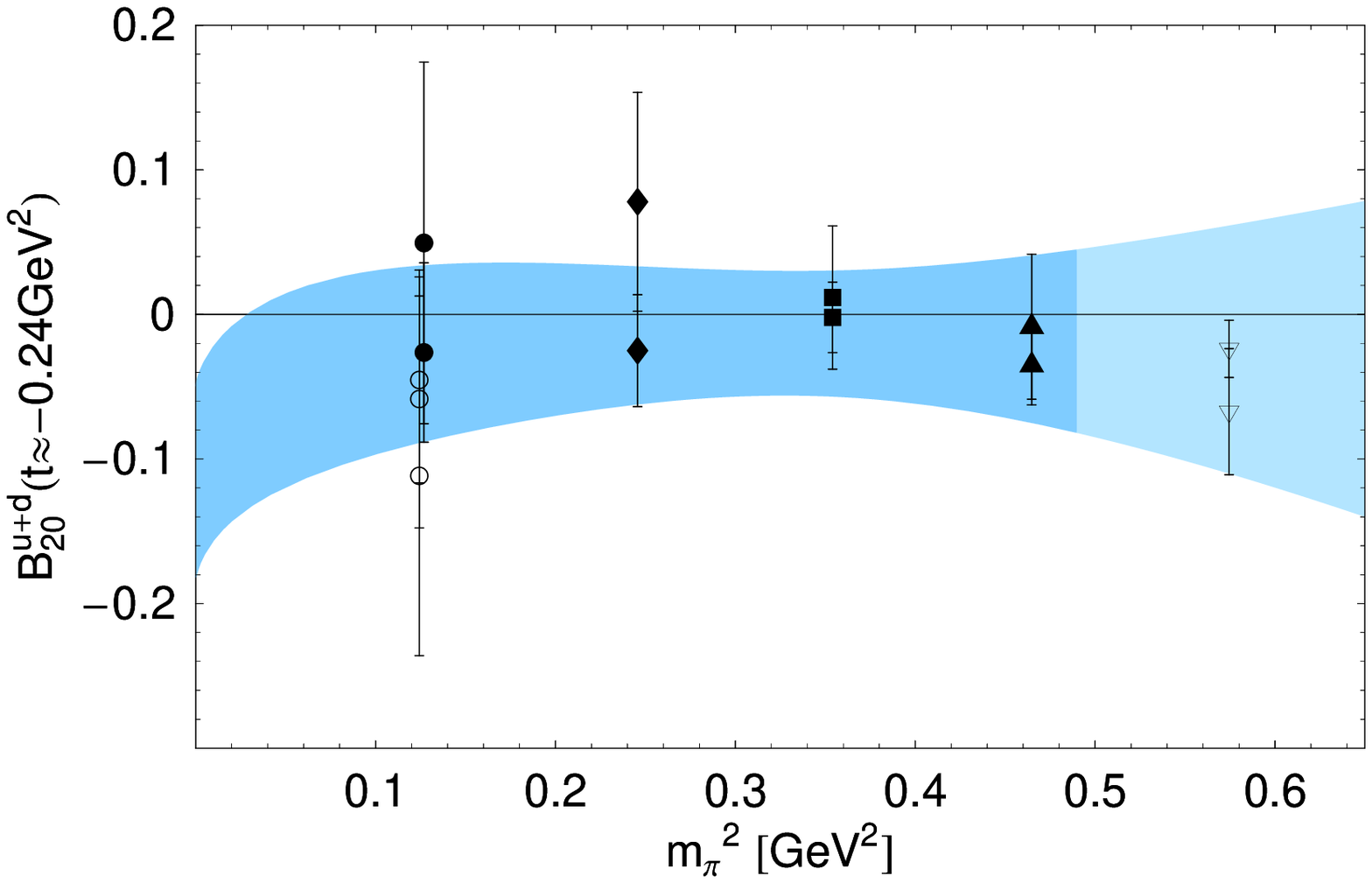}
 %         \includegraphics[scale=0.4,clip=true,angle=0]{B20IsoSinglet_relp4_tp24_w_CT_no_DeltaB20_v2}
%      \vspace*{-0.8cm}
  \caption{Lattice results for $B_{20}^{u+d}$ at $t\approx-0.24$ GeV$^2$ versus $m_\pi^2$ together 
  with the result of a global simultaneous chiral fit based on 
  Eqs.~(\ref{ChPTA20umdp4new}), (\ref{ChPTC20updp4}) and a variant of Eq.~(\ref{ChPTB20updp4}),
  as described in the text.}\label{B20Isosingletp4v1}
     \end{minipage}
     %\hfill
     \hspace{1cm}
     \begin{minipage}{0.4\textwidth}
      \centering
          \includegraphics[scale=0.4,clip=true,angle=0]{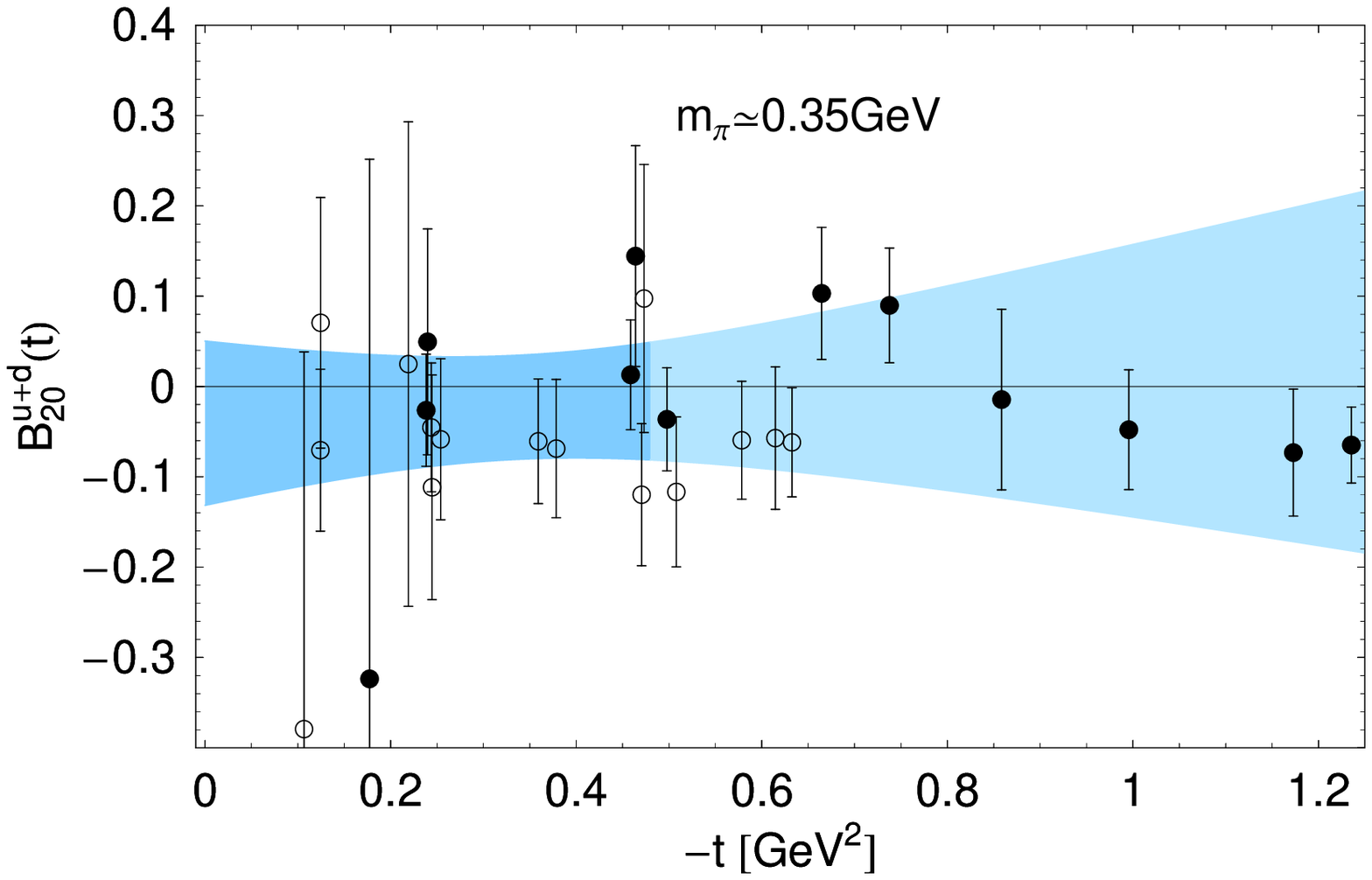}
%      \vspace*{-0.8cm}
  \caption{Lattice results for $B_{20}^{u+d}(t)$ at $m_\pi\approx350$ MeV versus $(-t)$
  together with the result of a global simultaneous chiral fit based on 
  Eqs.~(\ref{ChPTA20umdp4new}), (\ref{ChPTC20updp4}) and a variant of Eq.~(\ref{ChPTB20updp4}),
  as described in the text.}
  \label{B20Isosingletp4v2}
     \end{minipage}
   \end{figure}
Since the instability of the fit can be traced back to the term $\Delta B_{20}^{u+d}(t,m_\pi)$, 
we performed the final fit dropping this contribution but keeping the counter terms $\propto t$ and $\propto m^2_\pi$.
Based on this approach, we find that a global simultaneous fit to the GFFs  $A_{20}^{u+d}$, $B_{20}^{u+d}$ and $C_{20}^{u+d}$,
using Eqs.~(\ref{ChPTA20umdp4new}) and (\ref{ChPTC20updp4}) as described in the previous section,
leads to a stable chiral extrapolation of all three GFFs. 
In particular the counter term parameters $\delta_{B}^{t,u+d}$ and $\delta_{B}^{m_\pi,u+d}$ turn out to be
very small and fully compatible with zero within errors.
%Using $A_{20}^{0,u+d}=0.528$ from the chiral extrapolation above as an input parameter, w
We obtain $B_{20}^{0,u+d}=-0.140(84)$ and $B_{20}^{u+d,}(t=0,m_{\pi,\text{phys}})=-0.095(86)$ 
at the physical pion mass. Results of the fit are shown in Figs.~(\ref{B20Isosingletp4v1}) and (\ref{B20Isosingletp4v2}).
We cannot rule out that the inclusion of the full $\mathcal{O}(p^3)$ contributions to 
$B_{20}^{u+d}$ will lead to a qualitatively different dependence on $t$ and $m_\pi$ 
in the region where lattice results are available, so that the results above should to be taken with due caution.

In section \ref{EMIsoSingletextr}, we will study the combined $(t,m_\pi)$-dependence
of $B_{20}^{u+d}$ in HBChPT at $\mathcal{O}(p^2)$ \cite{Diehl:2006ya}. 
One-loop graphs with insertions of pion operators are fully included,
leading to a non-analytic dependence on the pion mass and the momentum transfer
that is quite different from that shown in Figs.~(\ref{B20Isosingletp4v1}) and (\ref{B20Isosingletp4v2}).

%%%%%%%%%%%%%%%%%%%%%%%%%%%%%%%%%%%%%%%%%%%%%%%%%%%%%%%%%%%%%%%%%%%%%%%%%%%%%%%%%%%%%%%%%%%%%%%%%%%%%%%%%%%%%%%%%%%%%%
%
\subsection{\label{sec:C20Isosinglet}CBChPT extrapolation of $C_{20}^{u+d}(t)$}
The $(t,m_\pi)$-dependence of the isosinglet GFF $C_{20}$  in CBChPT to  $\mathcal{O}(p^2)$ is
given by \cite{Dorati:2007bk}
\begin{equation}
C_{20}^{u+d}(t,m_\pi)=\frac{m_N(m_\pi)}{m_N} C_{20}^{0,u+d} + A_{20}^{0,u+d} h_C^{u+d}(t,m_\pi)  + 
\Delta C_{20}^{u+d}(t,m_\pi) + \mathcal{O}(p^3)\,,
\label{ChPTC20updp4}
\end{equation}
where $C_{20}^{0,u+d} \equiv C_{20}^{u+d}(t=0,m_\pi=0)$, and the term $\Delta C_{20}^{u+d} \propto \langle x\rangle^{\pi,0}_{u+d}$ is a part of the full $\mathcal{O}(p^3)$-corrections \cite{Dorati:2007bk}. 
In this counting scheme, counter terms of the form $\delta_{C}^{t,u+d}\,t$ and $\delta_{C}^{m_\pi,u+d}\,m_\pi^2$ 
first appear at $\mathcal{O}(p^4)$. In order to get a first idea about the possible   
$t$- and $m_\pi$-dependence of $C_{20}^{u+d}$ in CBChPT, 
we have included the formally higher order counter terms $\delta_{C}^{t,u+d}\,t$ and $\delta_{C}^{m_\pi,u+d}\,m_\pi^2$
in the fit to our lattice data, resulting in a stable chiral extrapolation. 
\begin{figure}[htbp]
     \begin{minipage}{0.4\textwidth}
      \centering
          \includegraphics[scale=0.4,clip=true,angle=0]{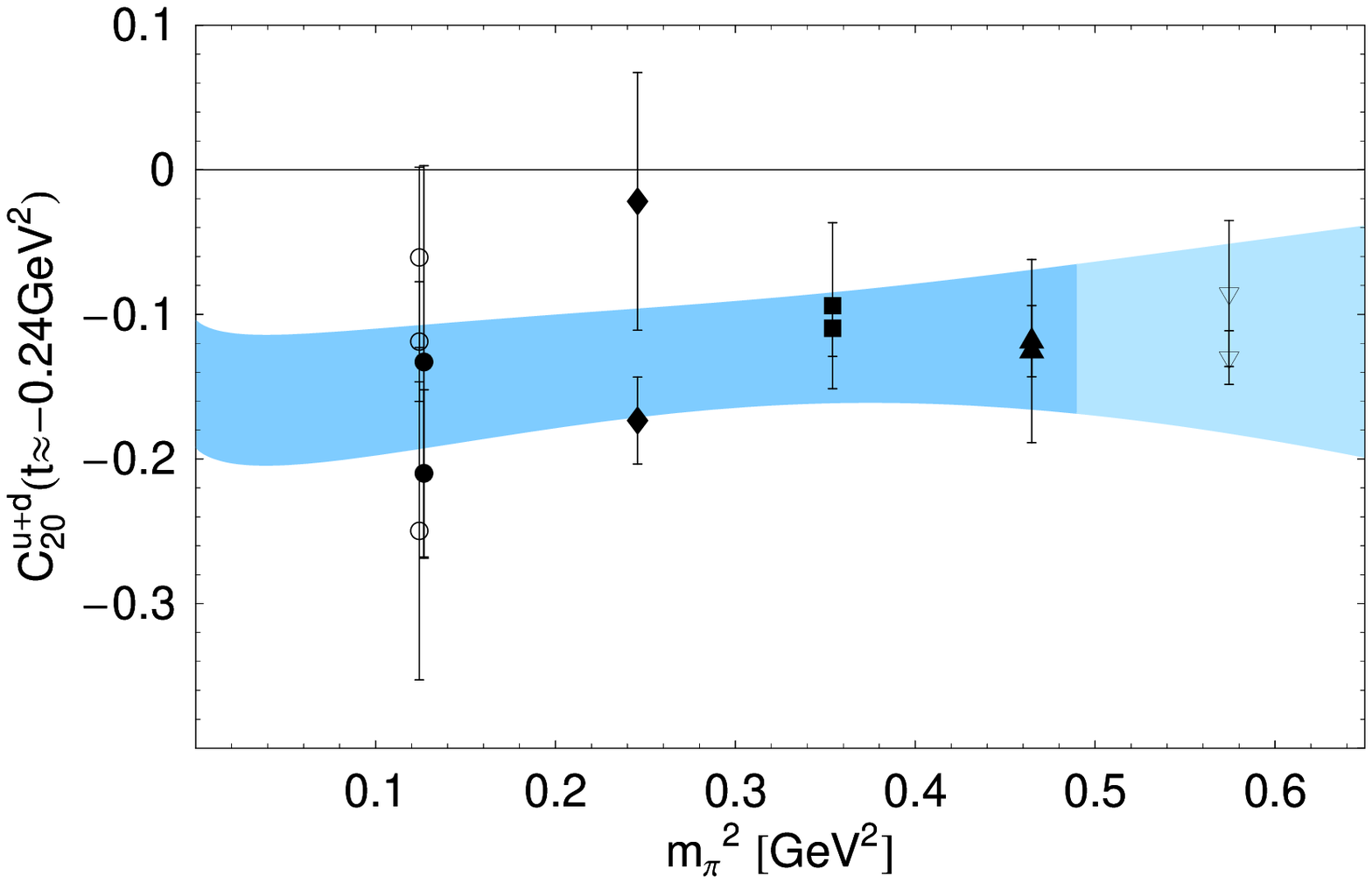}
%          \includegraphics[scale=0.4,clip=true,angle=0]{C20IsoSinglet_relp4_w_Delta_w_2CT_tp24_v2}
%      \vspace*{-0.8cm}
  \caption{Lattice results for $C_{20}^{u+d}$ at $t\approx-0.24$ GeV$^2$ versus $m_\pi^2$ together 
  with the result of a global simultaneous chiral fit using Eqs.~(\ref{ChPTA20umdp4new}), (\ref{ChPTB20updp4}) and (\ref{ChPTC20updp4}).}
  \label{C20Isosingletp4v1}
     \end{minipage}
     %\hfill
     \hspace{1cm}
     \begin{minipage}{0.4\textwidth}
      \centering
          \includegraphics[scale=0.4,clip=true,angle=0]{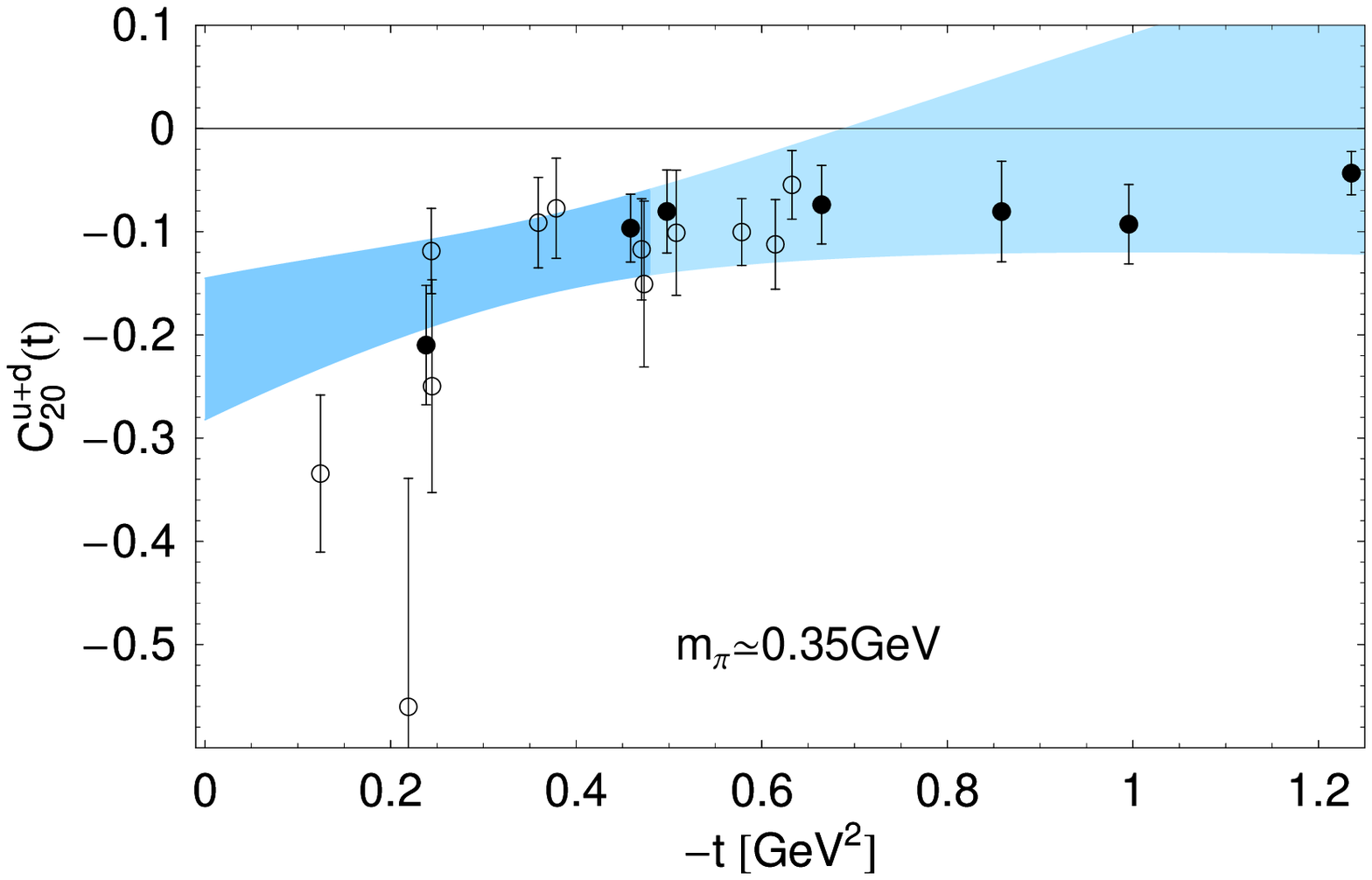}
%      \vspace*{-0.8cm}
  \caption{Lattice results for $C_{20}^{u+d}(t)$ at $m_\pi\approx350$ MeV versus $(-t)$ 
  together with the result of a global simultaneous chiral fit using Eqs.~(\ref{ChPTA20umdp4new}), (\ref{ChPTB20updp4}) and (\ref{ChPTC20updp4}).}
  \label{C20Isosingletp4v2}
     \end{minipage}
   \end{figure}
With $\langle x\rangle^{\pi,0}_{u+d}=0.5$ from Table~\ref{tab:LECs} as an input parameter,
and taking into account the CBChPT results for $A_{20}^{u+d}$ and $B_{20}^{u+d}$ discussed in sections \ref{sec:A20Isosinglet} 
and \ref{sec:B20Isosinglet}, respectively,
we obtain from a global simultaneous fit $C_{20}^{0,u+d}=-0.317(59)$ for the forward value in the chiral limit, 
and $C_{20}^{u+d}(t=0,m_{\pi,\text{phys}})=-0.267(62)$ at the physical pion mass. 
The parameters $\delta_{C}^{t,u+d}$ and $\delta_{C}^{m_\pi,u+d}$ turn out to be small.  
Changing $\langle x\rangle^{\pi,0}_{u+d}$ by $\pm$10\% results in a variation of $C_{20}^{0,u+d}(t=0)$ by $\pm5\%$, which is significantly smaller than the statistical error of 19\%.
The corresponding extrapolations are shown in Figs.~(\ref{C20Isosingletp4v1}) and (\ref{C20Isosingletp4v2}). These results indicate a non-trivial dependence of the $n=2$ moment of the isosinglet
GPDs $H$ and $E$ on the longitudinal momentum transfer $\xi$. 

\begin{figure}[h]
        \includegraphics[scale=0.5,clip=true,angle=0]{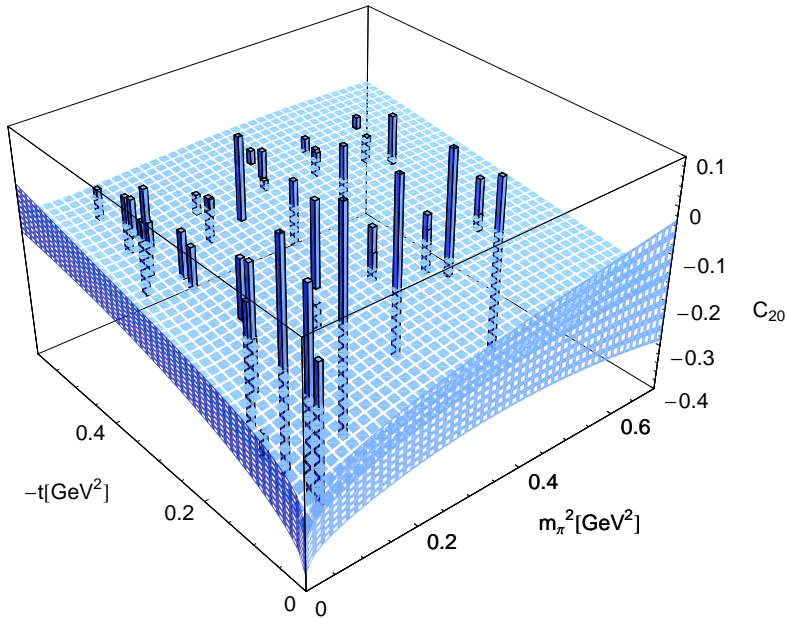}
%      \vspace*{-0.8cm}
  \caption{Combined $(t,m_\pi)$-dependence of $C^{u+d}_{20}$ from a global simultaneous chiral fit using Eqs.~(\ref{ChPTA20umdp4new}), (\ref{ChPTB20updp4}) and (\ref{ChPTC20updp4}) compared to lattice data.}
  \label{FigC20reltb}
\end{figure}

A three-dimensional plot showing the combined $(t,m_\pi)$-dependence of  $C_{20}^{u+d}$ is presented in Fig.~(\ref{FigC20reltb}), where the error bars of the lattice data points are illustrated by the stretched cuboids. 
The lattice data are superimposed with the result from the chiral fit discussed above, which is shown as a surface. The statistical error bars of the fit are shown for clarity only as bands for $t=0$ and $m_\pi=0$, respectively. In section \ref{subsecHBC20} below, we will compare the results based on CBChPT with a fit to $C_{20}^{u+d}$ in the framework of HBChPT. 
\subsection{\label{AMcovariant}Quark angular momentum $J$ in CBChPT}
The forward limit values of the isovector and isosinglet GFFs $A_{20}(t=0)$ and $B_{20}(t=0)$
we have studied in sections \ref{sec:A20IsoVecrel}, \ref{sec:B20IsoVecrel}, 
\ref{sec:A20Isosinglet} and \ref{sec:B20Isosinglet} 
allow us to compute the angular momentum
contributions of up- and down-quarks to the spin of the nucleon, 
$J^q=1/2( A^q_{20}(0) + B^q_{20}(0) )=1/2( \langle x\rangle^q + B^q_{20}(0) )$.
From the separate chiral extrapolations of the isosinglet $A_{20}$ and
$B_{20}$ in CBChPT, we find for the
total $u+d$ quark angular momentum at the physical pion mass 
$J^{u+d}(m_{\pi,\text{phys}})=0.213(44)$, corresponding to $43\%$ of the total
nucleon spin $S=1/2$. 
Together with the chiral extrapolations for the isovector $u-d$ combination, we obtain
the surprising result that the total quark angular momentum is
carried by the up-quarks, $J^{u}(m_{\pi,\text{phys}})=0.214(27)$ and
that the contribution from down quarks is zero $J^{d}(m_{\pi,\text{phys}})=-0.001(27)$.
From Fig.~\ref{Lq2}, we note that the cancellation of $\Sigma^d/2$ and  $L^d$  appears to be systematic for all $m_\pi$, and it will be interesting to understand whether this is accidental or has a physical origin. 
Taking into account preliminary results for the quark spin
$\widetilde A^q_{10}/2(t=0)=\Delta\Sigma^q /2$, 
as obtained from a ChPT extrapolation including the $\Delta$ resonance \cite{Hemmert:2003cb,Beane:2004rf} of $g_A=\Delta\Sigma^{u-d}$ and a self-consistently improved one-loop 
ChPT extrapolation of $\Delta\Sigma^{u+d}$ \cite{DrusPDFs},
we find that the quark orbital angular momentum $L^q=J^q-\Delta\Sigma^q/2$ contributions to the nucleon spin 
are $L^u=-0.195(44)$ and $L^d=0.200(44)$ at the physical pion mass. The nearly complete cancellation of up
and down quark OAM contributions that we observe for pion masses above $350$ MeV therefore also holds at 
at $m_{\pi,\text{phys}}$, where we find $L^{u+d}=0.005(52)$.
We emphasize again that no phenomenological values for $\Delta\Sigma=\langle 1\rangle_{\Delta q}$, $\langle x\rangle_q$ and $\langle x\rangle_{\Delta q}$ have been included in the extrapolations, and that we have omitted disconnected diagrams in the lattice calculations.
We will compare these CBChPT results with corresponding HBChPT results below in section \ref{subsecHBC20}.

\subsection{\label{EMIsoSingletextr}HBChPT extrapolation of  $E_{20}^{u+d}$ and $M_{20}^{u+d}$}
\begin{figure}[htbp]
     \begin{minipage}{0.4\textwidth}
      \centering
       \includegraphics[scale=0.4,clip=true,angle=0]{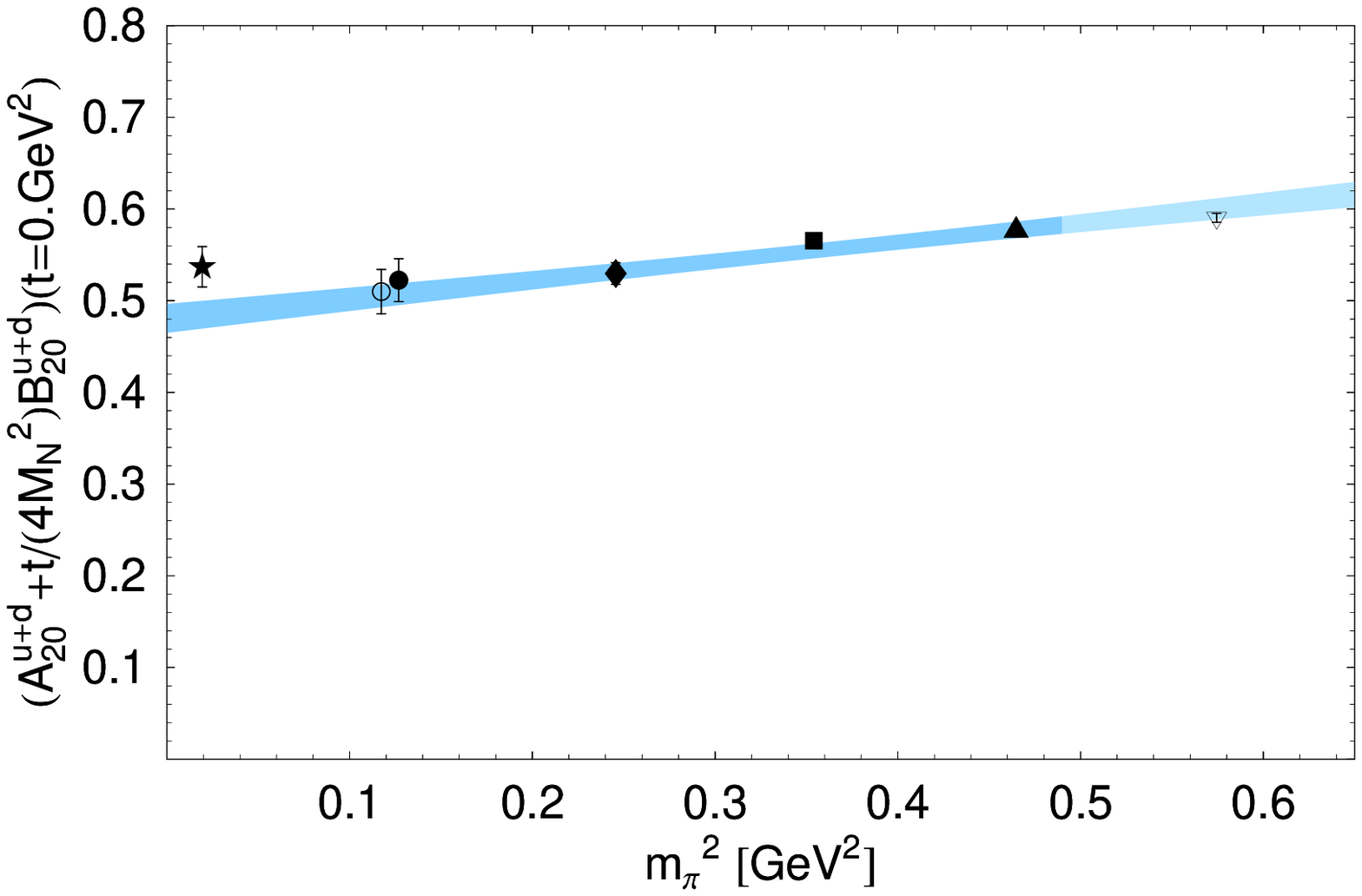}
  \caption{Lattice results for $E_{20}^{u+d}$ at $t=0$ versus $m_\pi^2$ together with the result of 
  a global chiral fit using Eq.~(\ref{ChPTE20}).}\label{E20v1}
     \end{minipage}
     %\hfill
     \hspace{1cm}
     \begin{minipage}{0.4\textwidth}
      \centering
          \includegraphics[scale=0.4,clip=true,angle=0]{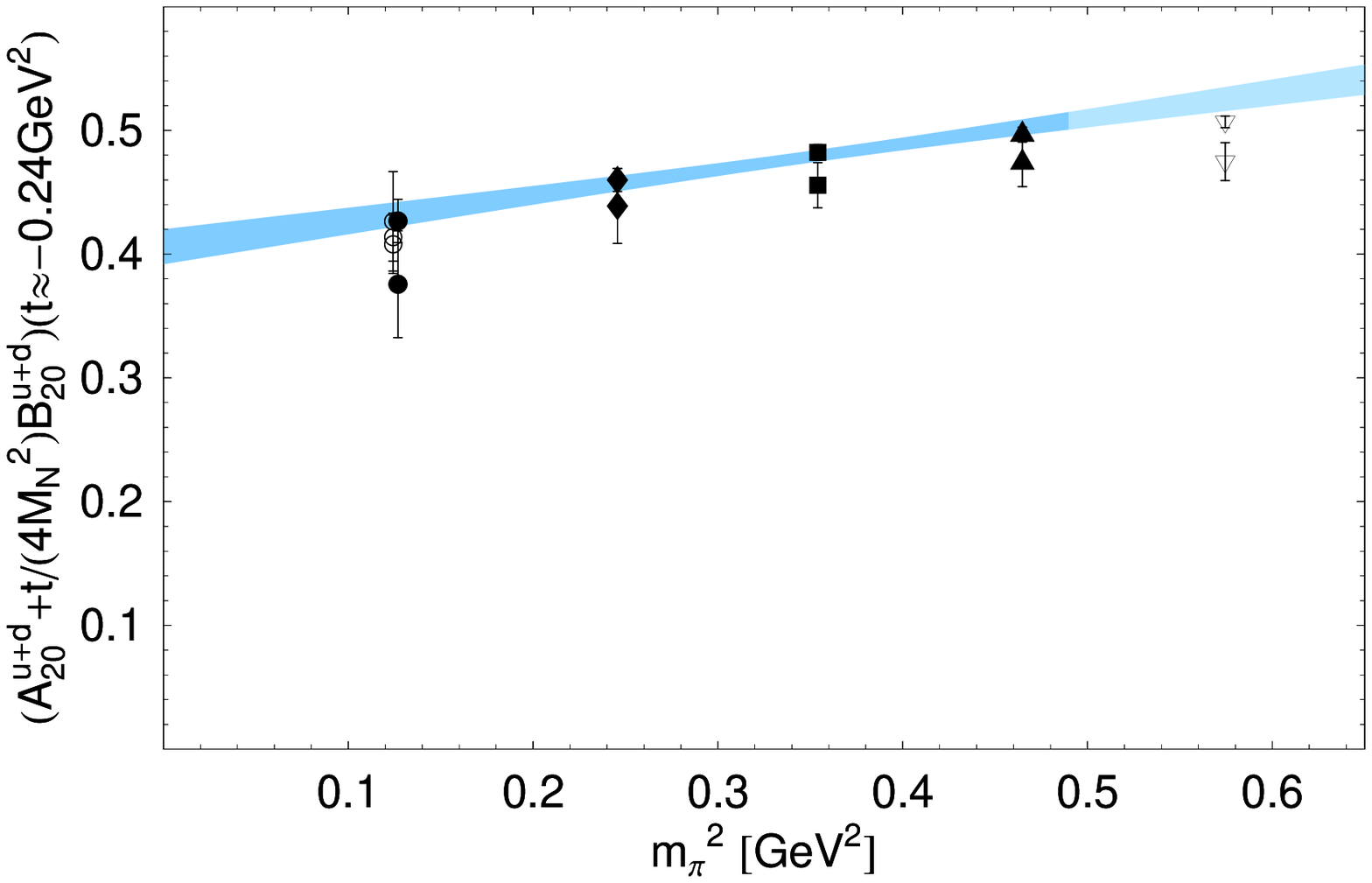}
%      \vspace*{-0.8cm}
  \caption{Lattice results for $E_{20}^{u+d}$ at $-t\approx 0.24$ GeV$^2$ versus $m_\pi^2$ together with the result of a global chiral fit using Eq.~(\ref{ChPTE20}).}\label{E20v2}
     \end{minipage}
   \end{figure}

\begin{figure}[htbp]
     \begin{minipage}{0.4\textwidth}
      \centering
          \includegraphics[scale=0.4,clip=true,angle=0]{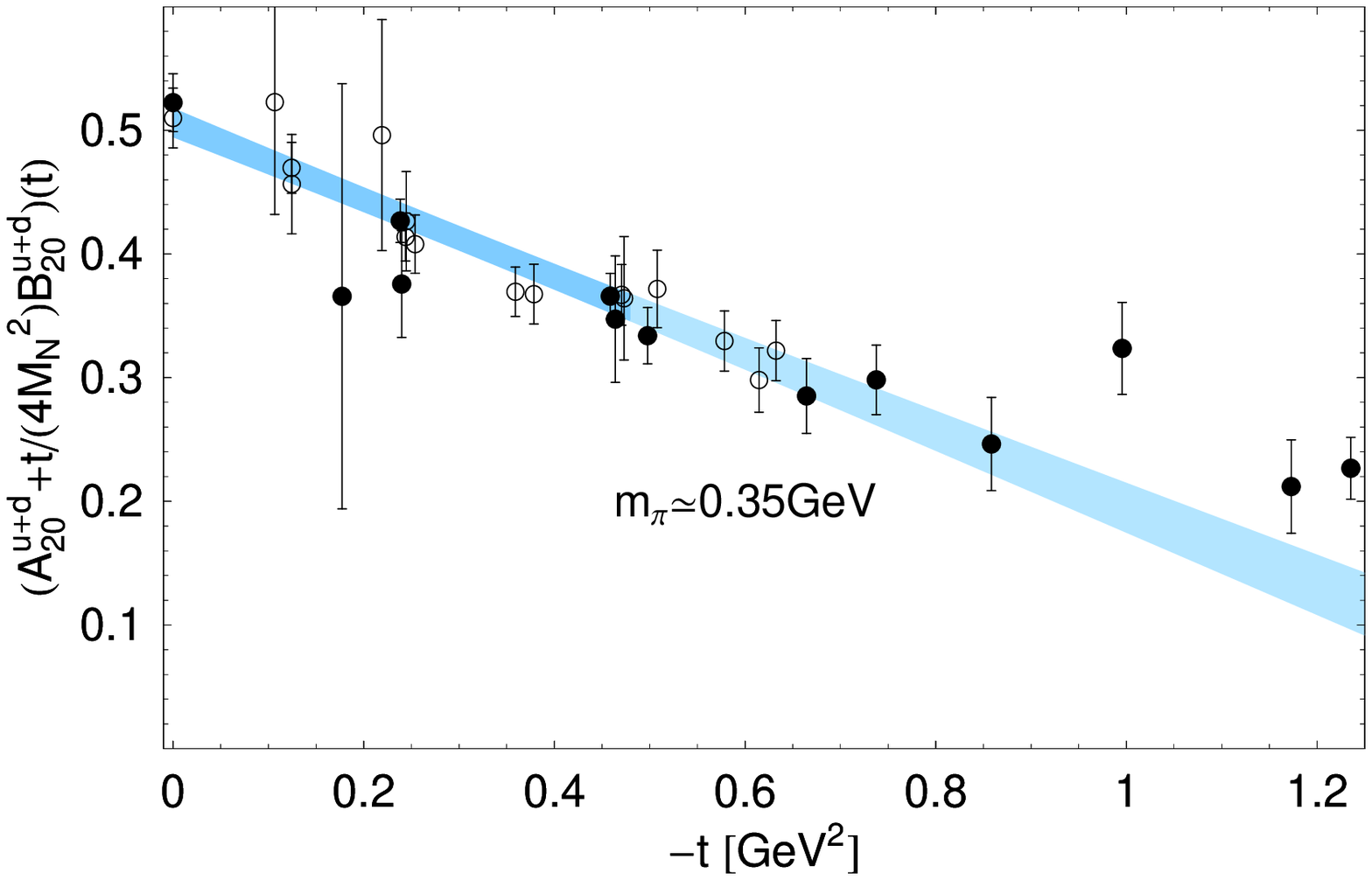}
%      \vspace*{-0.8cm}
  \caption{ Lattice results for $E_{20}^{u+d}$ at $m_\pi\approx 350$ MeV versus $-t$ together with the result of a global chiral fit using Eq.~(\ref{ChPTE20}).\newline}\label{E20v3}
     \end{minipage}
     %\hfill
     \hspace{1cm}
     \begin{minipage}{0.4\textwidth}
      \centering
          \includegraphics[scale=0.4,clip=true,angle=0]{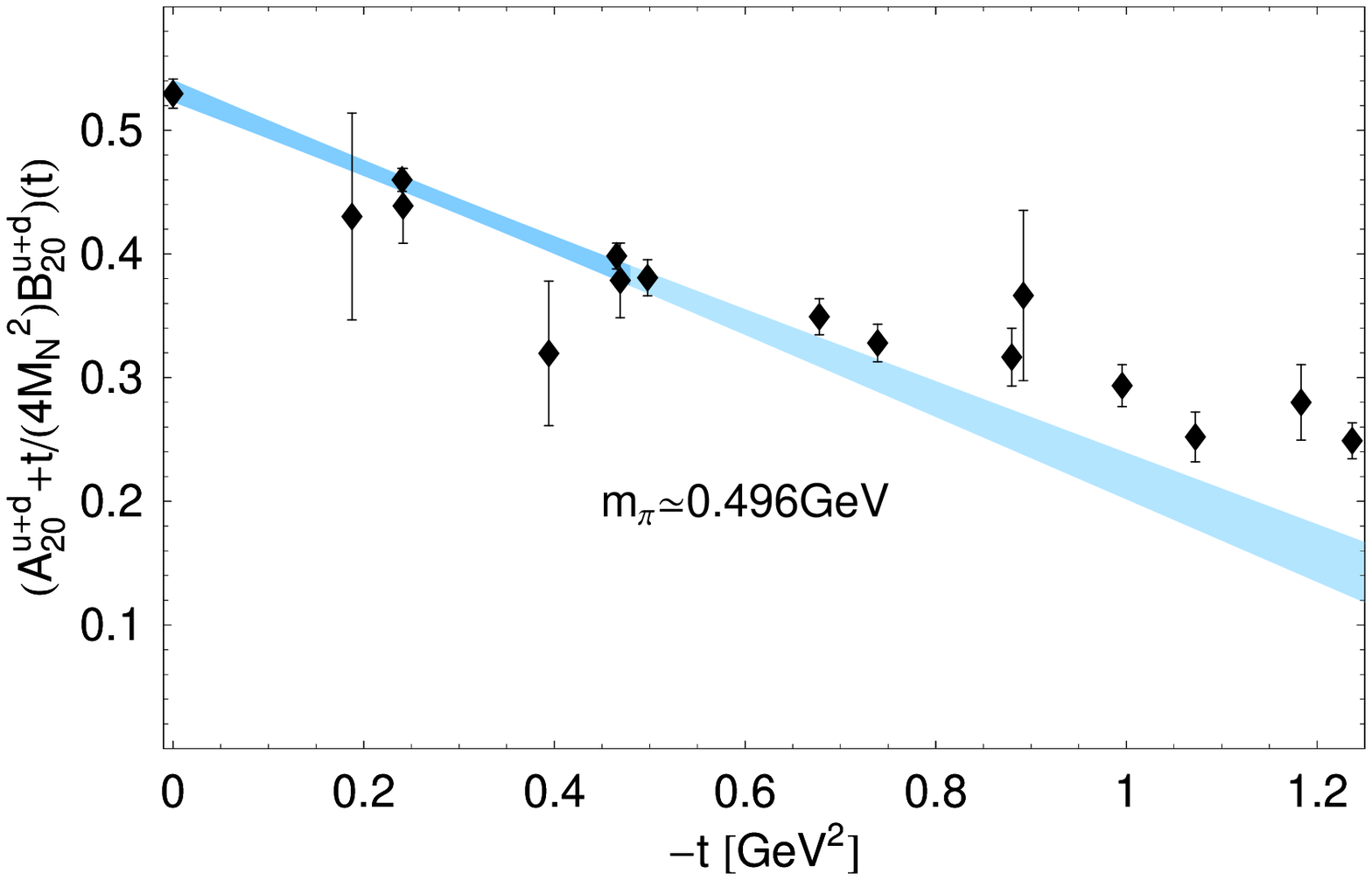}
%      \vspace*{-0.8cm}
  \caption{ Lattice results for $E_{20}^{u+d}$ at $m_\pi\approx 496$ MeV versus $-t$ together with the result of a global chiral fit using Eq.~(\ref{ChPTE20}).\newline}\label{E20v4}
     \end{minipage}
   \end{figure}

In heavy baryon chiral perturbation theory \cite{Diehl:2006ya,Diehl:2006js} 
to $\mathcal{O}(p^2)$, the combined ($t,m_\pi$)-dependence
of the GFF-combination $E_{20}^{u+d}(t)= A_{20}^{u+d}(t)+t/(4m_N^2)B_{20}^{u+d}(t)$ is quite different
from that of $M_{20}^{u+d}(t)=A_{20}^{u+d}(t)+B_{20}^{u+d}(t)$, which in the forward limit is equal to two times the
total quark angular momentum $2J_q=M_{20}^{u+d}(t=0)$.
While at this order $M_{20}^{u+d}$  shows a non-analytic dependence on $t$ and $m_\pi$ as discussed below,
$E_{20}^{u+d}$ is constant up to analytic tree-level contributions,
\begin{equation}
E_{20}^{u+d}(t,m_\pi)=E_{20}^{0,u+d} + E_{20}^{m_\pi,u+d} m_\pi^2 + E_{20}^{t,u+d} t.
\label{ChPTE20}
\end{equation}
A fit to our lattice results 
based on Eq.~(\ref{ChPTE20}) is shown in Figs.~(\ref{E20v1}), (\ref{E20v2}), (\ref{E20v3}) and (\ref{E20v4}).
The linear dependence on $t$ and $m_\pi^2$  works well even beyond the fitted range, i.e. for $-t\geq 0.48$ GeV$^2$. This is not surprising since $E_{20}^{u+d}$ is clearly dominated by the GFF $A_{20}^{u+d}$, which 
does not show a strong curvature in $t$ as seen in  Figs.~(\ref{ABC1st}) and (\ref{A123}). However, 
it is obvious that the HBChPT result in Eq.~(\ref{ChPTE20}) lacks structures which 
would allow for an upwards bending of $E_{20}^{u+d}(t=0)=A_{20}^{u+d}(t=0)$ at small pion masses
towards the phenomenological value, in contrast to the covariant approach studied in 
section \ref{sec:A20IsoVecrel}.
The fit gives $E_{20}^{0,u+d}=0.481(15)$ in the chiral limit, which we will use for the chiral
extrapolations based on HBChPT of the total angular momentum and the anomalous gravitomagnetic moment $B_{20}^{u+d}(t=0)$ below. At the physical pion mass, we find
$E_{20}^{u+d}(t=0, m_{\pi,\text{phys}})=\langle x\rangle_{u+d}=0.485(14)$, which is 
approximately 10\% below the phenomenological results, $\langle x\rangle_{u+d}^\text{CTEQ6}=0.537(22)$ and 
$\langle x\rangle_{u+d}^\text{MRST2001}=0.538(22)$ \cite{DurhamDatabase}.

\begin{figure}[htbp]
     \begin{minipage}{0.4\textwidth}
      \centering
          \includegraphics[scale=0.4,clip=true,angle=0]{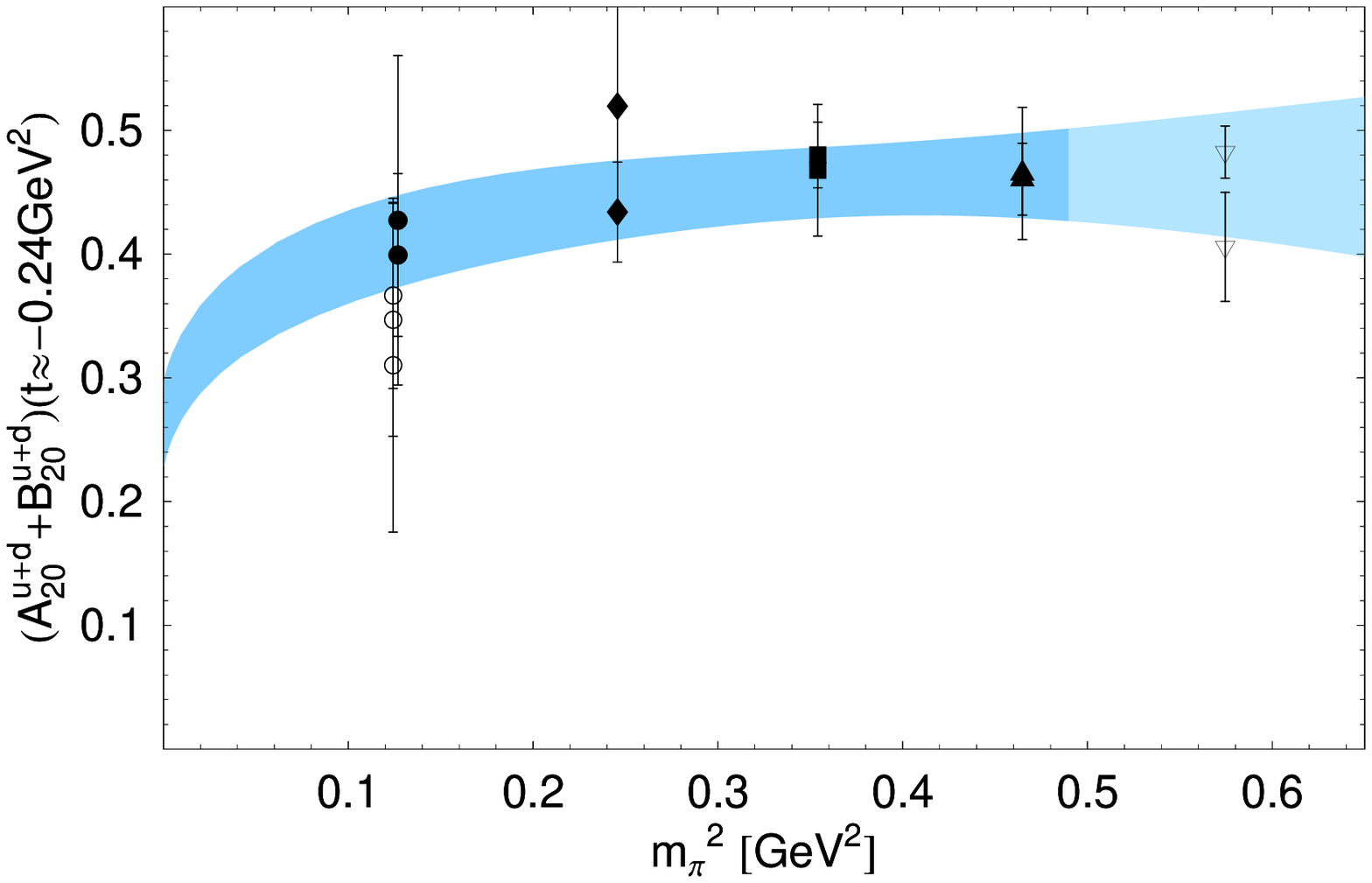}
%      \vspace*{-0.8cm}
  \caption{Lattice results for $M_{20}^{u+d}$ at $|t|\approx 0.24$ GeV$^2$ versus $m_\pi^2$ together with the result of a global chiral fit using Eq.~(\ref{M20chPT1}).}\label{M20v1}
     \end{minipage}
\hspace{1cm}
     \begin{minipage}{0.4\textwidth}
      \centering
          \includegraphics[scale=0.4,clip=true,angle=0]{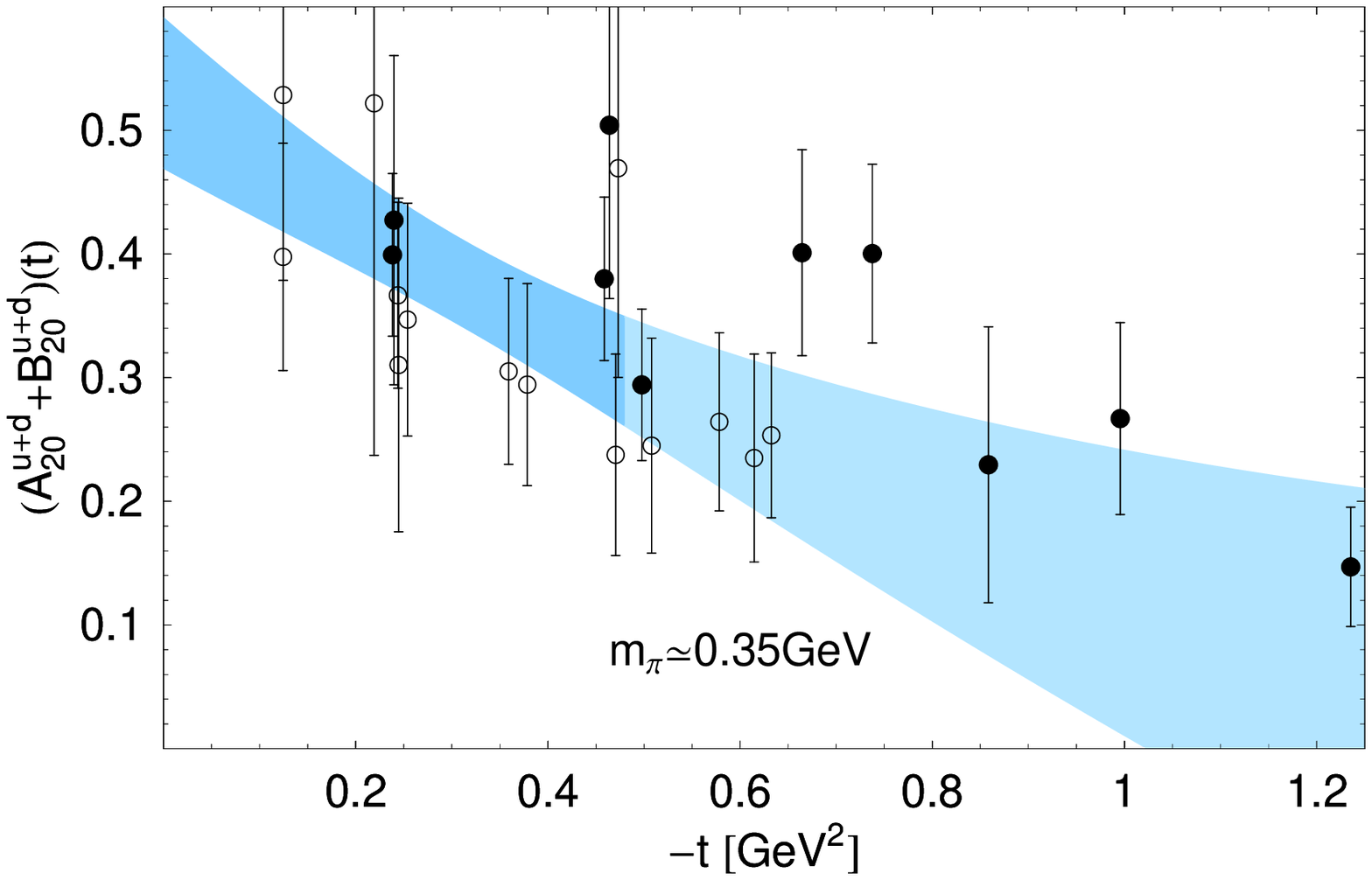}
%      \vspace*{-0.8cm}
  \caption{ Lattice results for $M_{20}^{u+d}$ at $m_\pi\approx 350$ MeV versus $(-t)$ together with the result of a global chiral fit using Eq.~(\ref{M20chPT1}).\newline}\label{M20v3}
     \end{minipage}
   \end{figure}

The pion mass dependence of $M_{20}^{u+d}(t)$ for non-zero $t$ is given by \cite{Diehl:2006ya,Diehl:2006js}
\begin{equation}
M_{20}^{u+d}(t,m_\pi)=M_{20}^{0,u+d}\bigg\{ 1-\frac{3g_A^2 m_\pi^2}{(4 \pi f_\pi)^2}\ln\left(\frac{m_\pi^2}{\Lambda_\chi^2}\right)\bigg\} + M_2^{(2,\pi)}(t,m_\pi)
+ M_{20}^{m_\pi,u+d} m_\pi^2 + M_{20}^{t,u+d} t,
\label{M20chPT1}
\end{equation}
with new counter terms $ M_{20}^{m_\pi,u+d}$ and $M_{20}^{t,u+d}$. The non-analytic dependence on $t$ and $m_\pi$
in $M_2^{(2,\pi)}(t,m_\pi)$ results from pion-operator insertions and is directly
proportional to the (isosinglet) momentum fraction of quarks in the pion in the chiral limit, 
$\langle x\rangle^{\pi,0}_{u+d}$.
We use $\langle x\rangle^{\pi,0}_{u+d}=0.5$ 
from Table~\ref{tab:LECs} for the fit.
No additional parameters are needed to this order. The results of chiral fits based on Eq.~(\ref{M20chPT1}) are presented in Figs.~(\ref{M20v1}) and (\ref{M20v3}).
We find $M_{20}^{0,u+d}=0.522(41)$ and $M_{20}^{u+d}(t=0, m_{\pi,\text{phys}})=0.526(48)$.

\subsection{HBChPT extrapolation of $B^{u+d}_{20}$}
\label{subsecB20}
Total momentum and angular momentum conservation implies that the total anomalous gravitomagnetic moment of 
quarks and gluons in the nucleon has to vanish, $\sum_{q,g} B_{20}(t=0)=0$. An interesting
question is whether the individual quark and gluon contributions to $B_{20}$ are separately zero or very small, as previously speculated \cite{Teryaev:1999su,Teryaev:2006fk}. 
The first lattice calculations \cite{Gockeler:2003jf,Hagler:2003jd}
showed that $B_{20}^{u+d}$ is compatible with zero 
for $(-t)\geq0.5$ GeV$^2$ at pion masses of $m_\pi\ge 600$ MeV.
Based on our new results and the ChPT fits performed above, we are now in a position to study
$B_{20}^{u+d}$ more carefully as a function of $t$ and $m_\pi$.
The GFF $B_{20}^{u+d}$ can be written as a linear combination
of Eq.~(\ref{ChPTE20}) and Eq.~(\ref{M20chPT1}).  A separate fit to the data with fixed $E_{20}^{0,u+d}=0.481(15)$ 
gives $B_{20}^{u+d}(t=0,m_{\pi,\text{phys}})=0.050(49)$, which is compatible with the fits to
$M_{20}^{u+d}$ and $E_{20}^{u+d}$ above that in combination give 
$(M-E)_{20}^{u+d}(t=0,m_{\pi,\text{phys}})=0.041(50)$. Although the absolute value of $B_{20}^{u+d}(t=0)$ is
again rather small, we note that the sign is different from that found in section \ref{sec:B20Isosinglet} 
based on the CBChPT fit. 
A $10\%$ variation of the input parameter $\langle x\rangle^{\pi,0}_{u+d}$
leads to change of 0.023 in $B_{20}^{u+d}(t,m_{\pi,\text{phys}})$ at $t=0$, which is
well below the statistical error of 0.049, and a change of 0.008 at a momentum 
transfer of $|t|\approx 0.24$ GeV$^2$, which is well below the statistical error of 0.031.

\begin{figure}[htbp]
     \begin{minipage}{0.4\textwidth}
      \centering
          \includegraphics[scale=0.4,clip=true,angle=0]{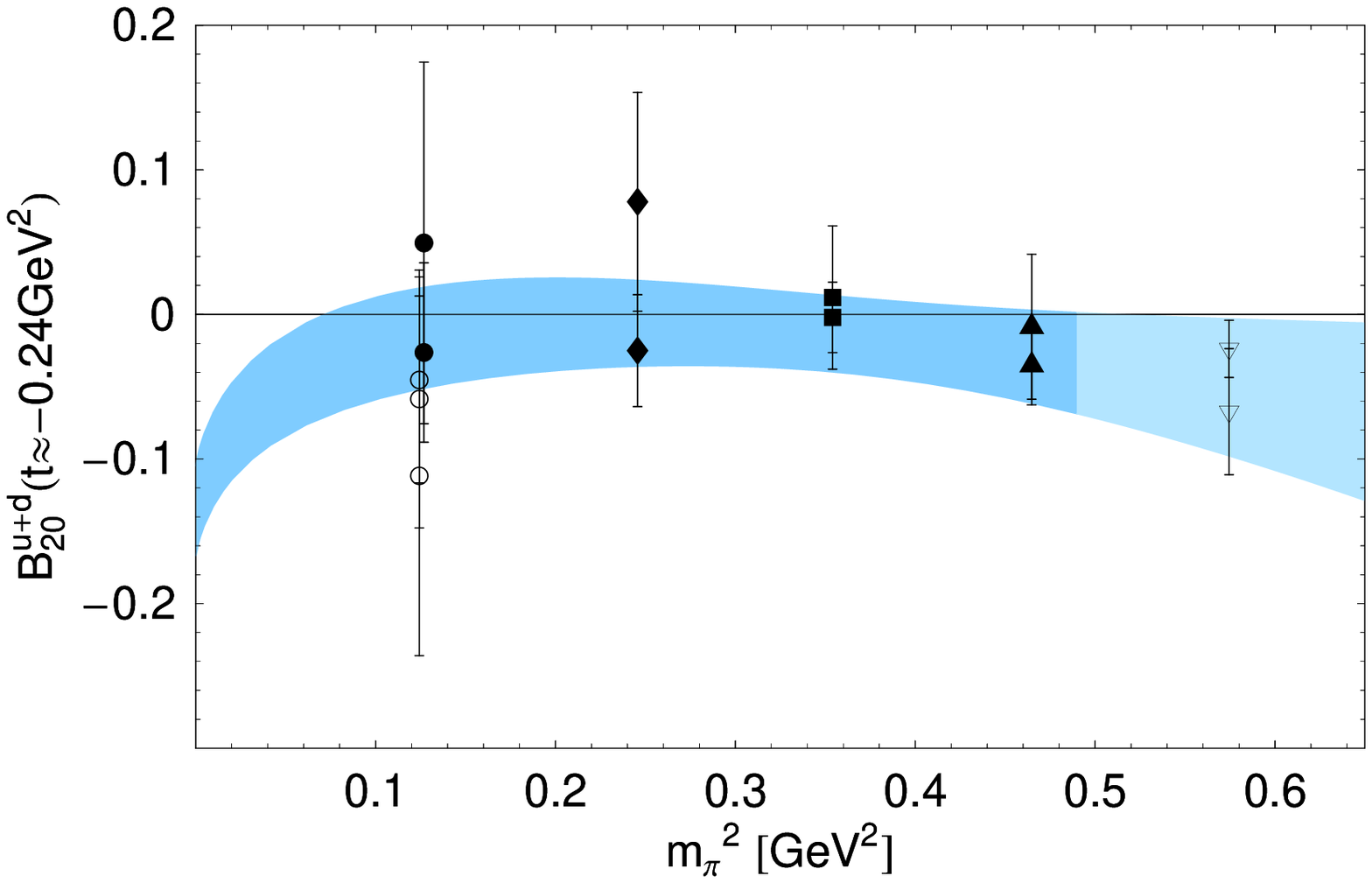}
%      \vspace*{-0.8cm}
  \caption{Lattice results for $B_{20}^{u+d}$ at $|t|\approx 0.24$ GeV$^2$ versus $m_\pi^2$ together with the result of a global chiral fit using a linear combination of Eqs.~(\ref{ChPTE20}) and (\ref{M20chPT1}).}\label{B20v1}
  \end{minipage}
   \hspace{1cm}
  \begin{minipage}{0.4\textwidth}
      \centering
          \includegraphics[scale=0.4,clip=true,angle=0]{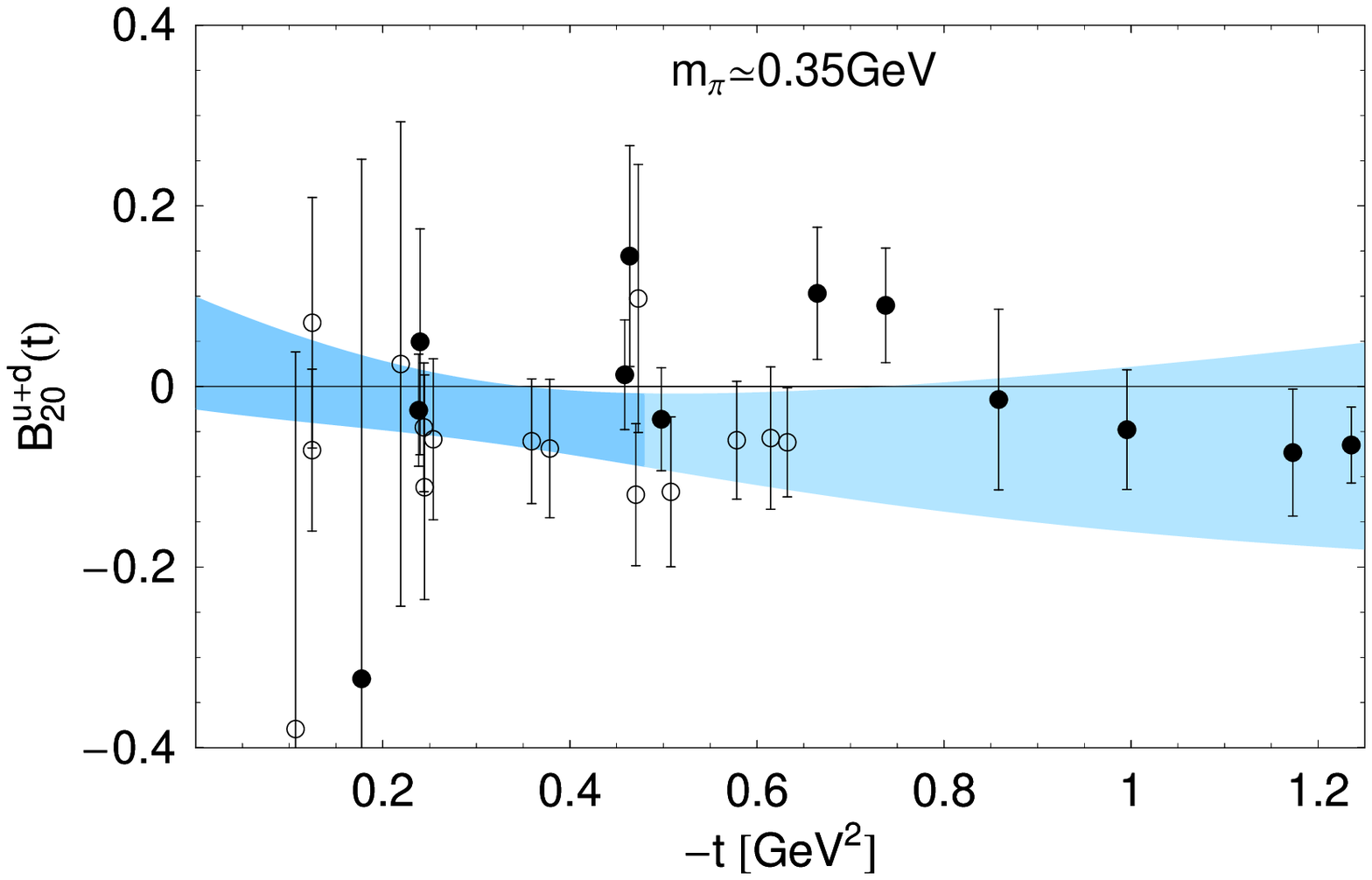}
%      \vspace*{-0.8cm}
  \caption{Lattice results for $B_{20}^{u+d}$ at $m_\pi\!\approx\!350$ MeV versus $(-t)$ together with the result of a global chiral fit using a linear combination of Eqs.~(\ref{ChPTE20}) and (\ref{M20chPT1}).\newline}\label{B20v2}
     \end{minipage}
     %\hfill
   \end{figure}

The dependence of $B_{20}^{u+d}$ on $t$ and on the pion mass is
shown in Figs.~(\ref{B20v1}) and (\ref{B20v2}). 
The dependence on the momentum
transfer squared turns out to be somewhat
different from the CBChPT result in Fig.~(\ref{B20Isosingletp4v2}) 
where contributions from pion operator insertions $\propto \langle x\rangle^{\pi,0}_{u+d}$ have not been included, but the two results are statistically consistent.
\begin{figure}[th]
% \vspace*{-0.75cm}
        \includegraphics[scale=0.5,clip=true,angle=0]{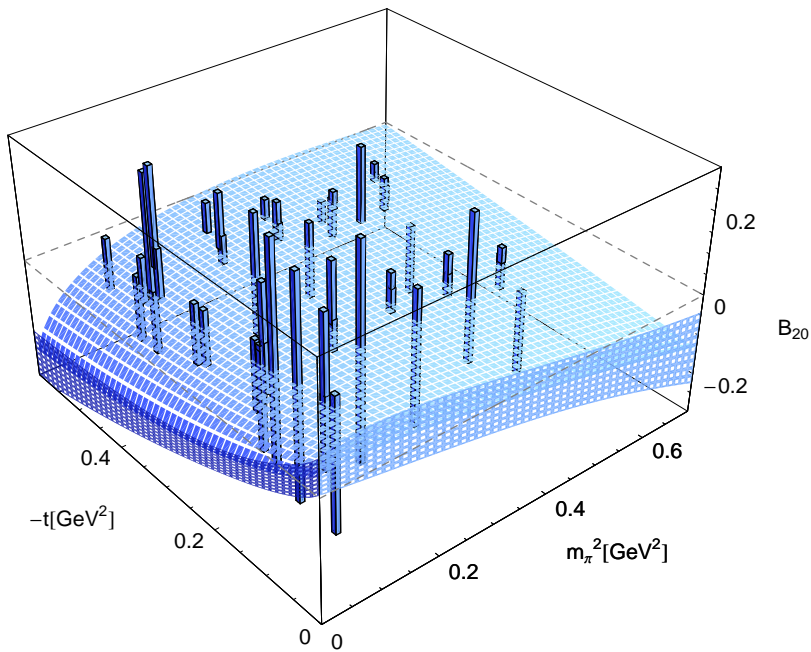}
%      \vspace*{-0.8cm}
  \caption{Combined $(t,m_\pi)$-dependence of the quark anomalous gravitomagnetic moment $B^{u+d}_{20}$ from a global chiral fit using a linear combination of Eqs.~(\ref{ChPTE20}) and (\ref{M20chPT1}) compared to lattice data.}
  \label{FigB20tb}
\end{figure}
The combined dependence of $B_{20}^{u+d}$ on $t$ and $m_\pi$ from
the HBChPT fit compared to lattice data points is presented in a 3d-plot in Fig.~(\ref{FigB20tb}).
It is interesting to note that the fit based on Eq.~(\ref{M20chPT1})
leads to a clearly visible non-analytic dependence of $B_{20}^{u+d}$
on the pion mass and the momentum transfer. In particular, we find
a non-zero, negative $B_{20}^{u+d}$ for $|t|>0.05\text{GeV}^2,m_\pi^2 \lessapprox 0.1\text{GeV}^2$
from the chiral extrapolation, which may be confirmed in future lattice calculations 
or experimental measurements.
\subsection{HBChPT extrapolation of $C_{20}^{u+d}$ }
\label{subsecHBC20}
\begin{figure}[htbp]
     \begin{minipage}{0.4\textwidth}
      \centering
    \includegraphics[scale=0.42,clip=true,angle=0]{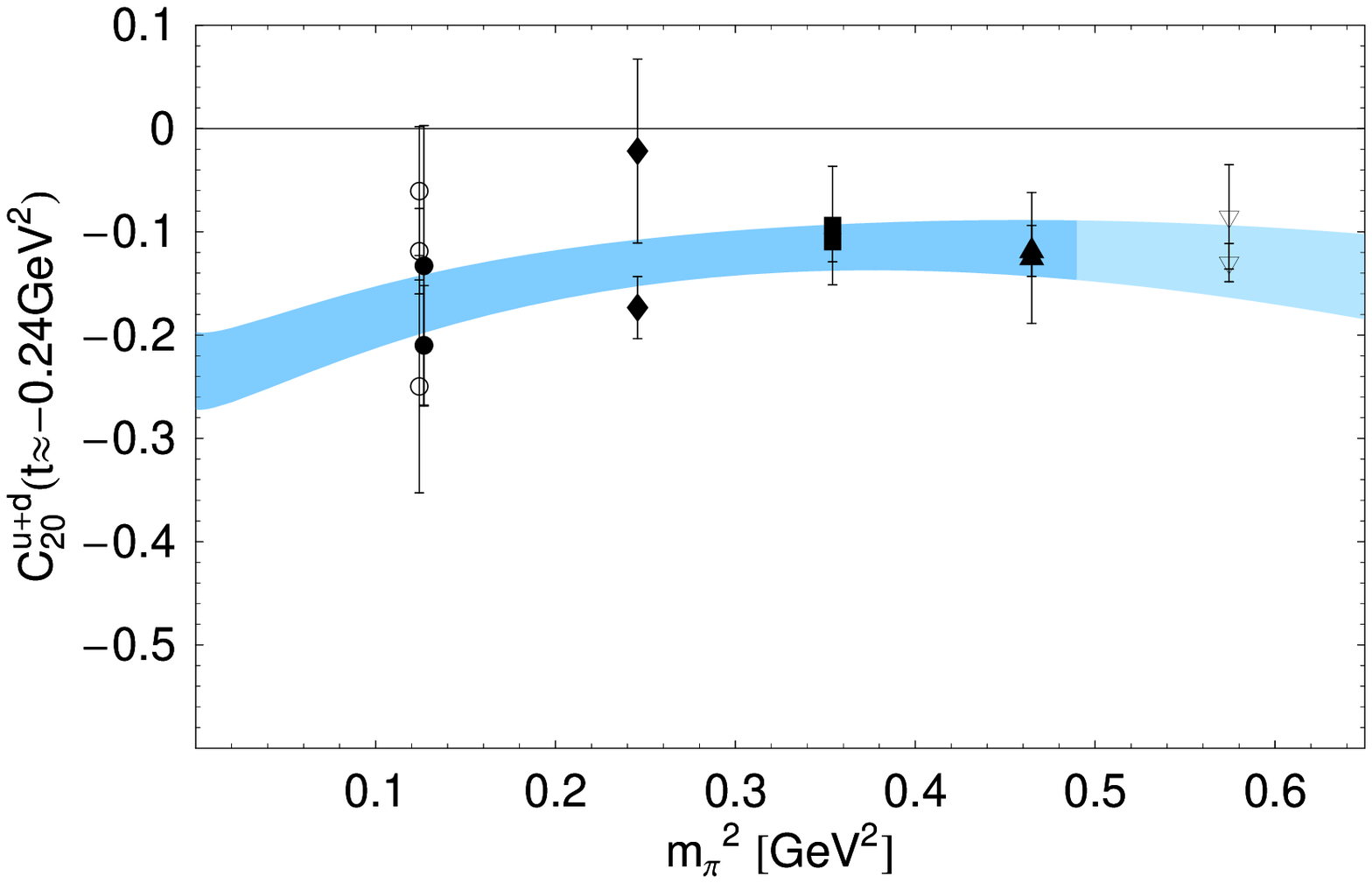}
%%      \vspace*{-0.8cm}
  \caption{Lattice results for $C^{u+d}_{20}$ at $|t|\approx 0.24$ GeV$^2$ versus $m_\pi^2$ together with the result of a global chiral fit using Eq.~(\ref{C20chPT1}). }
  \label{FigC20v1}
     \end{minipage}
     %\hfill
     \hspace{1cm}
  \begin{minipage}{0.4\textwidth}
      \centering
    \includegraphics[scale=0.42,clip=true,angle=0]{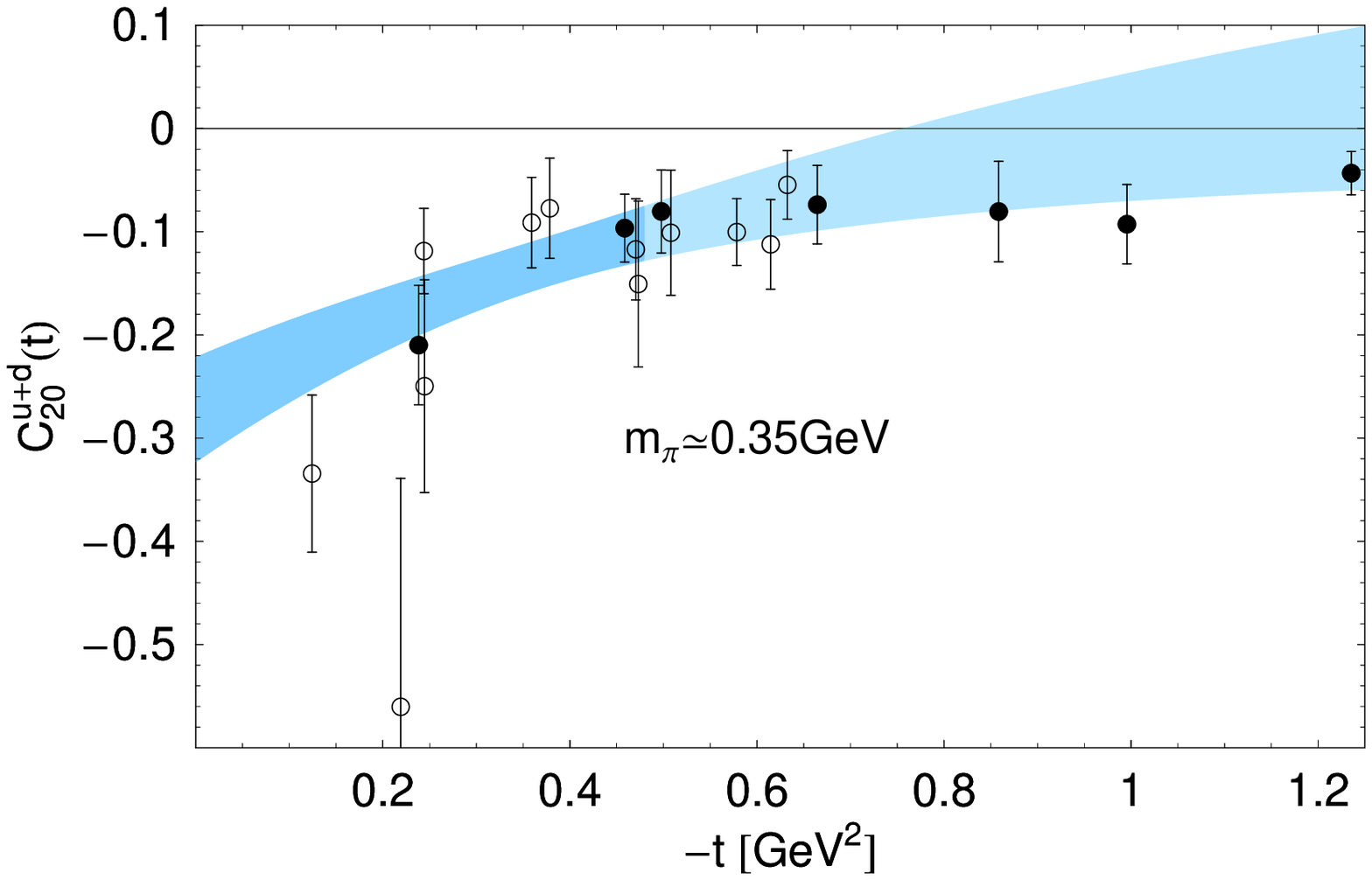}
%%      \vspace*{-0.8cm}
  \caption{Lattice results for $C^{u+d}_{20}$ at $m_\pi\!\approx\!350$ MeV versus $-t$ together with the result of a global chiral fit using  Eq.~(\ref{C20chPT1}).  }
  \label{FigC20v3}
     \end{minipage}
   \end{figure}

At order $\mathcal{O}(p^2)$, the pion mass dependence of the GFF $C^{u+d}_{20}(t)$ is
given by \cite{Ando:2006sk,Diehl:2006ya,Diehl:2006js}
\begin{equation}
C^{u+d}_{20}(t,m_\pi) = \frac{1}{1-t/(4m_N^2)}\bigg\{ C^{0,u+d}_{20} + E_2^{(1,\pi)}(t,m_\pi) + E_2^{(2,\pi)}(t,m_\pi)
+ C^{m_\pi,u+d} m_\pi^2 + C^{t,u+d} \,t \bigg\}\,,
\label{C20chPT1}
\end{equation}
where $C^{0,u+d}_{20} \equiv C^{u+d}_{20}(t=0,m_\pi=0)$. The terms $E_2^{(1,\pi)}(t,m_\pi)$ and $E_2^{(2,\pi)}(t,m_\pi)$ contain non-analytic terms in $t$ and $m_\pi$ that come from insertions of pion operators
proportional to $\langle x\rangle^{\pi,0}_{u+d}$.
Additionally, $E_2^{(2,\pi)}(t,m_\pi)$ depends on the low energy constants $c_1$, $c_2$ and $c_3$.
We fix these parameters to the values given in  Table~\ref{tab:LECs}.
The result of a fit to our lattice data as a function of the pion mass squared for fixed $t\approx -0.24\text{ GeV}^2$
is shown in Fig.~\ref{FigC20v1}. The $t$-dependence at 
a pion mass of $\approx 350$ MeV is presented in Fig.~(\ref{FigC20v3}).
We find $C^{0,u+d}_{20}=-0.507(55)$, and $C^{u+d}_{20}(t=0,m_{\pi,\text{phys}})=-0.421(54)$ at the physical pion mass.
We note that these values are approximately $60\%$ larger in magnitude than the
corresponding  CBChPT results in section \ref{sec:C20Isosinglet} based on covariant ChPT,
which is directly related to the stronger downwards bending of $C^{u+d}_{20}$
for $(-t)\rightarrow0$ in Fig.~\ref{FigC20v1} compared to the slight upwards bending 
in Fig.~(\ref{C20Isosingletp4v1}). 
As in the case of the CBChPT extrapolation, a variation of the input value 
$\langle x\rangle^{\pi,0}_{u+d}$ by $10\%$ 
results in a $\approx 5\%$ change of $C_{20}^{0,u+d}(t=0)$, which is below the 11\% statistical error.
\begin{figure}[h]
% \vspace*{-0.75cm}
        \includegraphics[scale=0.5,clip=true,angle=0]{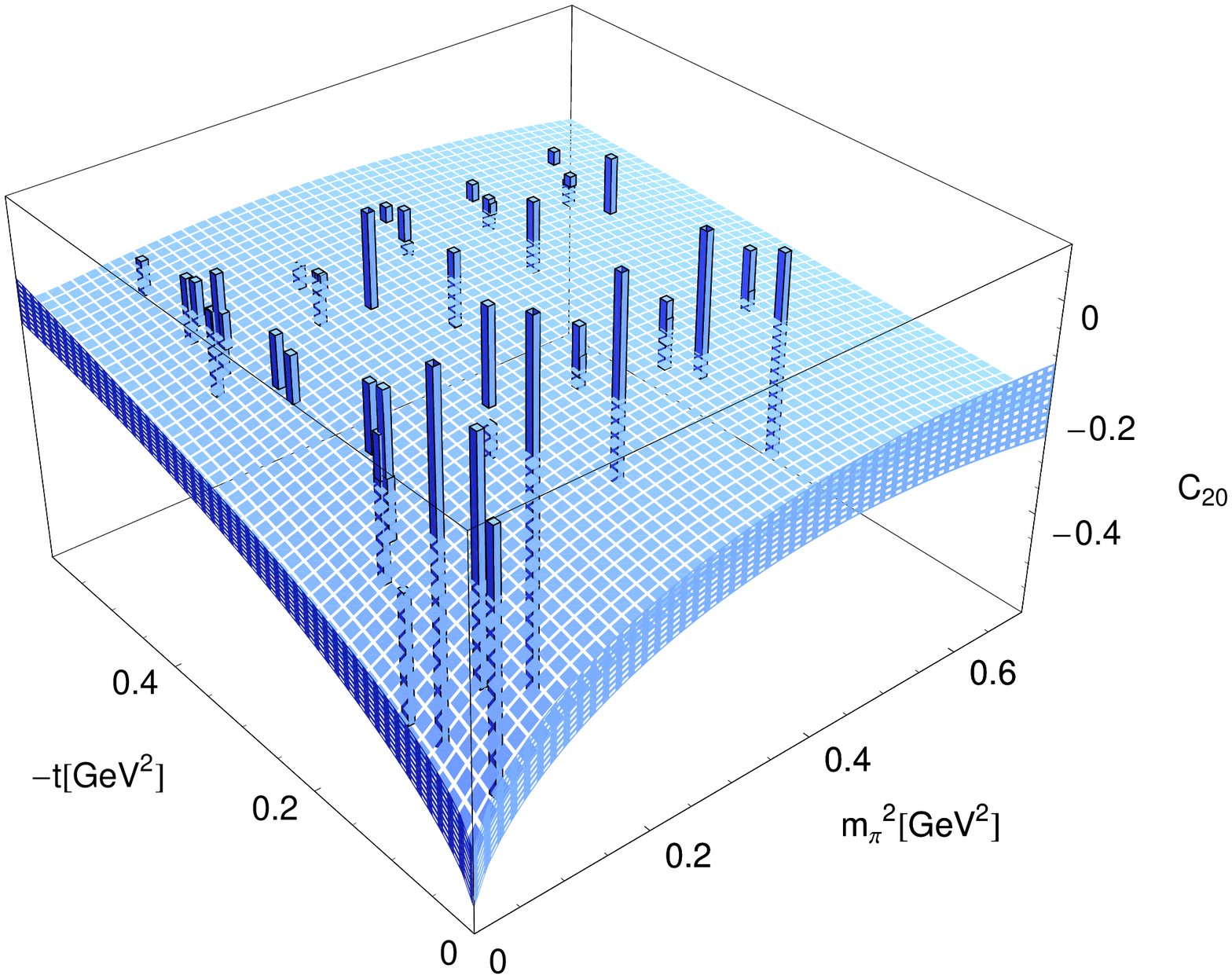}
%      \vspace*{-0.8cm}
  \caption{Combined $(t,m_\pi)$-dependence of $C^{u+d}_{20}$ from a global chiral fit using  Eq.~(\ref{C20chPT1}) compared to lattice data.}
  \label{FigC20tb}
\end{figure}

Fig.~(\ref{FigC20tb}) shows our combined lattice results for $C^{u+d}_{20}$ versus $(-t)$ and $m_\pi$ in a single
three-dimensional plot, together with the result from the HBChPT  fit discussed above, which is shown as a surface. The statistical error bars are shown for clarity as bands for $t=0$ GeV$^2$ and $m_\pi=0$ MeV, respectively. It is interesting to note that the overall shape of the extrapolation surface
is  similar to the CBChPT result in section \ref{sec:C20Isosinglet}. The only behavior that differs by more than the statistical errors is the slightly stronger bending towards negative values of $C_{20}^{u+d}$ at the origin in  Fig.~(\ref{FigC20tb}).
\subsection{\label{sec:OAMHB}HBChPT extrapolation of quark angular momentum $J$}
From our results for $M_{20}$ above, we find
a total quark angular momentum $J^{u+d}=0.263(24)$ at the physical pion mass, which is larger than
but statistically compatible with the CBChPT value in section \ref{AMcovariant}.

As an alternative, we can also calculate $J^{u+d}$ by first extrapolating $B_{20}(t,m_\pi)$ to $t=0$ and combining it with $A_{20}(t=0,m_\pi)$ to obtain $J(m_\pi)$, and then extrapolating the values of  $J(m_\pi)$ to the physical pion mass using HBChPT that explicitly includes the $\Delta$ resonance \cite{Chen:2001pv}.
Evaluating  Eq.~(\ref{M20chPT1}) at $t=0$ yields
\begin{equation}
J^{u+d}(m_\pi)=\frac{1}{2}\left\{(A+B)^{0,u+d}_{20} + 3\bigg(\langle x\rangle^{\pi,0}_{u+d}-(A+B)^{0,u+d}_{20}\bigg)\frac{g_A^2 m_\pi^2}{(4 \pi f_\pi)^2}\ln\left(\frac{m_\pi^2}{\Lambda_\chi^2}\right)\right\} + J^{m_\pi,u+d} m_\pi^2 \,,
\label{JqchPT1}
\end{equation}
which agrees with \cite{Chen:2001pv}. Note that in the notation of \cite{Chen:2001pv} we have $b_{qN}=(A+B)^{0,u+d}_{20}=M^{0,u+d}$.
\begin{figure}[htbp]
     \begin{minipage}{0.4\textwidth}
      \centering
          \includegraphics[scale=0.4,clip=true,angle=0]{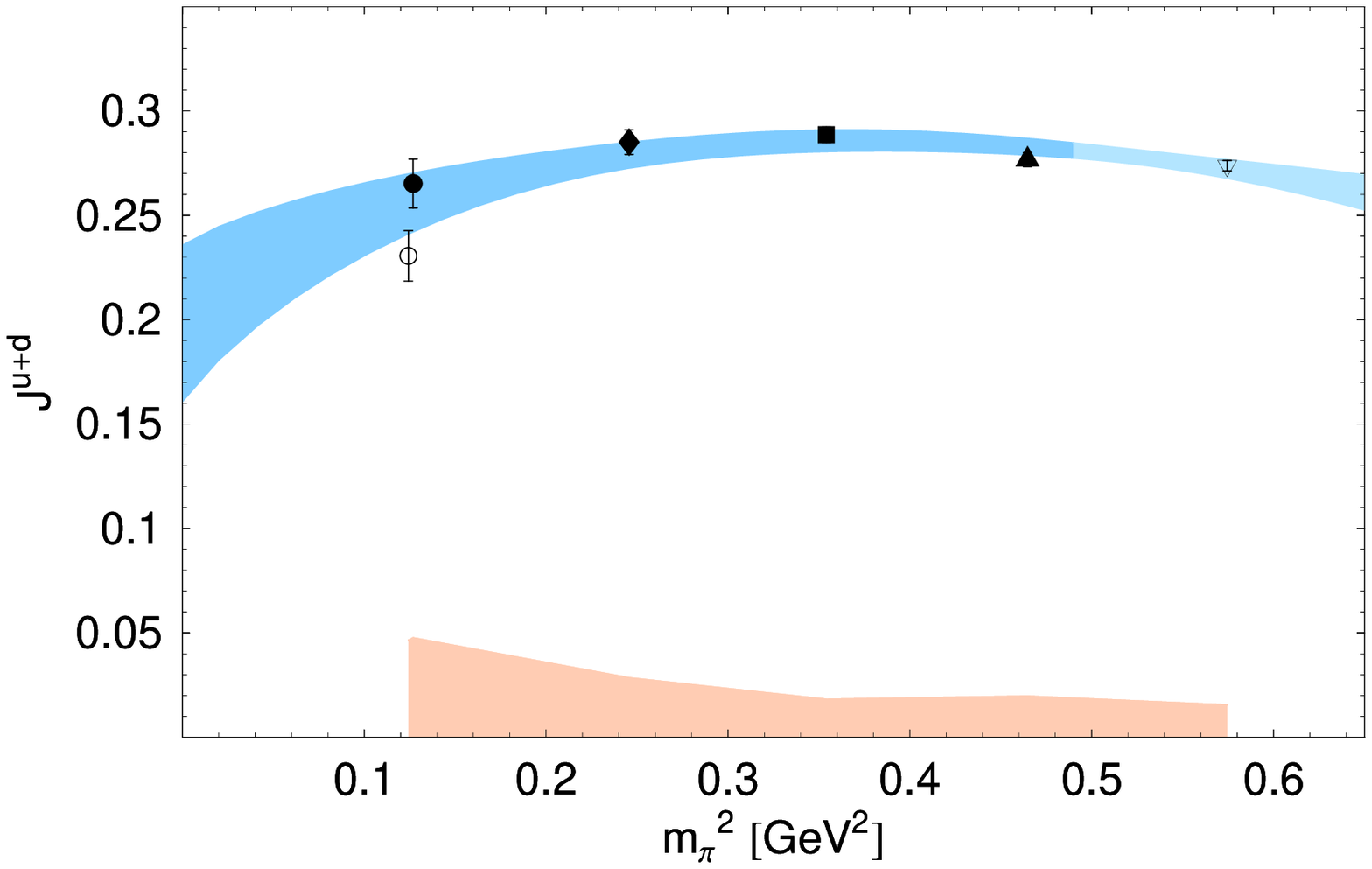}
%      \vspace*{-0.8cm}
  \caption{Chiral extrapolation of $J^{u+d}$ using HBChPT including the $\Delta$ resonance, Eq.~(\ref{ChPTJqDelta}).  The fit and error band on the axis are explained in the text.\newline}\label{Jq1}
     \end{minipage}
     %\hfill
     \hspace{1cm}
     \begin{minipage}{0.4\textwidth}
      \centering
         \includegraphics[scale=0.4,clip=true,angle=0]{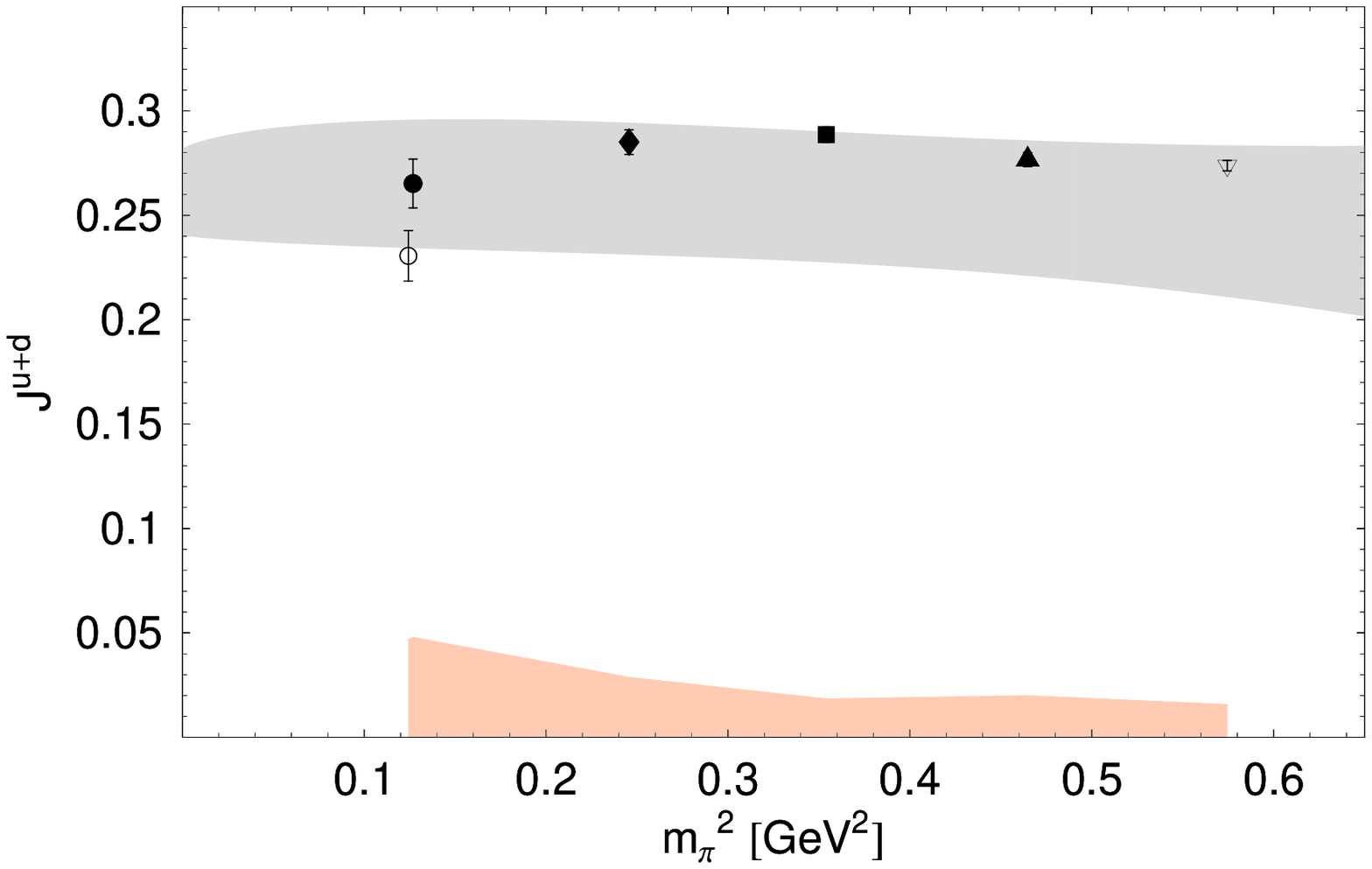}
%      \vspace*{-0.8cm}
  \caption{Comparison of a global fit to $A_{20}^{u+d}+B_{20}^{u+d}$ using Eq.~(\ref{M20chPT1})
  and the lattice results for $J^{u+d}$.  The  fit and error band on the axis are explained in the text.}\label{Jq2}
     \end{minipage}
   \end{figure}
%
%\end{document}
Including explicitly the $\Delta$ resonance in the calculation, the ChPT result then reads  \cite{Chen:2001pv}
\begin{eqnarray}
J^{u+d}(m_\pi;\Delta)&=& J^{u+d}(m_\pi) - \frac{1}{2}\left(\frac{9}{2} (A+B)^{0,u+d}_{20}+ 3 
\langle x\rangle^{\pi,0}_{u+d} - \frac{15}{2}b_{q\Delta}\right)\nonumber\\ &&\times
\frac{8 g_{\pi N\Delta}^2}{9(4 \pi f_\pi)^2}\left\{(m_\pi^2-2\Delta^2)\ln\left(\frac{m_\pi^2}{\Lambda_\chi^2}\right)
+2\Delta \sqrt{\Delta^2 - m_\pi^2} \ln\left(\frac{\Delta - \sqrt{\Delta^2 - m_\pi^2}}{\Delta + \sqrt{\Delta^2 - m_\pi^2}}\right)\right\}\label{ChPTJqDelta} \,,
\end{eqnarray}
where $\Delta=m_\Delta-m_N$ denotes the $\Delta$-nucleon mass difference. 
In order to reduce the number of free parameters in the fit, we use 
$\Delta=0.3$~GeV and the large-$N_c$ relation $g_{\pi N\Delta} = 3/(2^{3/2}) g_A$ from Table \ref{tab:LECs}. 
We first extrapolate $B_{20}^{u+d}(t, m_\pi)$ linearly in $t$ to $t=0$. A 
fit based on Eq.~(\ref{ChPTJqDelta}) to the lattice results for $(A+B)^{u+d}_{20}(t=0,m_\pi)$ with $m_\pi\le 700$ MeV 
then gives $(A+B)^{0,u+d}_{20}=b_{qN}=0.545(12)$, $b_{q\Delta}=0.427(51)$ and $J^{u+d}(m_{\pi,\text{phys}};\Delta)=0.212(32)$ 
at the physical pion mass, which is very close to the CBChPT result in section \ref{AMcovariant}, $J^{u+d}(m_{\pi,\text{phys}})=0.213(44)$.
 The result for $J^{u+d}(m_\pi;\Delta)$ as a function of the pion mass is shown in Fig.~(\ref{Jq1}), where
the error due to the linear extrapolation of $B(t,m_\pi)$  to $t = 0$ is indicated by the error band at the 
$m_\pi^2$-axis as explained  at the end of section \ref{sec:results} and the error bars on the lattice data points for $J$ only include the errors arising from $A_{20}(t=0)$.
In Fig.~(\ref{Jq2}), we compare the result of the chiral extrapolation of $M_{20}^{u+d}(t)$ for non-zero $t$ 
from the previous section with the lattice results for $J^{u+d}$ corresponding to the extrapolated $B_{20}^{u+d}(t=0)$.
The two different ways of fitting and extrapolating $J^{u+d}$ in $m_\pi$ are compatible within errors,
$J^{u+d}_{\text{from }t\not=0}=0.263(24)$ versus $J^{u+d}_{t\rightarrow 0}(m_{\pi,\text{phys}};\Delta)=0.212(32)$,
where the chiral fit including the $\Delta$ leads to a stronger curvature at small $m_\pi$ and
therefore to a smaller central value for $J^{u+d}$ at the physical point.
Together with preliminary results for the quark spin
$\widetilde A^{u+d}_{10}/2(t=0)=\Delta\Sigma^{u+d} /2$ 
as obtained from a self-consistently improved one-loop ChPT extrapolation \cite{DrusPDFs},
we find that the quark orbital angular momentum 
$L^{u+d}=J^{u+d}-\Delta\Sigma^{u+d}/2$ contribution to the nucleon spin 
is zero within errors: $L^{u+d}=0.005(43)$ for $J^{u+d}_{t\rightarrow 0}(m_{\pi,\text{phys}};\Delta)=0.212(32)$ and $L^{u+d}=0.056(37)$ for $J^{u+d}_{\text{from }t\not=0}=0.263(24)$ at the physical pion mass.

% --------------------------------------------------------------------------
%
%  Conclusions
%
% --------------------------------------------------------------------------

\section{\label{sec:Conclusions}Summary and Conclusions}
This work presents the first comprehensive full lattice QCD study of the lowest three moments
of unpolarized and polarized GPDs in the chiral regime with pion masses as low as $350$ MeV. 
We find good overall agreement with existing experimental results. We note, however, that
we have omitted disconnected diagrams, which in principle contribute to isoscalar observables.

As in our previous study of the axial vector coupling constant \cite{Edwards:2005ym},
the consistency of these moments in lattice volumes of $ (2.5\text{fm})^3$ and $(3.5\text{fm})^3$
at $m_\pi = 350$~MeV indicates that finite volume effects are negligible within statistical errors.

One significant result of this work is the clear indication that the transverse size of the nucleon, as
characterized by the transverse 2d rms radius, is a strongly decreasing function of the 
momentum fraction $x$ carried by the quarks. At the lightest quark mass,
the isovector transverse rms radius drops by almost $60 \%$ between the zeroth moment, which roughly corresponds to an average momentum fraction\cite{Negele:2004iu} $\overline{\langle x\rangle} \approx 0.2$ and the second moment, which roughly corresponds to an average momentum fraction $\overline{\langle x\rangle} \approx 0.4$. 
This decrease in the chiral regime is even stronger than our original observation of 
the decrease of the transverse size with momentum fraction in the ``heavy pion world''\cite{LHPC:2003is}.

In a first direct comparison with phenomenological parametrizations of the GPDs $H(x,\xi=0,t)$ 
and $\tilde H(x,\xi=0,t)$ constrained by structure function and form factor data\cite{Diehl:2004cx}, we
find qualitative consistency for the ratios of GFFs in both the isovector and isosinglet cases. 

Our results provide  insight into the contributions of the spin and orbital angular momentum of u and d quarks to the spin of the proton. Although the individual orbital angular momentum contributions of the u and d quarks are sizeable, $L^d\approx-L^{u}\approx 30\%$, they cancel within errors so that the total contribution is $L^{u+d} \approx 0$.  In addition, the spin and orbital contributions of the d quark also cancel within errors, so $J^{d} \approx 0$.  The total quark angular momentum contribution is 
$J^{u+d}\approx 40-50\%$ at our lowest pion masses.

\begin{table}[ht]
  \begin{tabular}{|c|c|c|c|c|c|c|c|c|c|}
    \hline 
     & $A^{u-d}_{20}$ & $B^{u-d}_{20}$ & $C^{u-d}_{20}$ & $A^{u+d}_{20}$ & $B^{u+d}_{20}$ & $C^{u+d}_{20}$ & $J^{u+d}$ & $J^{u}$ & $J^{d}$ \\ \hline
    covariant BChPT & $0.157(10)$ & $0.273(63)$ & $-0.017(41)$ & $0.520(24)$ & $-0.095(86)$ & $-0.267(62)$ & $0.213(44)$ & $0.214(27)$ & $-0.001(27)$\\
    HBChPT &  &  &  & $0.485(14)$ & $0.050(49)$ & $-0.421(54)$ & $0.263(24)$ & &  \\
    HBChPT with $\Delta$ & &  & &  &  &  & $0.212(32)$ &  &       \\
    phenomenology & $0.155(5)$ &  & & $0.537(22)$ &  &  &  &  &\\  
     \hline
  \end{tabular}
  \caption{Summary of proton observables at $t=0$ and $m_{\pi,\text{phys}}$ from chiral extrapolations, in the $\overline{\text{MS}}$ scheme at a scale of $\mu^2=4$ GeV$^2$.}
      \label{tab:chptres}
\end{table}

More quantitatively, we performed a variety of chiral fits to the unpolarized $n=2$ moments using covariant BChPT \cite{Dorati:2007bk} and HBChPT results \cite{Chen:2001pv,Ando:2006sk,Diehl:2006js,Diehl:2006ya}. A summary of our results
for various observables at the physical pion mass and vanishing momentum transfer $t=0$ 
is given in Table \ref{tab:chptres}, where the quoted errors are statistical only.
We note that a consistent inclusion of all $\mathcal{O}(p^3)$ terms in our fits, which may involve additional constants, would have the potential to increase the statistical errors on the physical quantities at the chiral limit.  However, for the reasons described in section \ref{sec:A20IsoVecrel}, the prescription of adding a single $m_\pi^3$ term with extremal values of ``natural" coefficients does not lead to reasonable $\chi^2$ fits, so we have not attempted to include quantitative estimates of these errors in the present work.

With the exception of a fit to $J_q$ based on HBChPT including the $\Delta$ resonance,
we have consistently extrapolated the GFFs  simultaneously in the pion mass and the
momentum transfer squared $t$. The simultaneous covariant BChPT fits to the GFFs $A_{20}$, $B_{20}$ and $C_{20}$, 
which include approximately 120 lattice data points and 
typically 9 unknown low energy constants, produce reasonable parametrizations of the 
$(t,m_\pi)$-dependences of the generalized form factors in the ranges $m_\pi\le700$MeV and $|t|\le0.5$ GeV$^2$. 
In particular, the covariant extrapolations for the isovector and isosinglet momentum 
fractions $\langle x\rangle$ yield values at the physical point remarkably close 
to phenomenology. This represents a significant advance in our understanding of the pion mass
dependence of these important observables. 
The first exploration of the combined non-analytic dependence of the isosinglet GFFs $B^{u+d}_{20}$ and $C^{u+d}_{20}$ on $t$ and $m_\pi$
was made using covariant BChPT and HBChPT, and visualizations of the resulting $(t,m_\pi)$-dependences of these GFFs 
in  3d-plots reveal interesting non-linear correlations in the pion mass and the momentum transfer squared.

In spite of the overall success of the chiral extrapolations, it is clear that the ChPT analysis has not yet been carried to sufficiently high order to be applicable to the full range of pion masses included in this work.  The facts that HBChPT fits to $A_{20}^{u-d}$ and $A_{20}^{u+d}$  cannot describe the behavior of the lattice data over  as large a range or as accurately as covariant BChPT, and that the fitted counter terms are so different indicate that the higher order terms in $1/m_N$  included in CBChPT are important for these observables.  Similarly, the fact that it was important to include some particular terms of  order $\mathcal{O}(p^3)$ and $\mathcal{O}(p^4)$ in some CBChPT fits indicates the need to 
fully determine these orders in ChPT so that they can be consistently included in fits to lattice data.
Finally, the significant effect of including the $\Delta$ in $J^{u+d}$, its known importance in the axial charge, and large $N_c$ arguments\cite{Dashen:1993jt} indicate the desirability of consistent inclusion of the $\Delta$. Thus, future progress requires both the extension of lattice calculations to lower pion mass and the inclusion of higher order terms and $\Delta$ degrees of freedom in ChPT.

\begin{acknowledgments}
  The authors wish to thank M.~Diehl, M.~Dorati, T.~Gail, T.~Hemmert, J.-W.~Chen and  A.~Manohar for stimulating discussions.
  This work was supported 
  in part by U.S. DOE Contract No. DE-AC05-06OR23177 under which JSA operates Jefferson Laboratory, 
  by the DOE Office of Nuclear Physics under grants DE-FG02-94ER40818, DE-FG02-04ER41302, 
  DE-FG02-96ER40965, the Jeffress Memorial Trust grant J-813, the DFG (Forschergruppe     Gitter-Hadronen-Ph{\"a}nomenologie) and in 
  part by the EU Integrated Infrastructure Initiative Hadron Physics (I3HP) 
  under contract number RII3-CT-2004-506078. Ph.~H. and B.~M. acknowledge support by the
  Emmy-Noether program of the DFG. Ph.~H. and W.~S. would like to thank 
  the A.v.~Humboldt-foundation for support by the Feodor-Lynen program.
  It is a pleasure to acknowledge the use of computer resources provided by the DOE through the USQCD project at Jefferson Lab and ORNL and through its support of the MIT Blue Gene/L.
  We are indebted to members of the MILC and SESAM Collaborations for providing the dynamical
  quark configurations which made our full QCD calculations possible.
\end{acknowledgments}

\clearpage

\appendix
\section{Tables}
\label{appendix1}
%\vspace{1mm}
%\hline

Lattice parameters of the datasets $1,\ldots,6$ are provided in Table~\ref{tab:parameters-summary}.
The GFFs are given in the $\overline{\text{MS}}$ scheme at a scale of $\mu^2=4$ GeV$^2$.

\begin{table}[htbp]
\begin{tabular}{@{}*{10}{l}}
 \hline
 $\left.\text{-t[}\text{GeV}^2\right]$ & $A_{10}^{u-d}$ & $B_{10}^{u-d}$ & $A_{20}^{u-d}$ & $B_{20}^{u-d}$ &
   $C_{20}^{u-d}$ & $A_{30}^{u-d}$ & $B_{30}^{u-d}$ \\
 \hline
 0.000 & \text{1.000(1)} & \text{  -} & \text{0.223(3)} & \text{  -} & \text{  -} & \text{0.072(4)} & \text{  -} \\
 0.203 & \text{0.731(73)} & \text{2.331(347)} & \text{0.179(17)} & \text{0.331(91)} & \text{-0.094(113)} & \text{0.072(10)} &
   \text{0.051(59)} \\
 0.243 & \text{0.796(6)} & \text{2.645(55)} & \text{0.201(3)} & \text{0.398(16)} & \text{-0.002(11)} & \text{0.071(2)} &
   \text{0.109(9)} \\
 0.243 & \text{0.735(29)} & \text{2.570(120)} & \text{0.187(7)} & \text{0.340(29)} & \text{-0.006(28)} & \text{0.066(3)} &
   \text{0.078(13)} \\
 0.420 & \text{0.606(64)} & \text{1.748(239)} & \text{0.160(16)} & \text{0.258(53)} & \text{-0.065(32)} & \text{0.055(7)} &
   \text{0.078(29)} \\
 0.475 & \text{0.659(9)} & \text{2.093(48)} & \text{0.180(3)} & \text{0.341(15)} & \text{-0.008(6)} & \text{0.066(3)} &
   \text{0.092(8)} \\
 0.477 & \text{0.605(29)} & \text{2.000(110)} & \text{0.170(7)} & \text{0.311(29)} & \text{-0.027(14)} & \text{0.062(3)} &
   \text{0.069(13)} \\
\hline
\end{tabular}
\caption{Results for the isovector unpolarized generalized form factors for dataset 1.}
  \label{GFFsUnpol1}
\end{table}

\begin{table}[htbp]
 \begin{tabular}{@{}*{10}{l}}
 \hline
 $\left.\text{-t[}\text{GeV}^2\right]$ & $A_{10}^{u-d}$ & $B_{10}^{u-d}$ & $A_{20}^{u-d}$ & $B_{20}^{u-d}$ &
   $C_{20}^{u-d}$ & $A_{30}^{u-d}$ & $B_{30}^{u-d}$ \\
 \hline
 0.000 & \text{1.000(2)} & \text{  -} & \text{0.223(4)} & \text{  -} & \text{  -} & \text{0.078(6)} & \text{  -} \\
 0.200 & \text{0.924(100)} & \text{3.454(490)} & \text{0.196(26)} & \text{0.461(125)} & \text{0.261(149)} & \text{0.054(15)} &
   \text{0.168(76)} \\
 0.242 & \text{0.798(8)} & \text{2.540(67)} & \text{0.201(4)} & \text{0.350(20)} & \text{-0.006(14)} & \text{0.070(3)} &
   \text{0.072(11)} \\
 0.243 & \text{0.804(38)} & \text{2.384(146)} & \text{0.196(9)} & \text{0.289(36)} & \text{-0.044(37)} & \text{0.072(5)} &
   \text{0.057(17)} \\
 0.414 & \text{0.782(85)} & \text{2.226(301)} & \text{0.189(22)} & \text{0.346(69)} & \text{0.005(43)} & \text{0.061(10)} &
   \text{0.137(34)} \\
 0.473 & \text{0.678(11)} & \text{2.034(57)} & \text{0.181(4)} & \text{0.314(18)} & \text{-0.001(7)} & \text{0.063(4)} &
   \text{0.073(10)} \\
 0.475 & \text{0.656(37)} & \text{1.697(133)} & \text{0.171(9)} & \text{0.221(35)} & \text{0.000(19)} & \text{0.066(5)} &
   \text{0.053(18)} \\
\hline
\end{tabular}
\caption{Results for the isovector unpolarized generalized form factors for dataset 2.}
 \label{GFFsUnpol2}
\end{table}

\begin{table}[htbp]
\begin{tabular}{@{}*{10}{l}}
 \hline
 $\left.\text{-t[}\text{GeV}^2\right]$ & $A_{10}^{u-d}$ & $B_{10}^{u-d}$ & $A_{20}^{u-d}$ & $B_{20}^{u-d}$ &
   $C_{20}^{u-d}$ & $A_{30}^{u-d}$ & $B_{30}^{u-d}$ \\
 \hline
 0.000 & \text{1.000(1)} & \text{  -} & \text{0.226(4)} & \text{  -} & \text{  -} & \text{0.074(5)} & \text{  -} \\
 0.193 & \text{0.674(83)} & \text{2.147(465)} & \text{0.176(24)} & \text{0.226(119)} & \text{-0.044(152)} & \text{0.046(13)} &
   \text{0.058(87)} \\
 0.241 & \text{0.770(8)} & \text{2.399(62)} & \text{0.200(4)} & \text{0.358(18)} & \text{-0.009(13)} & \text{0.074(3)} &
   \text{0.094(12)} \\
 0.242 & \text{0.724(35)} & \text{2.232(154)} & \text{0.187(8)} & \text{0.345(35)} & \text{0.052(35)} & \text{0.067(4)} &
   \text{0.075(17)} \\
 0.403 & \text{0.627(78)} & \text{1.723(305)} & \text{0.165(20)} & \text{0.243(68)} & \text{-0.010(45)} & \text{0.049(9)} &
   \text{0.118(38)} \\
 0.469 & \text{0.640(12)} & \text{1.890(54)} & \text{0.181(4)} & \text{0.300(17)} & \text{0.000(7)} & \text{0.069(4)} &
   \text{0.078(10)} \\
 0.472 & \text{0.584(34)} & \text{1.713(142)} & \text{0.171(9)} & \text{0.265(34)} & \text{0.003(20)} & \text{0.065(5)} &
   \text{0.075(17)} \\
\hline
\end{tabular} 
\caption{Results for the isovector unpolarized generalized form factors for dataset 3.}
 \label{GFFsUnpol3}
\end{table}

\begin{table}[htbp]
\begin{tabular}{@{}*{10}{l}}
\hline
 $\left.\text{-t[}\text{GeV}^2\right]$ & $A_{10}^{u-d}$ & $B_{10}^{u-d}$ & $A_{20}^{u-d}$ & $B_{20}^{u-d}$ &
   $C_{20}^{u-d}$ & $A_{30}^{u-d}$ & $B_{30}^{u-d}$ \\
 \hline0.000 & \text{1.000(2)} & \text{  -} & \text{0.214(7)} & \text{  -} & \text{  -} & \text{0.070(8)} & \text{  -} \\
 0.188 & \text{0.548(129)} & \text{3.044(876)} & \text{0.202(46)} & \text{0.505(205)} & \text{-0.059(262)} & \text{0.074(28)} &
   \text{0.045(165)} \\
 0.240 & \text{0.753(13)} & \text{2.281(96)} & \text{0.190(6)} & \text{0.307(30)} & \text{-0.015(21)} & \text{0.071(5)} &
   \text{0.059(20)} \\
 0.241 & \text{0.765(60)} & \text{2.335(250)} & \text{0.182(14)} & \text{0.309(56)} & \text{0.091(57)} & \text{0.066(6)} &
   \text{0.095(28)} \\
 0.394 & \text{0.529(110)} & \text{1.927(522)} & \text{0.143(31)} & \text{0.188(118)} & \text{0.119(81)} & \text{0.047(16)} &
   \text{0.001(71)} \\
 0.465 & \text{0.601(19)} & \text{1.696(83)} & \text{0.173(7)} & \text{0.271(26)} & \text{-0.020(12)} & \text{0.070(6)} &
   \text{0.072(16)} \\
 0.469 & \text{0.630(57)} & \text{2.034(223)} & \text{0.161(14)} & \text{0.311(52)} & \text{0.003(32)} & \text{0.066(7)} &
   \text{0.083(28)} \\
\hline
\end{tabular} 
\caption{Results for the isovector unpolarized generalized form factors for dataset 4.}
 \label{GFFsUnpol4}
\end{table}

\begin{table}[htbp]
\begin{tabular}{@{}*{10}{l}}
\hline
 $\left.\text{-t[}\text{GeV}^2\right]$ & $A_{10}^{u-d}$ & $B_{10}^{u-d}$ & $A_{20}^{u-d}$ & $B_{20}^{u-d}$ &
   $C_{20}^{u-d}$ & $A_{30}^{u-d}$ & $B_{30}^{u-d}$ \\
 \hline
  0.000 & \text{1.000(3)} & \text{  -} & \text{0.205(14)} & \text{  -} & \text{  -} & \text{0.063(16)} & \text{  -} \\
 0.177 & \text{0.387(251)} & \text{2.385(1.795)} & \text{0.095(79)} & \text{0.717(475)} & \text{0.129(518)} & \text{-0.033(63)}
   & \text{0.537(408)} \\
 0.238 & \text{0.723(24)} & \text{1.957(166)} & \text{0.172(11)} & \text{0.282(49)} & \text{-0.025(36)} & \text{0.049(11)} &
   \text{0.020(40)} \\
 0.240 & \text{0.587(97)} & \text{1.541(396)} & \text{0.135(20)} & \text{0.262(91)} & \text{-0.095(93)} & \text{0.063(12)} &
   \text{0.045(52)} \\
 0.378 & \text{  -} & \text{  -} & \text{  -} & \text{  -} & \text{  -} & \text{  -} & \text{  -} \\
 0.459 & \text{0.604(35)} & \text{1.566(137)} & \text{0.156(12)} & \text{0.221(44)} & \text{0.009(22)} & \text{0.043(12)} &
   \text{0.065(33)} \\
 0.464 & \text{0.502(100)} & \text{1.072(373)} & \text{0.131(25)} & \text{0.344(97)} & \text{0.047(65)} & \text{0.052(15)} &
   \text{0.069(52)} \\
\hline
\end{tabular} 
\caption{Results for the isovector unpolarized generalized form factors for dataset 5.}
 \label{GFFsUnpol5}
\end{table}

\begin{table}[htbp]
\begin{tabular}{@{}*{10}{l}}
\hline
 $\left.\text{-t[}\text{GeV}^2\right]$ & $A_{10}^{u-d}$ & $B_{10}^{u-d}$ & $A_{20}^{u-d}$ & $B_{20}^{u-d}$ &
   $C_{20}^{u-d}$ & $A_{30}^{u-d}$ & $B_{30}^{u-d}$ \\
 \hline
0.000 & \text{1.000(4)} & \text{  -} & \text{0.206(14)} & \text{  -} & \text{  -} & \text{0.078(16)} & \text{  -} \\
 0.107 & \text{1.035(192)} & \text{3.055(997)} & \text{0.190(45)} & \text{0.032(253)} & \text{-0.083(435)} & \text{0.060(32)} &
   \text{-0.303(221)} \\
 0.124 & \text{0.850(19)} & \text{2.756(233)} & \text{0.197(11)} & \text{0.331(59)} & \text{-0.097(54)} & \text{0.070(10)} &
   \text{0.099(46)} \\
 0.125 & \text{0.829(84)} & \text{2.262(376)} & \text{0.192(20)} & \text{0.284(110)} & \text{-0.017(113)} & \text{0.072(12)} &
   \text{0.094(62)} \\
 0.219 & \text{0.868(190)} & \text{2.749(823)} & \text{0.167(42)} & \text{0.182(169)} & \text{0.026(130)} & \text{0.063(24)} &
   \text{-0.084(109)} \\
 0.244 & \text{0.767(28)} & \text{2.330(193)} & \text{0.177(11)} & \text{0.298(49)} & \text{-0.015(28)} & \text{0.063(10)} &
   \text{0.091(34)} \\
 0.245 & \text{0.733(79)} & \text{1.883(328)} & \text{0.179(20)} & \text{0.137(93)} & \text{-0.164(73)} & \text{0.068(12)} &
   \text{0.099(64)} \\
\hline
\end{tabular} 
\caption{Results for the isovector unpolarized generalized form factors for dataset 6.}
 \label{GFFsUnpol6}
\end{table}

\begin{table}[htbp]
\begin{tabular}{@{}*{10}{l}}
 \hline
 $\left.\text{-t[}\text{GeV}^2\right]$ & $A_{10}^{u+d}$ & $B_{10}^{u+d}$ & $A_{20}^{u+d}$ & $B_{20}^{u+d}$ &
   $C_{20}^{u+d}$ & $A_{30}^{u+d}$ & $B_{30}^{u+d}$ \\
 \hline
 0.000 & \text{3.000(2)} & \text{  -} & \text{0.590(5)} & \text{  -} & \text{  -} & \text{0.176(7)} & \text{  -} \\
 0.203 & \text{2.068(180)} & \text{-0.828(581)} & \text{0.473(40)} & \text{-0.230(145)} & \text{-0.201(196)} & \text{0.178(19)}
   & \text{-0.123(95)} \\
 0.243 & \text{2.229(14)} & \text{0.067(67)} & \text{0.506(5)} & \text{-0.024(20)} & \text{-0.130(19)} & \text{0.163(4)} &
   \text{-0.009(13)} \\
 0.243 & \text{2.067(71)} & \text{0.083(150)} & \text{0.473(15)} & \text{-0.067(44)} & \text{-0.086(51)} & \text{0.158(6)} &
   \text{-0.024(20)} \\
 0.420 & \text{1.649(153)} & \text{-0.380(350)} & \text{0.398(35)} & \text{-0.178(91)} & \text{-0.056(53)} & \text{0.137(14)} &
   \text{-0.082(45)} \\
 0.475 & \text{1.756(17)} & \text{0.008(60)} & \text{0.437(5)} & \text{-0.042(19)} & \text{-0.090(10)} & \text{0.144(5)} &
   \text{-0.013(12)} \\
 0.477 & \text{1.633(63)} & \text{0.117(135)} & \text{0.415(15)} & \text{-0.044(43)} & \text{-0.116(23)} & \text{0.143(7)} &
   \text{-0.024(19)} \\
\hline
\end{tabular}
\caption{Results for the isosinglet unpolarized generalized form factors for dataset 1.}
  \label{GFFsUnpolIsoSingl1}
\end{table}

\begin{table}[htbp]
\begin{tabular}{@{}*{10}{l}}
 \hline
 $\left.\text{-t[}\text{GeV}^2\right]$ & $A_{10}^{u+d}$ & $B_{10}^{u+d}$ & $A_{20}^{u+d}$ & $B_{20}^{u+d}$ &
   $C_{20}^{u+d}$ & $A_{30}^{u+d}$ & $B_{30}^{u+d}$ \\
 \hline
 0.000 & \text{3.000(3)} & \text{  -} & \text{0.577(7)} & \text{  -} & \text{  -} & \text{0.170(9)} & \text{  -} \\
 0.200 & \text{2.286(222)} & \text{0.641(651)} & \text{0.478(51)} & \text{-0.010(188)} & \text{0.077(243)} & \text{0.131(26)} &
   \text{-0.029(131)} \\
 0.242 & \text{2.227(17)} & \text{0.049(88)} & \text{0.496(6)} & \text{-0.035(27)} & \text{-0.119(25)} & \text{0.152(5)} &
   \text{-0.033(17)} \\
 0.243 & \text{2.134(91)} & \text{0.186(183)} & \text{0.474(19)} & \text{-0.009(50)} & \text{-0.125(63)} & \text{0.153(8)} &
   \text{-0.022(25)} \\
 0.414 & \text{1.773(178)} & \text{0.352(392)} & \text{0.416(42)} & \text{0.083(95)} & \text{-0.087(67)} & \text{0.124(16)} &
   \text{0.064(49)} \\
 0.473 & \text{1.763(23)} & \text{0.055(75)} & \text{0.431(7)} & \text{0.000(23)} & \text{-0.081(11)} & \text{0.137(7)} &
   \text{-0.006(15)} \\
 0.475 & \text{1.666(86)} & \text{0.191(179)} & \text{0.403(19)} & \text{-0.048(50)} & \text{-0.093(30)} & \text{0.138(8)} &
   \text{-0.014(26)} \\
\hline
\end{tabular}
\caption{Results for the isosinglet unpolarized generalized form factors for dataset 2.}
  \label{GFFsUnpolIsoSingl2}
\end{table}

\begin{table}[htbp]
\begin{tabular}{@{}*{10}{l}}
 \hline
 $\left.\text{-t[}\text{GeV}^2\right]$ & $A_{10}^{u+d}$ & $B_{10}^{u+d}$ & $A_{20}^{u+d}$ & $B_{20}^{u+d}$ &
   $C_{20}^{u+d}$ & $A_{30}^{u+d}$ & $B_{30}^{u+d}$ \\
 \hline
 0.000 & \text{3.000(3)} & \text{  -} & \text{0.565(7)} & \text{  -} & \text{  -} & \text{0.171(9)} & \text{  -} \\
 0.193 & \text{2.036(204)} & \text{0.517(614)} & \text{0.429(47)} & \text{-0.111(175)} & \text{-0.034(247)} & \text{0.111(23)} &
   \text{-0.086(131)} \\
 0.241 & \text{2.172(16)} & \text{-0.034(82)} & \text{0.482(6)} & \text{-0.002(24)} & \text{-0.110(20)} & \text{0.159(5)} &
   \text{-0.014(18)} \\
 0.242 & \text{2.047(87)} & \text{0.024(170)} & \text{0.456(18)} & \text{0.012(50)} & \text{-0.094(57)} & \text{0.152(7)} &
   \text{-0.012(25)} \\
 0.403 & \text{1.578(169)} & \text{0.071(377)} & \text{0.382(41)} & \text{-0.040(95)} & \text{-0.113(71)} & \text{0.123(17)} &
   \text{0.001(54)} \\
 0.469 & \text{1.688(23)} & \text{-0.023(70)} & \text{0.420(7)} & \text{-0.019(22)} & \text{-0.092(11)} & \text{0.143(6)} &
   \text{-0.014(14)} \\
 0.472 & \text{1.608(80)} & \text{-0.156(161)} & \text{0.401(19)} & \text{0.005(50)} & \text{-0.049(29)} & \text{0.143(9)} &
   \text{-0.014(26)} \\
\hline
\end{tabular}
\caption{Results for the isosinglet unpolarized generalized form factors for dataset 3.}
  \label{GFFsUnpolIsoSingl3}
\end{table}
\begin{table}[htbp]
\begin{tabular}{@{}*{10}{l}}
 \hline
 $\left.\text{-t[}\text{GeV}^2\right]$ & $A_{10}^{u+d}$ & $B_{10}^{u+d}$ & $A_{20}^{u+d}$ & $B_{20}^{u+d}$ &
   $C_{20}^{u+d}$ & $A_{30}^{u+d}$ & $B_{30}^{u+d}$ \\
 \hline
 0.000 & \text{3.000(3)} & \text{  -} & \text{0.530(12)} & \text{  -} & \text{  -} & \text{0.149(14)} & \text{  -} \\
 0.188 & \text{1.817(355)} & \text{-1.870(1.160)} & \text{0.429(84)} & \text{-0.034(281)} & \text{-0.254(389)} &
   \text{0.126(39)} & \text{-0.019(243)} \\
 0.240 & \text{2.127(26)} & \text{-0.167(128)} & \text{0.459(9)} & \text{-0.025(39)} & \text{-0.173(30)} & \text{0.150(8)} &
   \text{-0.007(29)} \\
 0.241 & \text{2.098(153)} & \text{0.165(296)} & \text{0.442(30)} & \text{0.078(76)} & \text{-0.022(89)} & \text{0.134(11)} &
   \text{0.038(43)} \\
 0.394 & \text{1.382(262)} & \text{-0.670(677)} & \text{0.319(59)} & \text{-0.007(153)} & \text{0.038(110)} & \text{0.071(23)} &
   \text{-0.110(94)} \\
 0.465 & \text{1.632(35)} & \text{-0.085(113)} & \text{0.398(10)} & \text{-0.002(35)} & \text{-0.090(18)} & \text{0.131(9)} &
   \text{0.000(24)} \\
 0.469 & \text{1.611(130)} & \text{-0.008(263)} & \text{0.383(30)} & \text{0.067(70)} & \text{-0.121(49)} & \text{0.119(11)} &
   \text{0.007(40)} \\
\hline
\end{tabular}
\caption{Results for the isosinglet unpolarized generalized form factors for dataset 4.}
  \label{GFFsUnpolIsoSingl4}
\end{table}
\begin{table}[htbp]
\begin{tabular}{@{}*{10}{l}}
 \hline
 $\left.\text{-t[}\text{GeV}^2\right]$ & $A_{10}^{u+d}$ & $B_{10}^{u+d}$ & $A_{20}^{u+d}$ & $B_{20}^{u+d}$ &
   $C_{20}^{u+d}$ & $A_{30}^{u+d}$ & $B_{30}^{u+d}$ \\
 \hline
  0.000 & \text{3.000(5)} & \text{  -} & \text{0.522(23)} & \text{  -} & \text{  -} & \text{0.153(25)} & \text{  -} \\
 0.177 & \text{1.231(560)} & \text{-1.770(2.373)} & \text{0.355(168)} & \text{-0.324(575)} & \text{-0.461(825)} &
   \text{0.189(112)} & \text{0.033(574)} \\
 0.238 & \text{2.030(48)} & \text{-0.140(211)} & \text{0.426(18)} & \text{-0.026(62)} & \text{-0.210(58)} & \text{0.154(16)} &
   \text{-0.028(55)} \\
 0.240 & \text{1.767(225)} & \text{-0.592(516)} & \text{0.378(43)} & \text{0.049(125)} & \text{-0.133(136)} & \text{0.145(20)} &
   \text{0.055(80)} \\
 0.378 & \text{  -} & \text{  -} & \text{  -} & \text{  -} & \text{  -} & \text{  -} & \text{  -} \\
 0.459 & \text{1.503(55)} & \text{-0.013(184)} & \text{0.367(18)} & \text{0.013(61)} & \text{-0.097(33)} & \text{0.129(19)} &
   \text{-0.082(45)} \\
 0.464 & \text{1.328(202)} & \text{-0.943(508)} & \text{0.360(52)} & \text{0.144(122)} & \text{-0.090(93)} & \text{0.115(22)} &
   \text{0.012(73)} \\
\hline
\end{tabular}
\caption{Results for the isosinglet unpolarized generalized form factors for dataset 5.}
  \label{GFFsUnpolIsoSingl5}
\end{table}
\begin{table}[htbp]
\begin{tabular}{@{}*{10}{l}}
 \hline
 $\left.\text{-t[}\text{GeV}^2\right]$ & $A_{10}^{u+d}$ & $B_{10}^{u+d}$ & $A_{20}^{u+d}$ & $B_{20}^{u+d}$ &
   $C_{20}^{u+d}$ & $A_{30}^{u+d}$ & $B_{30}^{u+d}$ \\
 \hline
  0.000 & \text{3.000(6)} & \text{  -} & \text{0.510(24)} & \text{  -} & \text{  -} & \text{0.164(26)} & \text{  -} \\
 0.107 & \text{2.610(460)} & \text{3.374(1.535)} & \text{0.516(89)} & \text{-0.379(418)} & \text{-0.766(673)} & \text{0.139(50)}
   & \text{-0.325(359)} \\
 0.124 & \text{2.437(35)} & \text{-0.087(285)} & \text{0.468(20)} & \text{-0.071(90)} & \text{-0.334(76)} & \text{0.131(18)} &
   \text{0.091(71)} \\
 0.125 & \text{2.465(229)} & \text{-0.247(499)} & \text{0.458(40)} & \text{0.071(139)} & \text{-0.156(202)} & \text{0.140(20)} &
   \text{-0.060(90)} \\
 0.219 & \text{2.445(482)} & \text{1.221(1.054)} & \text{0.497(93)} & \text{0.025(268)} & \text{-0.560(221)} & \text{0.159(41)}
   & \text{-0.094(165)} \\
 0.244 & \text{2.067(42)} & \text{-0.102(238)} & \text{0.412(19)} & \text{-0.045(71)} & \text{-0.119(41)} & \text{0.110(19)} &
   \text{0.043(52)} \\
 0.245 & \text{2.113(213)} & \text{-0.505(455)} & \text{0.422(40)} & \text{-0.112(124)} & \text{-0.250(103)} & \text{0.130(20)}
   & \text{0.025(84)} \\
\hline
\end{tabular}
\caption{Results for the isosinglet unpolarized generalized form factors for dataset 6.}
  \label{GFFsUnpolIsoSingl6}
\end{table}

\begin{table}[htbp]
\begin{tabular}{@{}*{10}{l}}
\hline
 $\left.\text{-t[}\text{GeV}^2\right]$ & $\widetilde A_{10}^{u-d}$ & $\widetilde B_{10}^{u-d}$ & $\widetilde A_{20}^{u-d}$ & $\widetilde B_{20}^{u-d}$ & $\widetilde A_{30}^{u-d}$ & $\widetilde B_{30}^{u-d}$ \\
 \hline
0.000 & \text{1.141(12)} & \text{  -} & \text{0.280(4)} & \text{  -} & \text{0.100(2)} & \text{  -} \\
 0.203 & \text{0.963(91)} & \text{  -} & \text{0.244(22)} & \text{  -} & \text{0.087(10)} & \text{  -} \\
 0.243 & \text{0.973(12)} & \text{9.251(403)} & \text{0.256(4)} & \text{0.740(136)} & \text{0.093(2)} & \text{0.066(71)} \\
 0.243 & \text{0.930(37)} & \text{9.159(853)} & \text{0.234(10)} & \text{0.562(286)} & \text{0.086(4)} & \text{0.090(120)} \\
 0.420 & \text{0.786(80)} & \text{4.361(1.326)} & \text{0.216(21)} & \text{0.303(343)} & \text{0.081(9)} & \text{0.129(186)} \\
 0.475 & \text{0.849(13)} & \text{6.514(227)} & \text{0.231(4)} & \text{0.497(71)} & \text{0.085(2)} & \text{0.063(39)} \\
 0.477 & \text{0.780(37)} & \text{5.515(570)} & \text{0.208(10)} & \text{0.390(161)} & \text{0.078(4)} & \text{0.017(78)} \\
\hline
\end{tabular} 
\caption{Results for the isovector polarized generalized form factors for dataset 1.}
 \label{GFFspol1}
\end{table}

\begin{table}[t]
\begin{tabular}{@{}*{10}{l}}
\hline
 $\left.\text{-t[}\text{GeV}^2\right]$ & $\widetilde A_{10}^{u-d}$ & $\widetilde B_{10}^{u-d}$ & $\widetilde A_{20}^{u-d}$ & $\widetilde B_{20}^{u-d}$ & $\widetilde A_{30}^{u-d}$ & $\widetilde B_{30}^{u-d}$ \\
 \hline
 0.000 & \text{1.136(18)} & \text{  -} & \text{0.278(6)} & \text{  -} & \text{0.100(3)} & \text{  -} \\
 0.200 & \text{1.115(124)} & \text{  -} & \text{0.236(27)} & \text{  -} & \text{0.097(14)} & \text{  -} \\
 0.242 & \text{0.967(16)} & \text{9.891(469)} & \text{0.253(5)} & \text{0.562(177)} & \text{0.095(3)} & \text{-0.028(93)} \\
 0.243 & \text{0.974(51)} & \text{10.370(1.088)} & \text{0.237(12)} & \text{0.277(298)} & \text{0.087(5)} & \text{0.007(146)} \\
 0.414 & \text{0.902(107)} & \text{6.917(1.552)} & \text{0.217(26)} & \text{-0.199(414)} & \text{0.082(11)} & \text{-0.091(246)}
   \\
 0.473 & \text{0.848(16)} & \text{6.654(253)} & \text{0.231(6)} & \text{0.442(96)} & \text{0.087(3)} & \text{-0.053(50)} \\
 0.475 & \text{0.819(49)} & \text{6.279(672)} & \text{0.220(13)} & \text{0.391(195)} & \text{0.080(5)} & \text{0.061(91)} \\
\hline
\end{tabular} 
\caption{Results for the isovector polarized generalized form factors for dataset 2.}
 \label{GFFspol2}
\end{table}

\begin{table}[t]
\begin{tabular}{@{}*{10}{l}}
\hline
 $\left.\text{-t[}\text{GeV}^2\right]$ & $\widetilde A_{10}^{u-d}$ & $\widetilde B_{10}^{u-d}$ & $\widetilde A_{20}^{u-d}$ & $\widetilde B_{20}^{u-d}$ & $\widetilde A_{30}^{u-d}$ & $\widetilde B_{30}^{u-d}$ \\
 \hline
0.000 & \text{1.171(18)} & \text{  -} & \text{0.280(7)} & \text{  -} & \text{0.101(3)} & \text{  -} \\
 0.193 & \text{0.926(117)} & \text{  -} & \text{0.212(27)} & \text{  -} & \text{0.068(15)} & \text{  -} \\
 0.241 & \text{0.967(16)} & \text{9.646(437)} & \text{0.250(5)} & \text{0.601(166)} & \text{0.092(3)} & \text{0.102(89)} \\
 0.242 & \text{0.934(49)} & \text{9.656(1.001)} & \text{0.238(12)} & \text{0.684(315)} & \text{0.089(5)} & \text{0.171(139)} \\
 0.403 & \text{0.784(99)} & \text{5.562(1.440)} & \text{0.185(25)} & \text{0.581(359)} & \text{0.074(11)} & \text{0.324(245)} \\
 0.469 & \text{0.854(18)} & \text{6.667(257)} & \text{0.223(6)} & \text{0.484(73)} & \text{0.083(3)} & \text{0.067(49)} \\
 0.472 & \text{0.827(52)} & \text{6.440(675)} & \text{0.214(14)} & \text{0.495(189)} & \text{0.079(6)} & \text{0.044(101)} \\
\hline
\end{tabular} 
\caption{Results for the isovector polarized generalized form factors for dataset 3.}
 \label{GFFspol3}
\end{table}

\begin{table}[t]
\begin{tabular}{@{}*{10}{l}}
\hline
 $\left.\text{-t[}\text{GeV}^2\right]$ & $\widetilde A_{10}^{u-d}$ & $\widetilde B_{10}^{u-d}$ & $\widetilde A_{20}^{u-d}$ & $\widetilde B_{20}^{u-d}$ & $\widetilde A_{30}^{u-d}$ & $\widetilde B_{30}^{u-d}$ \\
 \hline
0.000 & \text{1.162(32)} & \text{  -} & \text{0.257(10)} & \text{  -} & \text{0.091(6)} & \text{  -} \\
 0.188 & \text{0.931(204)} & \text{  -} & \text{0.224(49)} & \text{  -} & \text{0.082(31)} & \text{  -} \\
 0.240 & \text{0.945(26)} & \text{9.410(668)} & \text{0.228(8)} & \text{0.357(248)} & \text{0.088(5)} & \text{0.034(152)} \\
 0.241 & \text{0.934(80)} & \text{9.110(1.700)} & \text{0.222(21)} & \text{-0.132(487)} & \text{0.078(9)} & \text{-0.411(227)}
   \\
 0.394 & \text{0.710(166)} & \text{4.255(2.373)} & \text{0.184(43)} & \text{-0.140(627)} & \text{0.087(21)} & \text{0.741(418)}
   \\
 0.465 & \text{0.815(27)} & \text{6.498(379)} & \text{0.214(9)} & \text{0.457(135)} & \text{0.086(5)} & \text{0.236(82)} \\
 0.469 & \text{0.894(86)} & \text{7.418(1.076)} & \text{0.218(23)} & \text{0.616(293)} & \text{0.078(10)} & \text{0.100(154)} \\
\hline
\end{tabular} 
\caption{Results for the isovector polarized generalized form factors for dataset 4.}
 \label{GFFspol4}
\end{table}

\begin{table}[t]
\begin{tabular}{@{}*{10}{l}}
\hline
 $\left.\text{-t[}\text{GeV}^2\right]$ & $\widetilde A_{10}^{u-d}$ & $\widetilde B_{10}^{u-d}$ & $\widetilde A_{20}^{u-d}$ & $\widetilde B_{20}^{u-d}$ & $\widetilde A_{30}^{u-d}$ & $\widetilde B_{30}^{u-d}$ \\
 \hline
 0.000 & \text{1.208(60)} & \text{  -} & \text{0.244(19)} & \text{  -} & \text{0.090(11)} & \text{  -} \\
 0.177 & \text{0.557(398)} & \text{  -} & \text{0.172(96)} & \text{  -} & \text{0.189(90)} & \text{  -} \\
 0.238 & \text{0.954(46)} & \text{11.827(1.139)} & \text{0.227(14)} & \text{-0.112(427)} & \text{0.099(10)} & \text{-0.019(302)}
   \\
 0.240 & \text{1.105(152)} & \text{14.903(3.016)} & \text{0.239(37)} & \text{0.651(848)} & \text{0.088(19)} & \text{0.389(475)}
   \\
 0.378 & \text{  -} & \text{  -} & \text{  -} & \text{  -} & \text{  -} & \text{  -} \\
 0.459 & \text{0.810(48)} & \text{6.659(597)} & \text{0.208(15)} & \text{0.430(217)} & \text{0.082(10)} & \text{-0.087(168)} \\
 0.464 & \text{0.974(168)} & \text{7.646(1.750)} & \text{0.214(43)} & \text{0.533(527)} & \text{0.095(23)} & \text{0.539(307)}
   \\
\hline
\end{tabular} 
\caption{Results for the isovector polarized generalized form factors for dataset 5.}
 \label{GFFspol5}
\end{table}

\begin{table}[t]
\begin{tabular}{@{}*{10}{l}}
\hline
 $\left.\text{-t[}\text{GeV}^2\right]$ & $\widetilde A_{10}^{u-d}$ & $\widetilde B_{10}^{u-d}$ & $\widetilde A_{20}^{u-d}$ & $\widetilde B_{20}^{u-d}$ & $\widetilde A_{30}^{u-d}$ & $\widetilde B_{30}^{u-d}$ \\
 \hline
0.000 & \text{1.187(64)} & \text{  -} & \text{0.252(20)} & \text{  -} & \text{0.083(12)} & \text{  -} \\
 0.107 & \text{1.203(223)} & \text{  -} & \text{0.262(51)} & \text{  -} & \text{0.064(37)} & \text{  -} \\
 0.124 & \text{1.070(53)} & \text{17.234(1.595)} & \text{0.237(17)} & \text{0.623(598)} & \text{0.083(10)} & \text{0.551(391)}
   \\
 0.125 & \text{1.150(117)} & \text{19.686(3.278)} & \text{0.251(28)} & \text{2.059(839)} & \text{0.080(13)} & \text{-0.454(561)}
   \\
 0.219 & \text{1.333(295)} & \text{16.441(5.731)} & \text{0.276(58)} & \text{1.696(1.095)} & \text{0.086(28)} &
   \text{-0.340(878)} \\
 0.244 & \text{1.037(55)} & \text{13.098(1.010)} & \text{0.230(16)} & \text{0.808(299)} & \text{0.080(10)} & \text{0.369(186)}
   \\
 0.245 & \text{1.055(113)} & \text{13.658(2.080)} & \text{0.250(28)} & \text{1.879(558)} & \text{0.091(14)} & \text{0.745(381)}
   \\
\hline
\end{tabular} 
\caption{Results for the isovector polarized generalized form factors for dataset 6.}
 \label{GFFspol6}
\end{table}

\begin{table}[htbp]
\begin{tabular}{@{}*{10}{l}}
\hline
 $\left.\text{-t[}\text{GeV}^2\right]$ & $\widetilde A_{10}^{u+d}$ & $\widetilde B_{10}^{u+d}$ & $\widetilde A_{20}^{u+d}$ & $\widetilde B_{20}^{u+d}$ & $\widetilde A_{30}^{u+d}$ & $\widetilde B_{30}^{u+d}$ \\
 \hline
  0.000 & \text{0.623(17)} & \text{  -} & \text{0.170(6)} & \text{  -} & \text{0.061(3)} & \text{  -} \\
 0.203 & \text{0.497(79)} & \text{  -} & \text{0.138(22)} & \text{  -} & \text{0.057(14)} & \text{  -} \\
 0.243 & \text{0.537(14)} & \text{5.234(429)} & \text{0.158(5)} & \text{0.657(161)} & \text{0.058(3)} & \text{0.186(96)} \\
 0.243 & \text{0.485(31)} & \text{4.941(958)} & \text{0.144(10)} & \text{0.676(308)} & \text{0.052(4)} & \text{0.046(152)} \\
 0.420 & \text{0.425(75)} & \text{2.380(1.559)} & \text{0.136(22)} & \text{0.624(461)} & \text{0.056(10)} & \text{0.210(277)} \\
 0.475 & \text{0.477(15)} & \text{4.086(272)} & \text{0.145(6)} & \text{0.545(107)} & \text{0.055(3)} & \text{0.096(54)} \\
 0.477 & \text{0.414(38)} & \text{3.397(726)} & \text{0.122(11)} & \text{0.304(243)} & \text{0.047(5)} & \text{0.014(113)} \\
\hline
\end{tabular} 
\caption{Results for the isosinglet polarized generalized form factors for dataset 1.}
 \label{GFFspol1}
\end{table}

\begin{table}[t]
\begin{tabular}{@{}*{10}{l}}
\hline
 $\left.\text{-t[}\text{GeV}^2\right]$ & $\widetilde A_{10}^{u+d}$ & $\widetilde B_{10}^{u+d}$ & $\widetilde A_{20}^{u+d}$ & $\widetilde B_{20}^{u+d}$ & $\widetilde A_{30}^{u+d}$ & $\widetilde B_{30}^{u+d}$ \\
 \hline
   0.000 & \text{0.627(23)} & \text{  -} & \text{0.169(8)} & \text{  -} & \text{0.063(4)} & \text{  -} \\
 0.200 & \text{0.663(107)} & \text{  -} & \text{0.125(29)} & \text{  -} & \text{0.065(18)} & \text{  -} \\
 0.242 & \text{0.531(18)} & \text{5.042(498)} & \text{0.157(7)} & \text{0.471(205)} & \text{0.062(4)} & \text{0.073(123)} \\
 0.243 & \text{0.568(44)} & \text{5.822(1.170)} & \text{0.146(14)} & \text{0.163(373)} & \text{0.053(7)} & \text{0.053(179)} \\
 0.414 & \text{0.543(94)} & \text{2.378(1.841)} & \text{0.122(29)} & \text{-0.367(566)} & \text{0.048(13)} & \text{0.096(365)}
   \\
 0.473 & \text{0.480(20)} & \text{4.096(332)} & \text{0.145(8)} & \text{0.392(129)} & \text{0.054(4)} & \text{0.008(66)} \\
 0.475 & \text{0.535(47)} & \text{4.970(889)} & \text{0.131(15)} & \text{0.255(294)} & \text{0.047(7)} & \text{0.012(124)} \\
\hline
\end{tabular} 
\caption{Results for the isosinglet polarized generalized form factors for dataset 2.}
 \label{GFFspol2}
\end{table}

\begin{table}[t]
\begin{tabular}{@{}*{10}{l}}
\hline
 $\left.\text{-t[}\text{GeV}^2\right]$ & $\widetilde A_{10}^{u+d}$ & $\widetilde B_{10}^{u+d}$ & $\widetilde A_{20}^{u+d}$ & $\widetilde B_{20}^{u+d}$ & $\widetilde A_{30}^{u+d}$ & $\widetilde B_{30}^{u+d}$ \\
 \hline
 0.000 & \text{0.580(24)} & \text{  -} & \text{0.181(8)} & \text{  -} & \text{0.072(4)} & \text{  -} \\
 0.193 & \text{0.526(105)} & \text{  -} & \text{0.131(31)} & \text{  -} & \text{0.046(20)} & \text{  -} \\
 0.241 & \text{0.515(18)} & \text{5.666(508)} & \text{0.158(7)} & \text{0.409(199)} & \text{0.062(4)} & \text{0.071(114)} \\
 0.242 & \text{0.514(43)} & \text{7.032(1.094)} & \text{0.153(13)} & \text{0.299(371)} & \text{0.064(6)} & \text{0.017(183)} \\
 0.403 & \text{0.450(97)} & \text{4.042(1.708)} & \text{0.132(29)} & \text{0.480(491)} & \text{0.041(15)} & \text{-0.056(362)}
   \\
 0.469 & \text{0.431(20)} & \text{3.438(295)} & \text{0.138(7)} & \text{0.245(108)} & \text{0.055(4)} & \text{0.016(67)} \\
 0.472 & \text{0.352(48)} & \text{2.286(768)} & \text{0.150(16)} & \text{0.592(275)} & \text{0.063(8)} & \text{0.123(155)} \\
\hline
\end{tabular} 
\caption{Results for the isosinglet polarized generalized form factors for dataset 3.}
 \label{GFFspol3}
\end{table}

\begin{table}[t]
\begin{tabular}{@{}*{10}{l}}
\hline
 $\left.\text{-t[}\text{GeV}^2\right]$ & $\widetilde A_{10}^{u+d}$ & $\widetilde B_{10}^{u+d}$ & $\widetilde A_{20}^{u+d}$ & $\widetilde B_{20}^{u+d}$ & $\widetilde A_{30}^{u+d}$ & $\widetilde B_{30}^{u+d}$ \\
 \hline
  0.000 & \text{0.598(38)} & \text{  -} & \text{0.159(14)} & \text{  -} & \text{0.064(8)} & \text{  -} \\
 0.188 & \text{0.393(194)} & \text{  -} & \text{0.177(56)} & \text{  -} & \text{0.088(40)} & \text{  -} \\
 0.240 & \text{0.474(30)} & \text{4.326(808)} & \text{0.143(11)} & \text{0.200(289)} & \text{0.059(7)} & \text{-0.008(204)} \\
 0.241 & \text{0.467(71)} & \text{6.043(1.796)} & \text{0.134(22)} & \text{-0.240(548)} & \text{0.054(11)} & \text{-0.291(284)}
   \\
 0.394 & \text{0.522(177)} & \text{5.751(2.875)} & \text{0.172(52)} & \text{0.146(821)} & \text{0.049(26)} & \text{-0.101(625)}
   \\
 0.465 & \text{0.427(33)} & \text{3.081(455)} & \text{0.136(12)} & \text{0.076(174)} & \text{0.057(7)} & \text{0.106(110)} \\
 0.469 & \text{0.419(81)} & \text{3.280(1.215)} & \text{0.148(27)} & \text{0.494(432)} & \text{0.064(12)} & \text{0.118(225)} \\
\hline
\end{tabular} 
\caption{Results for the isosinglet polarized generalized form factors for dataset 4.}
 \label{GFFspol4}
\end{table}

\begin{table}[t]
\begin{tabular}{@{}*{10}{l}}
\hline
 $\left.\text{-t[}\text{GeV}^2\right]$ & $\widetilde A_{10}^{u+d}$ & $\widetilde B_{10}^{u+d}$ & $\widetilde A_{20}^{u+d}$ & $\widetilde B_{20}^{u+d}$ & $\widetilde A_{30}^{u+d}$ & $\widetilde B_{30}^{u+d}$ \\
 \hline  
 0.000 & \text{0.519(68)} & \text{  -} & \text{0.123(26)} & \text{  -} & \text{0.049(16)} & \text{  -} \\
 0.177 & \text{0.003(440)} & \text{  -} & \text{-0.092(122)} & \text{  -} & \text{0.268(131)} & \text{  -} \\
 0.238 & \text{0.469(51)} & \text{7.022(1.388)} & \text{0.126(19)} & \text{-0.146(558)} & \text{0.065(14)} & \text{0.182(426)}
   \\
 0.240 & \text{0.604(146)} & \text{8.708(3.368)} & \text{0.094(42)} & \text{0.199(1.052)} & \text{0.031(25)} &
   \text{-0.209(618)} \\
 0.378 & \text{  -} & \text{  -} & \text{  -} & \text{  -} & \text{  -} & \text{  -} \\
 0.459 & \text{0.321(56)} & \text{2.498(733)} & \text{0.113(22)} & \text{0.069(314)} & \text{0.053(13)} & \text{-0.143(228)} \\
 0.464 & \text{0.536(170)} & \text{3.063(2.143)} & \text{0.113(56)} & \text{0.183(767)} & \text{0.033(29)} & \text{-0.064(436)}
   \\
\hline
\end{tabular} 
\caption{Results for the isosinglet polarized generalized form factors for dataset 5.}
 \label{GFFspol5}
\end{table}

\begin{table}[t]
\begin{tabular}{@{}*{10}{l}}
\hline
 $\left.\text{-t[}\text{GeV}^2\right]$ & $\widetilde A_{10}^{u+d}$ & $\widetilde B_{10}^{u+d}$ & $\widetilde A_{20}^{u+d}$ & $\widetilde B_{20}^{u+d}$ & $\widetilde A_{30}^{u+d}$ & $\widetilde B_{30}^{u+d}$ \\
 \hline
  0.000 & \text{0.484(72)} & \text{  -} & \text{0.143(25)} & \text{  -} & \text{0.040(15)} & \text{  -} \\
 0.107 & \text{0.732(209)} & \text{  -} & \text{0.184(59)} & \text{  -} & \text{0.045(41)} & \text{  -} \\
 0.124 & \text{0.493(54)} & \text{11.813(1.845)} & \text{0.139(20)} & \text{0.832(668)} & \text{0.049(14)} & \text{0.673(521)}
   \\
 0.125 & \text{0.722(109)} & \text{16.409(3.651)} & \text{0.154(34)} & \text{0.730(1.118)} & \text{0.026(18)} &
   \text{-0.717(732)} \\
 0.219 & \text{0.867(246)} & \text{10.112(5.404)} & \text{0.226(61)} & \text{0.704(1.398)} & \text{0.046(35)} &
   \text{0.104(1.269)} \\
 0.244 & \text{0.500(58)} & \text{7.939(1.175)} & \text{0.141(20)} & \text{1.075(372)} & \text{0.048(13)} & \text{0.348(271)} \\
 0.245 & \text{0.544(118)} & \text{6.702(2.649)} & \text{0.176(38)} & \text{1.910(815)} & \text{0.067(18)} & \text{1.500(526)}
   \\
\hline
\end{tabular} 
\caption{Results for the isosinglet polarized generalized form factors for dataset 6.}
 \label{GFFspol6}
\end{table}

% --------------------------------------------------------------------------
%
%  Bibliography
%
% --------------------------------------------------------------------------

\bibliography{LHPC_GPD_2007_v2}

% --------------------------------------------------------------------------
% 
% End of document
% 
% --------------------------------------------------------------------------

\end{document}